\documentclass[a4paper,noindent]{JHEP3}

\usepackage{epsfig, subfigure}
\usepackage{graphicx}
\usepackage{amsmath}

\usepackage{cite}

\usepackage{feynarts}

\newcommand{\nn}{\nonumber}

\renewcommand{\Re}{\mbox{Re}}
\providecommand{\Li}[1]{\text{Li}\,_2\left(#1\right)}

\def \drbar{\overline{\text{DR}}}
\def \os{\text{OS}}

\renewcommand{\eqref}[1]{Eq.~(\ref{#1})}

\newcommand{\rb}[1]{\raisebox{1.5ex}[-1.5ex]{#1}}
\newcommand{\cf}{\textit{cf. }}
\newcommand{\ie}{\textit{i.\,e. }}

\graphicspath{{pictures/}}
\newcommand{\image}[3]{
		\includegraphics[width=#1,#2]{#3} 
}

\newcommand{\gluino}{\tilde{g}}
\newcommand{\squark}{\tilde{q}}
\newcommand{\squarka}{\tilde{q}_a}

\newcommand{\ul}{\,\tilde{u}_L}
\newcommand{\ur}{\,\tilde{u}_R}
\newcommand{\dl}{\,\tilde{d}_L}
\newcommand{\dr}{\,\tilde{d}_R}

\providecommand{\neut}[1]{\tilde{\chi}_{#1}^0}

\makeatletter
\renewcommand\appendix{\par
  \setcounter{section}{0}%
  \setcounter{subsection}{0}%
  \renewcommand\thesection{\@Alph\c@section}
  \setcounter{figure}{0}%
  \setcounter{table}{0}%
  \renewcommand\thefigure{\thesection.\@arabic\c@figure}
  \renewcommand\thetable{\thesection.\@arabic\c@table}
  }
\makeatother

\title{Electroweak contributions to squark--gluino \\ production at the LHC}

\author{Wolfgang Hollik, Edoardo Mirabella, and Maike K. Trenkel \\
Max-Planck-Institut f\"ur Physik,
 F\"ohringer Ring 6, D-80805 M\"unchen, Germany }

\abstract{We calculated the electroweak contributions to the hadronic production 
of a squark in association with a gluino within the Minimal Supersymmetric 
Standard Model (MSSM). Presented are complete next-to-leading order 
electroweak (NLO EW) corrections at $\mathcal{O}(\alpha_s^2 \alpha)$, 
which include real photon and real quark radiation processes. Also considered are 
photon induced tree level $\mathcal{O}(\alpha_s \alpha)$ contributions.
}

\preprint{MPP-2008-54}

\begin{document}

\section{Introduction}
Supersymmetry (SUSY)~\cite{Wess, Wess2} is one of the most appealing scenarios 
for physics beyond the Standard Model (SM). Besides the well known features 
such as naturalness of the mass hierarchy, the possibility of grand unification and
the existence of a cold dark matter candidate, SUSY can explain, in contrast to the SM,  
the measured value of the anomalous magnetic moment of the muon~\cite{MuonEX, MuonEX2} 
and the observed cold dark matter density~\cite{WMAP}. 
In particular, the minimal supersymmetric extension  of the SM (MSSM)~\cite{MSSM, MSSM2, MSSM3} 
provides a fit to precision electroweak data and $B$-physics observables with a $\chi^2$ 
probability comparable to that of the SM~\cite{Ellis:2007fu,
  Buchmueller:2007zk} and compatible with a light Higgs boson in a natural way.
\medskip

If SUSY is an answer to the hierarchy problem then at least some of the SUSY 
particles have to be discovered at the Large Hadron Collider (LHC).
In a recent analysis~\cite{Buchmueller} it has been shown that the $95\%$ confidence level 
area in the $(m_0, m_{1/2})$ plane of the constrained MSSM (CMSSM) 
lies in the region that will be explored with $1~\mbox{fb}^{-1}$ of integrated luminosity 
at $14$~TeV. 
Among the potential SUSY discovery channels, certainly the direct production of  
strongly interacting SUSY particles with their large cross sections will play a key role.
Many searches for squarks and gluinos, the SUSY partners of SM quarks and gluons, 
have thus been performed at high energy colliders (recent results from D0 and CDF 
are reported e.\,g. in \cite{Shamim:2007yy,D'Onofrio:2007pq, CDFnote9229}). 
Studies for the LHC see the possibility of an early SUSY discovery 
with $1~\mbox{fb}^{-1}$ for inclusive multijet plus missing energy 
final states~\cite{SUSY@lhc,SUSY@lhc2}, provided that squark and gluino masses 
are not too heavy (\ie below 2~TeV). Also complementary approaches that avoid signatures 
involving missing energy have been proposed~\cite{Randall}. 
\medskip

Owing to the large interest in squarks and gluinos, theoretical predictions of
the leading order (LO) 
production cross sections were already published in the 1980's~\cite{Tree, Tree2, Tree3, Tree4, Tree5}. 
Important next-to-leading order (NLO) 
QCD calculations~\cite{Beenakker:1996ch, Beenakker1997} could reduce 
the dependence on factorization and renormalization scale and revealed 
large corrections of typically 20-30\%. They can be included in analyses via the public code 
{\tt Prospino}~\cite{Prospino}.
Only recently also NLO EW corrections were considered, for top-squark pair production 
in~\cite{Hollik:2007wf,Beccaria:2008mi} and for pair production of squarks of the first generations 
in~\cite{Hollik:2008}. Both processes receive further EW contributions from tree-level processes,
photon induced at $\mathcal{O}(\alpha_s \alpha)$ and 
from $q\bar{q}$ annihilation at $\mathcal{O}(\alpha^2)$
\cite{Bornhauser:2007bf,Bozzi:2007me,Hollik:2007wf,Hollik:2008}. For light flavor squarks, the 
tree-level QCD and EW amplitudes can interfere and give sizable contributions to the cross sections.
\medskip

In this paper, we consider the associated production of
squarks and gluinos and study the EW contribution.
We restrict the discussion to (anti-)squarks of the first two generations,
\begin{equation}
	PP \to \tilde{g} \tilde{q}_{a} +X ,~~%
	PP \to \tilde{g} \tilde{q}_{a}^{*} +X ,~~~~%
	q=u,d,c,s;~~a=L,R.
\end{equation}
The production of $\tilde{t}\,\gluino$ is suppressed due to the vanishing parton density of
top-quarks inside protons, while  $\tilde{b}\,\gluino$ production is suppressed by 
the  bottom-quark parton distribution function. 
Furthermore, bottom-squarks (resp. their decay products) will be experimentally 
distinguishable from squarks of the first two generations~\cite{MassR1, MassR2,Kawagoe:2004rz}.
%
The outline of our paper is as follows. In Section~2, we recall the LO cross 
section at the partonic and the hadronic level and introduce some basic notations.
The EW contribution is discussed in detail in Section~3.
In Section~4, we present numerical results for the hadronic cross sections 
and distributions for squark--gluino production at the LHC. 
Finally, a list of the Feynman diagrams and the input parameters used 
in the numerical analysis are given in the Appendix.

\section{LO cross section and conventions}

At hadron colliders, the LO contribution to the production of a gluino in association 
with an (anti)squark $\tilde{q}_{a}^{(*)}$ is QCD based
and is related to the following partonic processes:
\begin{align}
\begin{split}
	g(p_1)~q(p_2)~\rightarrow~\tilde{g}(k_1)~\tilde{q}_{a}(k_2),
\qquad &
	g(p_1)~\bar{q}(p_2)~\rightarrow~\tilde{g}(k_1)~\tilde{q}_{a}^{*}(k_2).
\label{eq_processes}
\end{split}
\end{align}
Due to CP symmetry the unpolarized cross sections of these two processes are equal; so in the following we will 
refer to the first partonic process only. The corresponding Feynman diagrams are shown in Fig.~\ref{fig_borndiags} in Appendix~\ref{app_Fey}.

Since the quarks of the first two generations are treated as massless, 
in the case of the squarks of the first two generations
 weak eigenstates are also mass eigenstates and we will distinguish squarks with same flavor 
by means of their chiralities, $\tilde{q}_a=\tilde{q}_L,\tilde{q}_R$.
Furthermore, the masses of the squarks 
of the second generation coincide with those in the first generation. 
We denote the mass of squark $\tilde{q}_a$ by $m_{\squarka}$, and
the gluino mass by $m_{\gluino}$.

We parameterize the cross sections in terms of the following kinematical variables, 
\begin{align}
\begin{split}
	\hat{s}  = (p_1 + p_2)^2, 
	\qquad \hat{t} &= (p_1 - k_{1})^2,
	\qquad \hat{u}  = (p_1 - k_{2})^2,
\\
	\hat{t}_{\gluino/\squarka} &= \hat{t} - m_{\gluino/\squarka}^2, \quad
	\hat{u}_{\gluino/\squarka} = \hat{u} - m_{\gluino/\squarka}^2,
\label{eq_mandelstams}
\end{split}
\end{align}
with $\hat{s} + \hat{t}_{\gluino/\squarka} + \hat{u}_{\squarka/\gluino} = 0$.
As a notation, we introduce the convention $d\hat \sigma^{a,b}_{X}$ for a cross section of
the partonic process $X$ at a given order $\mathcal{O}(\alpha_s^a \alpha^b)$ in the 
strong and electroweak coupling constants.

The differential partonic cross section for the process $gq \to \gluino\squarka$,
\begin{align}
        d\hat \sigma^{2,0}_{gq\to\gluino\tilde{q}_a} (\hat s) = 
        \frac{d\hat{t}}{16 \pi \hat s^2}\,\overline{\sum} 
        \bigl\lvert{\cal M}_{gq \to \gluino \tilde{q}_a}^0 (\hat s,\hat t,\hat u) \bigr\rvert^2 \; , 
\end{align}
expressed in terms of the squared spin- and color-averaged lowest-order
matrix element (Fig.~\ref{fig_borndiags} in Appendix B),
can be written as follows~\cite{Beenakker:1996ch},
\begin{align}
\begin{split}
        \overline{\sum} \bigl\lvert{\cal M}_{gq \to \gluino \squarka}^0\bigr\rvert^2 & =
                \frac{1}{4} \cdot \frac{1}{24}\cdot 16 \pi^2 \alpha_s^2
                \left[ C_0\left( 1-2\frac{\hat{s}\,\hat{u}_{\squarka}}{\hat{t}_{\gluino}^2}\right)-C_K  \right]
\\
        &\times        \left[ - \frac{\hat{t}_{\gluino}}{\hat s}
			+ \frac{ 2 (m_{\gluino}^2 - m_{\squarka}^2 ) \hat{t}_{\gluino}}{\hat{s} \hat{u}_{\squarka}}
		        \left( 1+ \frac{m_{\gluino}^2}{\hat{t}_{\gluino}}
				+ \frac{m_{\squarka}^2}{\hat{u}_{\squarka}}
                              \right)\right] ,
\end{split}
\end{align}
with the $SU(3)$ color factors $C_0 = N(N^2-1)=24$ and 
$C_K = (N^2-1)/N=8/3$ for $N=3$.
At the hadronic level, the cross section 
is obtained from the partonic cross section by the convolution
\begin{align}
	d\sigma^{2,0}_{AB\to\gluino\squarka}(S) =  
	 \int_{\tau_0}^1  \! d\tau \,  
	\frac{d \mathcal{L}^{AB}_{gq}}{d \tau}  \,
	 d\hat{\sigma}^{2,0}_{gq\to\gluino\squarka}(\hat s), 
\label{eq_sigmaLO}
\end{align}
where $\tau = \hat{s}/S$, $S$ $(\hat s)$ is the hadronic (partonic) center-of-mass energy squared,
and $\tau_0= (m_{\gluino}+m_{\squarka})^2/ S$ is the production threshold.
The parton luminosity 
\begin{align}
       \frac{d\mathcal{L}^{AB}_{ab}}{d\tau} = 
                \frac{1}{1+\delta_{ab}} \,  
		    \int_{\tau}^1 \! \frac{dx}{x} 
                \, \biggl[      f_{a/A}\bigl(x, \mu_F\bigr) \,
                                f_{b/B}\Bigl(\frac{\tau}{x}, \mu_F\Bigr)\,
                +               f_{b/A}\Bigl(\frac{\tau}{x}, \mu_F\Bigr) \,
                                f_{a/B}\bigl(x, \mu_F\bigr) \biggr].
\label{eq_luminosity}
\end{align}
contains the parton distribution functions (PDFs), where
$ f_{a/A}(x, \mu_F)$  gives the probability of finding a parton $a$
in the hadron $A$ carrying a fraction $x$ of the hadron's momentum at the factorization scale $\mu_F$.
At the LHC, both $A$ and $B$ are protons.

\section{Electroweak contributions}
\label{sec_nlocontributions}

In contrast to squark pair and top-squark pair production~\cite{Bornhauser:2007bf,Hollik:2008,Hollik:2007wf},
which allow for $q\bar{q}$ initial states at LO, gluino--squark final states cannot be produced 
at $\mathcal{O}(\alpha^2)$.
At EW NLO, gluino--squark production comprises virtual corrections 
and real photon radiation at $\mathcal{O}(\alpha_s^2 \alpha)$.
Further $\mathcal{O}(\alpha_s^2 \alpha)$ contributions arise from
interference of EW and QCD real quark radiation diagrams.  
We also consider photon induced gluino--squark production at the tree level 
(Fig.~\ref{fig_quarkphoton} of Appendix B), formally of different order, 
but expected to be comparable in size~\cite{Hollik:2007wf,Hollik:2008}.

The complete EW contribution to the hadronic cross section 
is obtained from the corresponding partonic cross sections by convolution and summation as follows,
\begin{align}
\begin{split}
	d\sigma^{\rm EW }_{PP\to\gluino\squarka+X}(S) 
	= &  \int_{\tau_0}^1  \! d\tau \,  \Bigg\lbrace 
	\frac{d \mathcal{L}^{PP}_{gq}}{d \tau}  \,
	\Bigl[ d\hat{\sigma}^{2,1}_{gq\to\gluino\squarka}(\hat s) + 
	d\hat{\sigma}^{2,1}_{gq\to\gluino\squarka\gamma}(\hat s) \Bigr]
	+
	\frac{d \mathcal{L}^{PP}_{\gamma q}}{d \tau}  \,
	 d\hat{\sigma}^{1,1}_{\gamma q\to\gluino\squarka}(\hat s)
\\[.5ex]
	& +  \sum_{q_i=u,d,c,s,\bar{d},\bar{c},\bar{s}}
	\frac{d \mathcal{L}^{PP}_{q q_i}}{d \tau}  \,
	 d\hat{\sigma}^{2,1}_{q q_i\to\gluino\squarka q_i}(\hat s)  
	+
	\sum_{q_i= u,d,c,s}
	\frac{d \mathcal{L}^{PP}_{q_i \bar{q}_i}} {d \tau}  \,
	  d\hat{\sigma}^{2,1}_{q_i \bar{q_i}\to\gluino\squarka \bar{q}}(\hat s)
	\Bigg\rbrace \, ,
\label{eq_masterformula}
\end{split}
\end{align} 
where the respective parton luminosities refer to \eqref{eq_luminosity}. 
We will discuss all of the partonic cross sections in the following subsections.

For the treatment of the Feynman diagrams and corresponding amplitudes 
we make use of
{\tt FeynArts~3.3}~\cite{Kublbeck:1990xc,Hahn:2000kx,Hahn:2001rv} and
{\tt FormCalc~5.3}
with {\tt LoopTools~2.2}~\cite{Hahn:1998yk,Hahn:2006qw}.
Infrared (IR) and collinear singularities are treated using mass regularization, {\it i.\,e.}
IR singularities are regularized by a small photon mass $\lambda$, and 
the masses of the light quarks are kept in collinearly singular integrals.

\subsection{Virtual corrections}

The first class of NLO contributions of EW origin are the virtual corrections,
\begin{align}
d\hat \sigma^{2,1}_{gq\to\gluino\squarka} (\hat s) 
= 
	\frac{d \hat t }{16 \pi \hat{s}^2}\; 
	 \overline{\sum}\; 2\, \mathfrak{Re}\, 
	\Big\lbrace \mathcal{M}^{0}_{gq \to \gluino\squarka}\,
	\mathcal{M}^{1*}_{gq \to \gluino\squarka}  \Big\rbrace,
\end{align} 
where $\mathcal{M}^1$ is the one-loop amplitude with EW insertions in the 
QCD-based $gq$ diagrams, leading to the self energy, vertex, box, 
and counter term diagrams shown in the Appendix~\ref{app_Fey}, 
Fig.~\ref{fig_vertexdiags} and Fig.~\ref{fig_counterterms}.
The explicit expressions of the counter terms and the required renormalization 
constants can be found in Ref.~\cite{Hollik:2008}. 
Both the quark and the squark sector require
renormalization.~\footnote{Different to 
Ref.~\cite{Hollik:2008}, 
we do not need to renormalize the gluon here. 
Gluino--squark production at LO can only proceed via QCD diagrams and 
thus no interference of EW born and QCD one-loop diagrams arises.}
The renormalization of the quark sector is performed
in the on-shell scheme as described in Ref.~\cite{Denner:1991kt}; 
squark renormalization is done
in close analogy to \cite{Hollik:2003jj,Heinemeyer:2004xw}. 
Here, in the limit of no L--R mixing, the independent parameters 
for a given squark isospin doublet are the masses of the two up-type 
squarks $\tilde{u}_{L,R}$ and  the mass of the right handed down-type squark $\tilde{d}_R$ 
(see also the discussion in Appendix~\ref{app_input}).

\subsection{Real photon radiation}

To compensate IR singularities in the virtual corrections,
 we have to include the tree level photon bremsstrahlung process, \cf the diagrams in 
Fig.~\ref{fig_realcorrs},
\begin{align}
\begin{split}
	g(p_1)~q(p_2)~\rightarrow~\tilde{g}(k_1)~\tilde{q}_{a}(k_2)~\gamma(k_3).
\label{eq_process23}
\end{split}
\end{align}
The integral over the photon phase space is divergent in the soft region 
($k^0_3\to 0$) and in the collinear region ( $k_3\!\cdot\! p_2 \to 0 $). 
The extraction of such singularities has been performed using two methods,   
phase space slicing~\cite{Denner:1991kt,Harris:2001sx} and dipole
 subtraction~\cite{Catani:1996jh,Catani:1996jh2,Catani:1996jh3,Dittmaier:1999mb,Dittmaier:2008md}. 
\medskip

In the phase space slicing approach,
 the phase space regions where the squared amplitude becomes singular are excluded 
from the numerical integration  by applying a cut $\Delta E =  \delta_s \sqrt{\hat s}/2$
 on the photon energy 
and a cut $\delta_\theta$ on the cosine of the angle between the photon and the quark.
The integral over the singular regions can be performed analytically in the eikonal approximation. 

In the soft region, we can exploit the results quoted in Ref.~\cite{Denner:1991kt}.
Written in an analogous way to \cite{Hollik:2007wf}, the soft part of the differential cross section is
\begin{align}
        d\hat{\sigma}^{2,1}_{gq\to\gluino\squarka\gamma} (\hat s) \bigg\lvert_{\rm soft} & = 
		\frac{\alpha}{\pi}\,\Big(
                 e_q^2\,\delta_{\rm{soft}}^{\rm{in}}  
		+ e_q^2\,\delta_{\rm{soft}}^{\rm{fin}} 
                + 2 e_q^2 \, \delta_{\rm{soft}}^{\rm{int}} \Big) 
                        \,d\hat{\sigma}^{2,0}_{gq\to\gluino\squarka} (\hat s)\,,
\label{eq_sigma_soft_phot}
\end{align}
with universal factors, $\delta_{\rm{soft}}^{\rm{in}, \rm{fin}, \rm{int}}$ that
 refer to the initial state radiation, final state radiation, or
interference of initial and final state radiation, respectively, 
\begin{align}
\begin{split}
        \delta_{\rm soft}^{\rm in} = & 
		 \ln\frac{\lambda^2}{\hat{s}} - \ln \delta_s^2 
		+ \ln\frac{\hat{s}}{m_q^2},
\\[.5ex]
        \delta_{\rm soft}^{\rm fin} = & 
		 \ln\frac{\lambda^2}{\hat{s}} - \ln \delta_s^2 
                + \frac{1}{\beta}\, \ln \left(\frac{1+\beta}{1-\beta}\right),
 \\[1.5ex]
        \delta_{\rm soft}^{\rm int} = &  
		\left[ \ln\frac{\lambda^2}{\hat{s}} - \ln \delta_s^2 \right] \,
                \ln\left( \frac{- \hat{t}_{\squarka} }{m_q \, m_{\squarka}} \right)
		+ \frac{1}{4} \ln^2 \frac{\hat{s}}{m_{\squarka}^2}
                + \Li{1-\frac{\hat s}{m^2_{q}}}
\\
		&- \frac{1}{4} \ln^2 \frac{1-\beta}{1+\beta}
                - \Li{1-\frac{p_2^0\, k_2^0}{p_2 k_2}  (1+\beta)}
                - \Li{1+\frac{p_2^0\, k_2^0}{p_2 k_2} (1-\beta)} .
\label{eq_delta_soft}
\end{split}
\end{align}
Here, $e_q$ is the electric charge of the quark and the squark, and
        $\beta = \sqrt{ 1 - m_{\squarka}^2/(k_2^0)^2}$.

In the collinear region the differential cross section reads~\cite{Baur:1998kt,Dittmaier:2001ay}
\begin{align}
\begin{split}
	d\hat \sigma^{2,1}_{gq\to\gluino\squarka\gamma} (\hat s) \bigg\lvert_{\rm coll.} =
	 \frac{\alpha}{\pi}\, e^2_q \int^{1-\delta_s}_{z_0} d z \,\,
	 \kappa_{\rm coll.}(z, \hat s) \,\,
	d\hat \sigma^{2,0}_{gq\to\gluino\squarka} (z \hat s),
\label{eq_coll}
\end{split}
\end{align}
where $z_0 = (m_{\gluino}+m_{\squarka})^2/ \hat s$ 
and
\begin{align}
\begin{split}
	\kappa_{\rm coll.}(z, \hat s) = 
	\frac{1}{2} P_{qq}(z) \left[
		\ln \left( \frac{\hat{s}}{m_q^2} \frac{ \delta_\theta}{2} \right) -1\right]
	 + \frac{1}{2}(1-z),
\end{split}
\end{align}
with the splitting function $P_{qq}(z) = (1+z^2)/ (1-z)$. 
\medskip

The basic subtraction method is to add to and subtract from 
the squared amplitude a function with the same behavior in the singular 
region but simple enough to be analytically integrated over the photon phase space. 
General expressions for those functions are available in literature; 
we use the expressions of the dipole subtraction formalism in Ref.~\cite{Dittmaier:1999mb}, 
within mass regularization.
The integral over the subtracted cross section is regular and can be performed numerically. 
  
The comparison between the two methods is illustrated in Fig.~\ref{fig_comparison}; 
the two methods are in mutual numerical agreement. 

\FIGURE[t]{ 
	\image{7.31cm}{trim= 0 15 0 0}{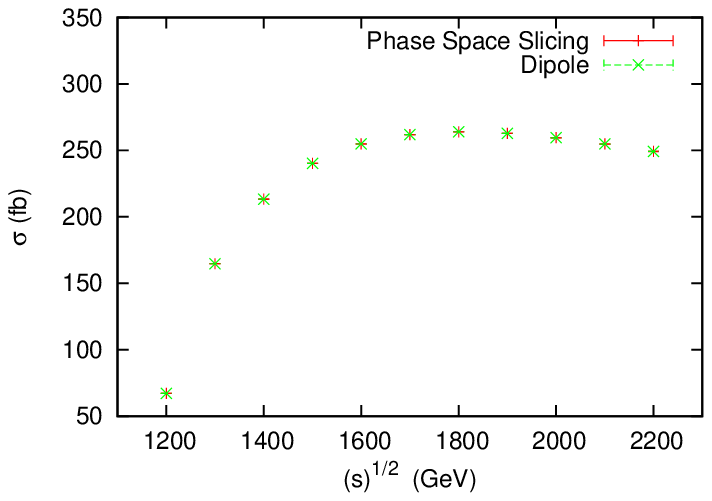}%
	\image{7.31cm}{trim= 0 15 0 0}{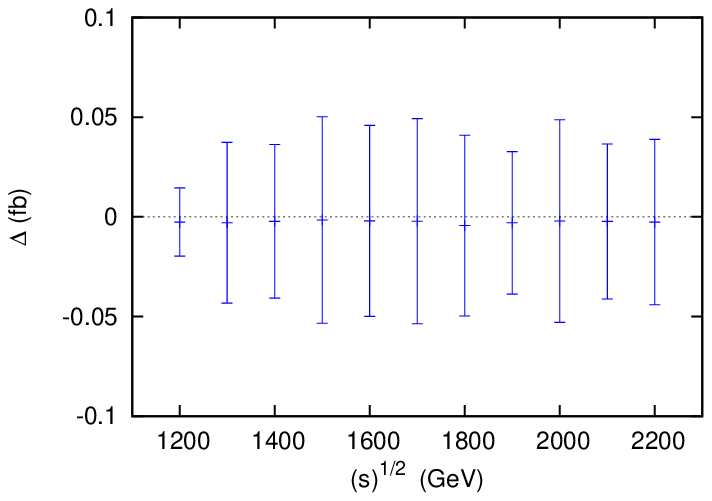}%
        \caption{Left: Comparison of the lowest order partonic cross section for the process 
	$gu \to \gluino \ul \gamma$ using the phase space slicing or dipole subtraction method.
	Right: Difference $\Delta = \sigma^{\rm dipole} - \sigma^{\rm slicing}$ as a function of 
	the partonic energy. The error bars represent the integration uncertainty. The SUSY 
	parameters are those of the SPS1a$'$ point~\cite{AguilarSaavedra:2005pw}, see also 
	Appendix~\ref{app_input}.}
        \label{fig_comparison}
}

\medskip

After adding virtual and real corrections, the mass singularity related to \eqref{eq_coll} 
does not cancel and has to be absorbed into the quark parton density function (PDF) 
choosing a factorization scheme. 
This can formally be achieved by relating the lowest order PDF $f_{a/A}(x)$ 
for parton $a$ in hadron $A$ to the experimentally accessible distribution $f_{a/A}(x,\mu_F)$
at NLO QED as~\cite{Baur:1998kt,Wackeroth:1996hz,Diener:2003ss}
\begin{align}
        f_{a/A}(x)  \rightarrow  \quad&
        f_{a/A}(x,\mu_F) \, \Big(1+\frac{\alpha}{\pi}\,e_q^2\, \kappa_{\rm soft}^{\rm PDF}\Big)
        + \frac{\alpha}{\pi}\,e_q^2\, \int_x^{1-\delta_s}\! \frac{dz}{z} \,
                 f_{a/A}\Bigl( \frac{x}{z}, \mu_F\Bigr) \,\kappa_{\rm coll.}^{\rm PDF}(z) \,,
\label{eq_pdfdef}
\end{align}
where, in the $DIS$ factorization scheme,
\begin{align}
\begin{split}
        \kappa_{\rm soft}^{\rm PDF} &=  \frac{5}{4} + \frac{\pi^2}{6}
		 + \frac{7}{4}\ln \delta_s + \frac{1}{2} \ln^2\delta_s 
                + \ln\biggl(\frac{m_q^2}{\mu_F^2}\biggr)\,
                        \left[\frac{3}{4} + \ln\delta_s \right],         
\\[1ex]
        \kappa_{\rm coll.}^{\rm PDF}(z) &=  \frac{1}{2} P_{qq}(z) \biggl[
	            \ln \biggl( \frac{m_q^2\, (1-z) z}{\mu_F^2} \biggr) + 1 \biggr] 
	          +\frac{3}{4 (1-z) } - z -\frac{3}{2}.
\label{eq_kappafact}
\end{split}
\end{align}
The actual effect of the redefinition (\ref{eq_pdfdef}) is to induce an extra term in
\eqref{eq_masterformula} via \eqref{eq_luminosity}. 
This term exactly cancels the mass singularity owing to collinear photon radiation,
as can be seen following the guideline of Ref.~\cite{Hollik:2007wf}.

\subsection{Real quark radiation}
\label{subsec_qqchannels}

For each production process of a gluino in association with a squark $\squarka$ of a given chirality and flavor,
there are eleven (quark--quark or quark--anti-quark induced) subprocesses with an additional 
real quark or anti-quark in the final state:
\begin{align}
\begin{split} 
	q(p_1)~q_i(p_2)~& \rightarrow~\tilde{g}(k_1)~\tilde{q}_{a}(k_2)~q_i(k_3)
	\qquad \text{for} \quad q_i= u, d, c, s, \bar{d}, \bar{c}, \bar{s};
\\
	q_i(p_1)~\bar{q}_i(p_2)~& \rightarrow~\tilde{g}(k_1)~\tilde{q}_{a}(k_2)~\bar{q}(k_3)
	\qquad\, \text{for} \quad q_i= u, d, c, s.
\end{split}
\label{eq_processesqq}
\end{align}
These tree level processes give an IR and collinear finite contribution of order  
$\mathcal{O}(\alpha_s^2\alpha)$ through the interference between 
the EW diagrams in Fig.~\ref{fig_qqbarchannels}~a and the QCD diagrams
 in Fig.~\ref{fig_qqbarchannels}~b
and between those in Fig.~\ref{fig_qqchannels}~a and Fig.~\ref{fig_qqchannels}~b.

In specific SUSY scenarios, internal gauginos or squarks can be on-shell.
The poles are regularized introducing the particle width in the corresponding propagator.
%
If both EW and QCD diagrams provide intermediate on-shell squarks, 
the non-vanishing interference contribution corresponds to the production of a squark pair 
at order $\mathcal{O}(\alpha_s \alpha)$  with subsequent decay of one of the two squarks,
\begin{align}
\begin{split} 
	q~q_i~& \rightarrow~\tilde{q}_a~\tilde{q}_i, \quad\,  
	\tilde{q}_i \rightarrow \tilde{g}~q_i  \,;
\\
	q_i~\bar{q}_i~& \rightarrow~\tilde{q}_a~\tilde{q}_a^*,  \quad
	 \tilde{q}_a^{*} \rightarrow \tilde{g}~\bar{q}   \, .
\end{split}
\label{eq_processesqq2}
\end{align}
To avoid double counting, these 
resonating squark contributions have to be subtracted~\cite{Beenakker:1996ch}. 
The pole term has thereby been isolated in the narrow width approximation.

\subsection{Photon induced gluino--squark production}

As an independent production channel, we also consider the photon--gluon induced subclass of 
gluino--squark production, as shown in Fig.~\ref{fig_quarkphoton}.
Photon induced processes do not contribute at leading order at the hadronic level, 
owing to the non-existence of a photon distribution inside the proton. 
But the inclusion of NLO QED effects in the evolution of the PDFs leads to a
non-zero photon density in the proton and thus to non-zero hadronic contributions. 
These are formally of different order than the $\mathcal{O}(\alpha_s^2 \alpha)$
corrections in \eqref{eq_masterformula}, but the diagrams contribute at tree level 
to the same final state and can be important~\cite{Hollik:2007wf,Hollik:2008}.

The partonic differential cross section for the photon induced gluino--squark production reads
\begin{align}
 	d\hat \sigma_{\gamma q \to \gluino \squarka}^{1,1} (\hat s) &=
        \frac{d\hat{t}}{16 \pi \hat s^2}\,\overline{\sum} 
        \bigl\lvert{\cal M}_{\gamma q \to \gluino \squarka}^0 (\hat s,\hat{t},\hat{u}) \bigr\rvert^2 \; ,
\label{eq_sigma_phot_ind}
\\[1ex]
        \overline{\sum} \bigl\lvert{\cal M}_{\gamma q \to \gluino \squarka}^0\bigr\rvert^2  & = 
                \frac{1}{4} \cdot \frac{1}{3}\cdot 32 \pi^2 \alpha_s \alpha \, e_q^2 
                \, N C_K\,
	      \left[ - \frac{\hat{t}_{\gluino}}{\hat s}
			+ \frac{ 2 (m_{\gluino}^2 - m_{\squarka}^2 ) \hat{t}_{\gluino}}{\hat{s} \hat{u}_{\squarka}}
		        \left( 1+ \frac{m_{\gluino}^2}{\hat{t}_{\gluino}}
				+ \frac{m_{\squarka}^2}{\hat{u}_{\squarka}} \right)\right],
\nn
\end{align}
expressed in terms of the reduced Mandelstam variables, \eqref{eq_mandelstams}.

Due to color conservation, photon--gluon induced partonic processes are only possible in combination 
with an additionally radiated quark and thus represent contributions of higher order. 
Since they are suppressed by the PDF of the photon compared to the bremsstrahlung 
processes \eqref{eq_process23} and \eqref{eq_processesqq}, 
we do not include them in our discussion here.

\section{Numerical Results}
\label{sec_Numerics}

We illustrate the numerical results in terms of the SPA mSUGRA scenario SPS1a$'$ 
 \cite{AguilarSaavedra:2005pw}. The input parameters are listed in 
Appendix \ref{app_input}. 
We present results both for the production of left- and right-handed, up- and down-type squarks separately
and for the inclusive production. 
We have chosen a common scale for 
factorization and renormalization, $\mu_F = \mu_R = 1$~TeV. 
We introduce the following conventions for the discussion of the results.
\begin{itemize}
	\item We will analyze the three different gauge invariant, IR and collinear finite subsets of the 
	EW contributions described in the previous section. 
	The sum of the virtual corrections and of the $\mathcal{O}(\alpha^2_s \alpha)$ contributions to 
	real photon radiation will be labeled as ``$gq$ channel contributions''. 
	The contributions of real quark emission processes will be referred to 
        as  ``$qq$ channel contributions'', the photon induced gluino--squark production processes
	as ``$\gamma q$ channel contributions''.
	\item The sum of the three channels will be labeled as ``the EW contribution''.
	\item The relative EW contribution is defined as 
	$\delta = (\mathcal{O}_{\rm NLO}- \mathcal{O}_{\rm LO}) / \mathcal{O}_{\rm LO}$,
	where $\mathcal{O}$ is a generic observable and $\mathcal{O}_{\rm NLO}$ is
	the sum of the LO contribution~(\ref{eq_sigmaLO}) and the EW contribution.  
\end{itemize}


\subsection{Hadronic cross sections}
\label{subsec_fullCS}

\TABULAR[t]{cc|c|rcl|r}{
\hline\hline&&&&&\\[-2.5ex]
&&&  \multicolumn{3}{c|}{EW contr. per channel} & 
\\
\rb{name} &\rb{process} &
 \multicolumn{1}{c|}{\rb{LO}}
& \multicolumn{1}{c}{$gq$} 
&\multicolumn{1}{c}{\hspace*{5mm}$qq$\hspace*{5mm}} 
&\multicolumn{1}{c|}{$\gamma q$} 
& \multicolumn{1}{c}{\rb{$ \delta $}}
\\[.5ex]
\hline&&&&&&
\\[-1.5ex]
$\gluino\ul$& $\tilde{g}\, \big(\tilde{u}^{}_L+\tilde{u}^*_L+\tilde{c}^{}_L+\tilde{c}^*_L\big)$
	 & 5340 & $-123$ & 4.03 &  3.98 & $-2.2\%$ \\[.5ex]
$\gluino\dl$& $\tilde{g}\, \big(\tilde{d}_L^{} +\tilde{d}_L^* +\tilde{s}_L^{} +\tilde{s}_L^*\big)$ 
	& 2880 & $-81.2$ & 2.94 & 0.636 & $-2.7 \%$ \\[1.ex]
$\gluino\ur$ & $\tilde{g}\, \big( \tilde{u}_R^{} +\tilde{u}_R^* +\tilde{c}_R^{} +\tilde{c}_R^* \big)$
	 & 5690 & 11.9 & 0.716 & 4.32 & $0.30\%$ \\[.5ex]
$\gluino\dr$& $\tilde{g}\, \big(\tilde{d}_R^{} +\tilde{d}_R^* +\tilde{s}_R^{} +\tilde{s}_R^*\big)$ 
	& 3210 & 1.71 & 0.259 &  0.730& $0.08\%$ \\[1.5ex]
\cline{2-7}
& \textbf{inclusive} $\mathbf{\gluino\, \squark}$
	 & \textbf{17120} & $\mathbf{-191}$ & \textbf{7.95} & \textbf{9.67} & $\mathbf{-1.0 \%}$
\\[1.5ex]
\hline\hline
}{ Integrated cross sections
 for squark--gluino production at the LHC within
  the SPS1a$'$ scenario \cite{AguilarSaavedra:2005pw}.
Shown are the leading order results, the EW contributions from the distinct channels, and the relative 
corrections $\delta$, as defined in the text. All cross sections are given in fb.
\label{tab_results}
}
We show in Table~\ref{tab_results} the results for the hadronic cross sections
for squark--gluino production at the LHC.  
We consider left- and right-handed, up- and down-type squark production separately.
Since light quark masses are negligible, squarks of the first two
generations are mass degenerate and cannot be distinguished experimentally. 
The cross sections for e.\,g.
$\gluino\tilde u_L$, $\gluino\tilde c_L$ (and by CP symmetry also for $\gluino\tilde u_L^*$,
$\gluino\tilde c_L^*$)~production
differ only through the parton luminosity;  we present in the following always their sum,
although denoted by the dominant contribution, e.\,g. $\gluino \tilde u_L$.
The last line in Table~\ref{tab_results} contains the inclusive ('$\tilde g \tilde q$') results. 

Being of QCD origin, the LO cross section of the partonic process
$gq \to \gluino \tilde{q}_a$ is independent of the chirality and of the flavor of the 
produced squark $\squarka$. 
Since all considered squark masses are of the same order ($\sim550$~GeV), 
the LO hadronic cross sections for up-type squark production are about twice as large as 
the cross sections for down-type squark production. 
In contrast, the EW contributions depend strongly on 
the chirality of the squarks and, to a less extent, on the squark flavors. 
The MSSM is a chiral theory and for the production of right-handed squarks 
some of the one-loop and $qq$~channel diagrams are suppressed by the couplings. 
The EW contribution to all left-handed squarks, \ie to $\gluino \tilde u_L$ and 
$\gluino \tilde d_L$~production, is dominated by the (negative) $gq$~channel contributions,
and alters the LO cross section by about $-2\%$.
For right-handed squarks, \ie for $\gluino \tilde u_R$ and $\gluino \tilde d_R$~production,
the $qq$ and $\gamma q$~channels contribute at almost the same order of magnitude 
than the (positive) $gq$~channel and the full EW contribution ranges at the $0.5\%$~level.

Summing up all processes for the inclusive $\gluino\squark$~production, 
the $gq$~channel corrections to right-handed squarks are negligible compared 
to those to left-handed squarks and the size of the relative contribution is roughly halved.
The $qq$ and $\gamma q$~channels give both positive contributions
at the permille level.  The full EW contribution to 
gluino--squark production amounts $-1\%$ within the  SPS1a$'$ scenario.


\subsection{Differential distributions}

\FIGURE[t]{ 
        \image{7.31cm}{trim=0 25 0 0}{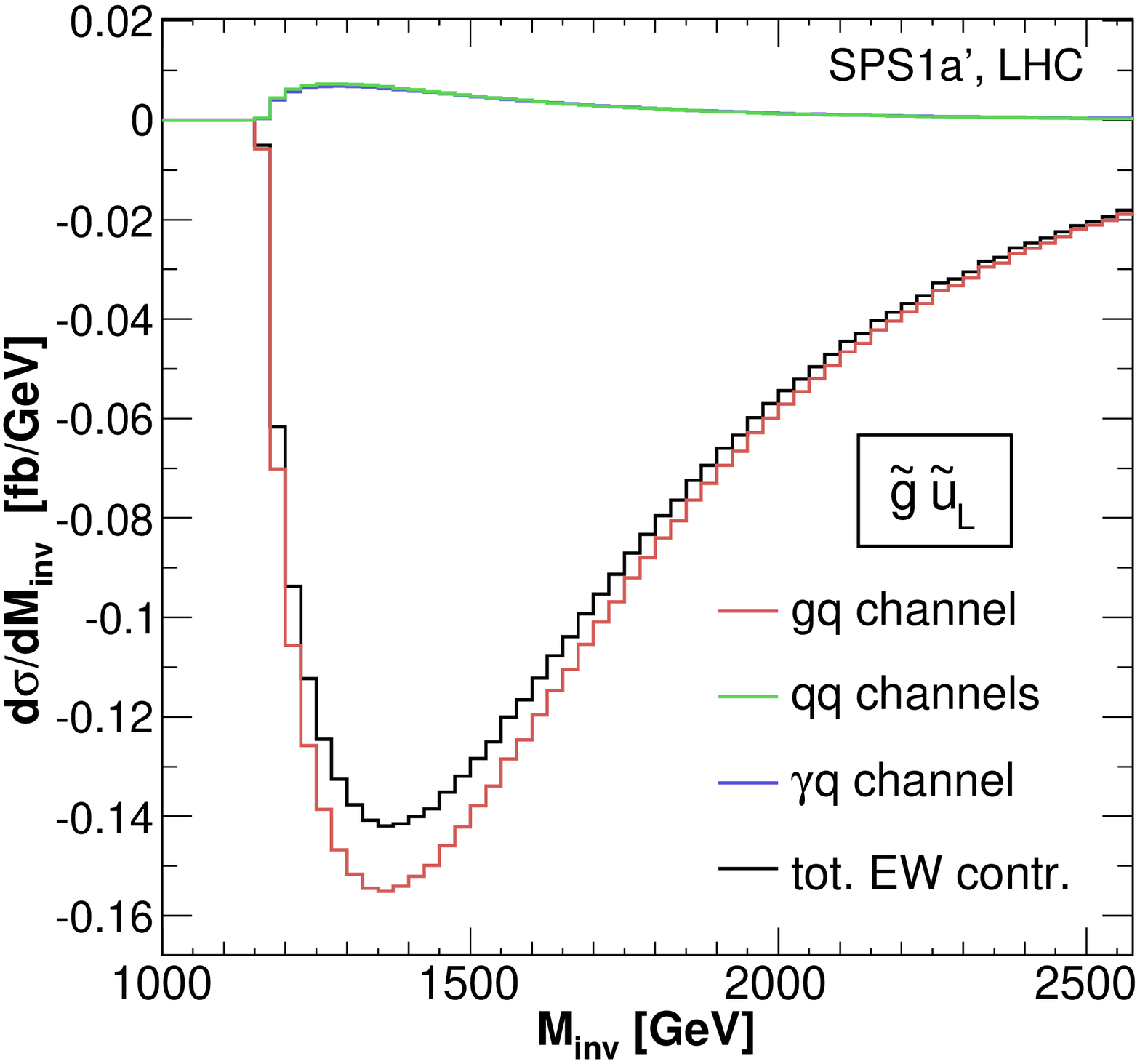}%
        \image{7.31cm}{trim=0 25 0 0}{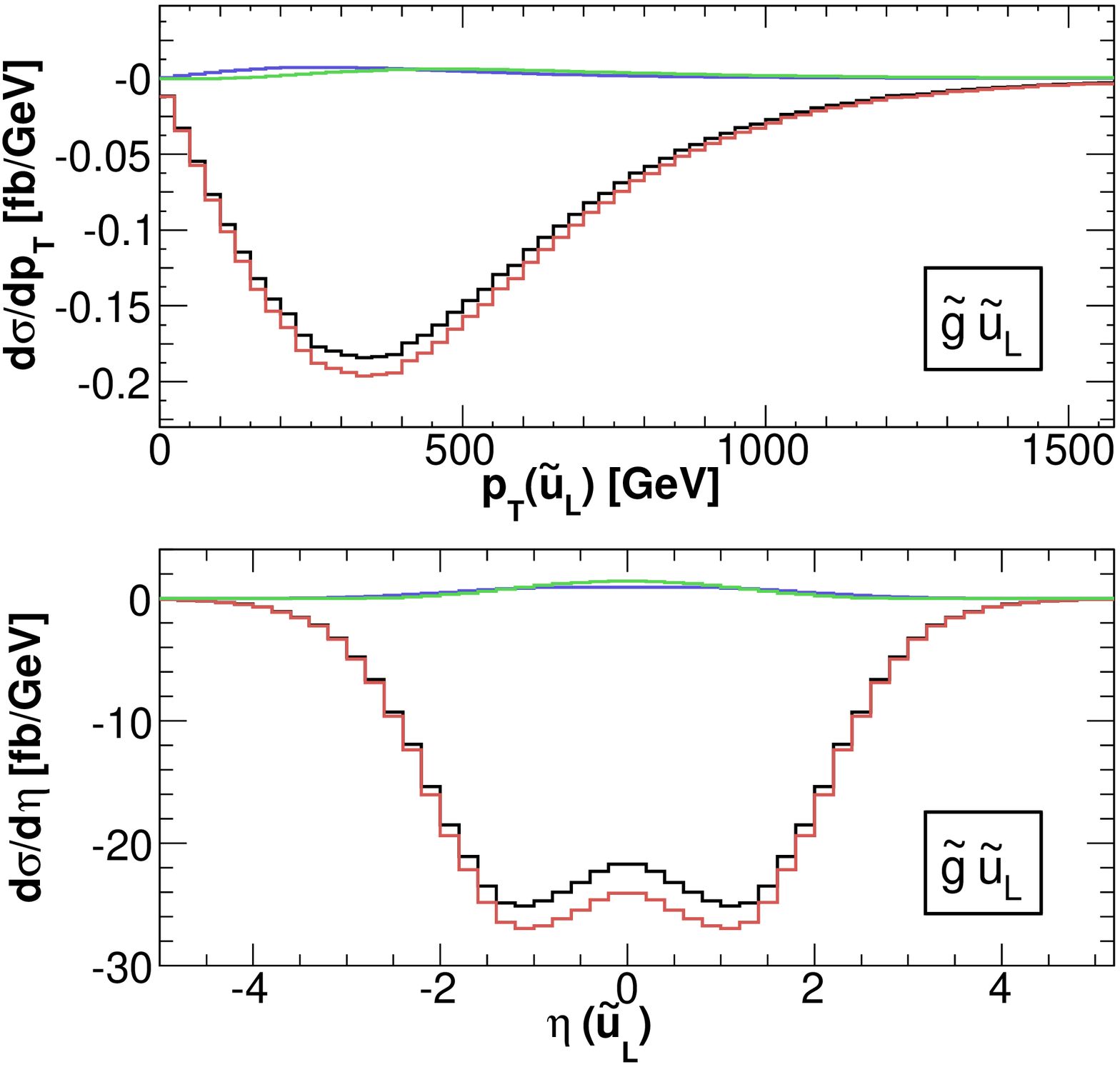}%
        \caption{Comparison of $gq$, $qq$, and $\gamma q$ channel contributions
	to $\gluino\ul$ production.
 	The total EW contribution is also given. Shown are the invariant mass distributions
	(left panel), and the transverse momentum and pseudo rapidity distributions 
	(right panels).}
        \label{fig_abscorrs_upL}
}

\FIGURE[t]{ 
        \image{7.31cm}{trim=0 35 0 0}{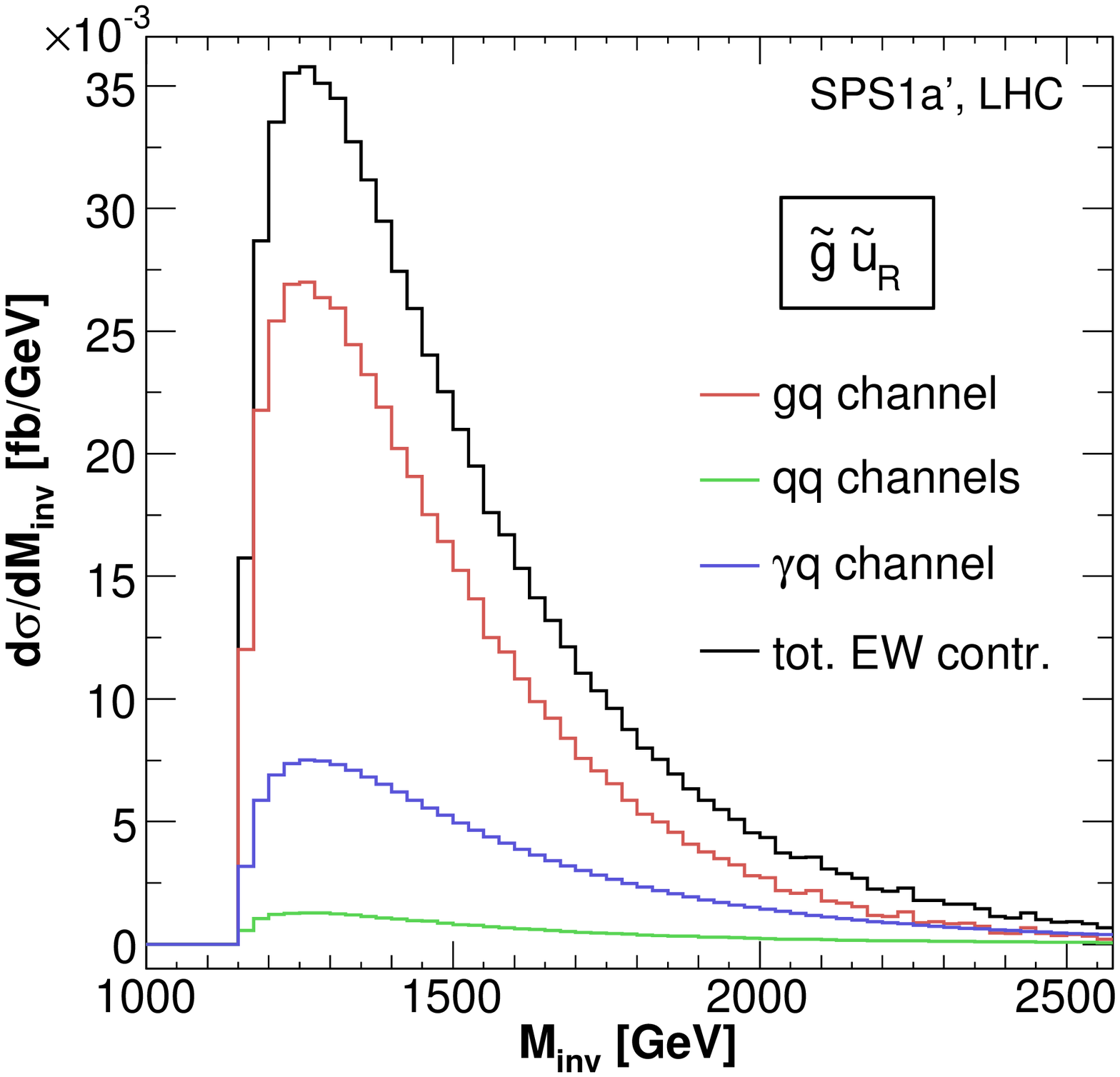}%
        \image{7.31cm}{trim=0 35 0 0}{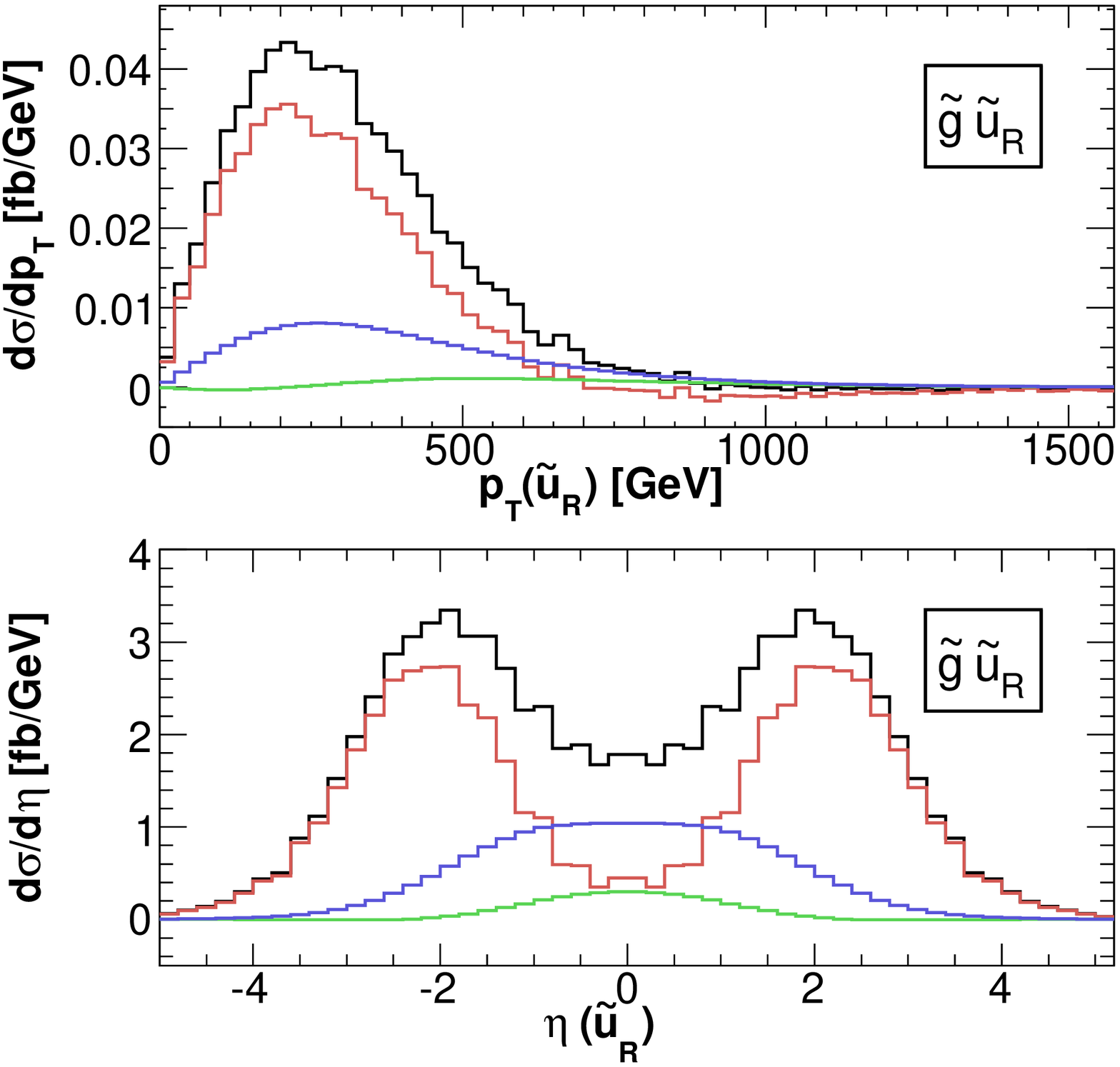}%
        \caption{Same as Fig.~\ref{fig_abscorrs_upL} for $\gluino\ur$ production.}
        \label{fig_abscorrs_upR}
}

The interplay of the various EW contributions is illustrated in
Figs.~\ref{fig_abscorrs_upL} and~\ref{fig_abscorrs_upR} for $\gluino\tilde{u}_L$ 
and $\gluino\tilde{u}_R$~production, respectively, where 
the absolute contributions from the three partonic channels  are given as distributions
with respect to the invariant mass $M_{\rm inv}$ of the squark and the gluino, 
as well as the transverse momentum $p_T$ and the pseudo rapidity $\eta$ of the squark. 
The plots for down-type squark 
production reveal a similar behavior and are not shown explicitly, here.
In Fig.~\ref{fig_abscorrs_upL}, one clearly sees that for left-handed squark production
the virtual and real photon corrections to the $gq$~channel dominate the EW contributions 
over the whole phase space. For right-handed squark production, Fig.~\ref{fig_abscorrs_upR},
the situation is more involved; in particular in the central region ($\lvert \eta \rvert <1$) 
the $\gamma q$~channel contribution is the leading while the other two are comparable.

\medskip


\FIGURE[t]{ 
        \image{7.31cm}{}{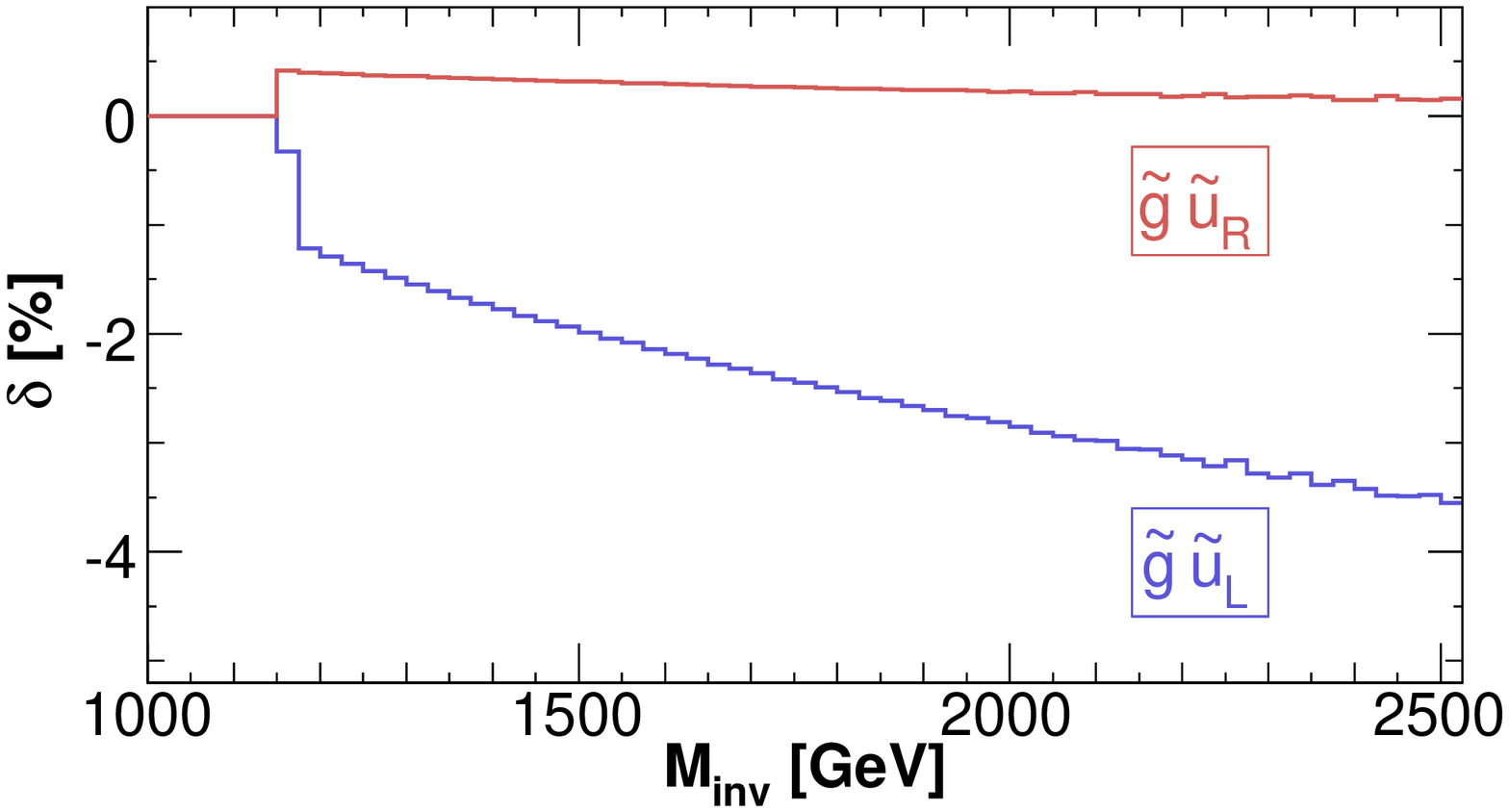}%
        \image{7.31cm}{}{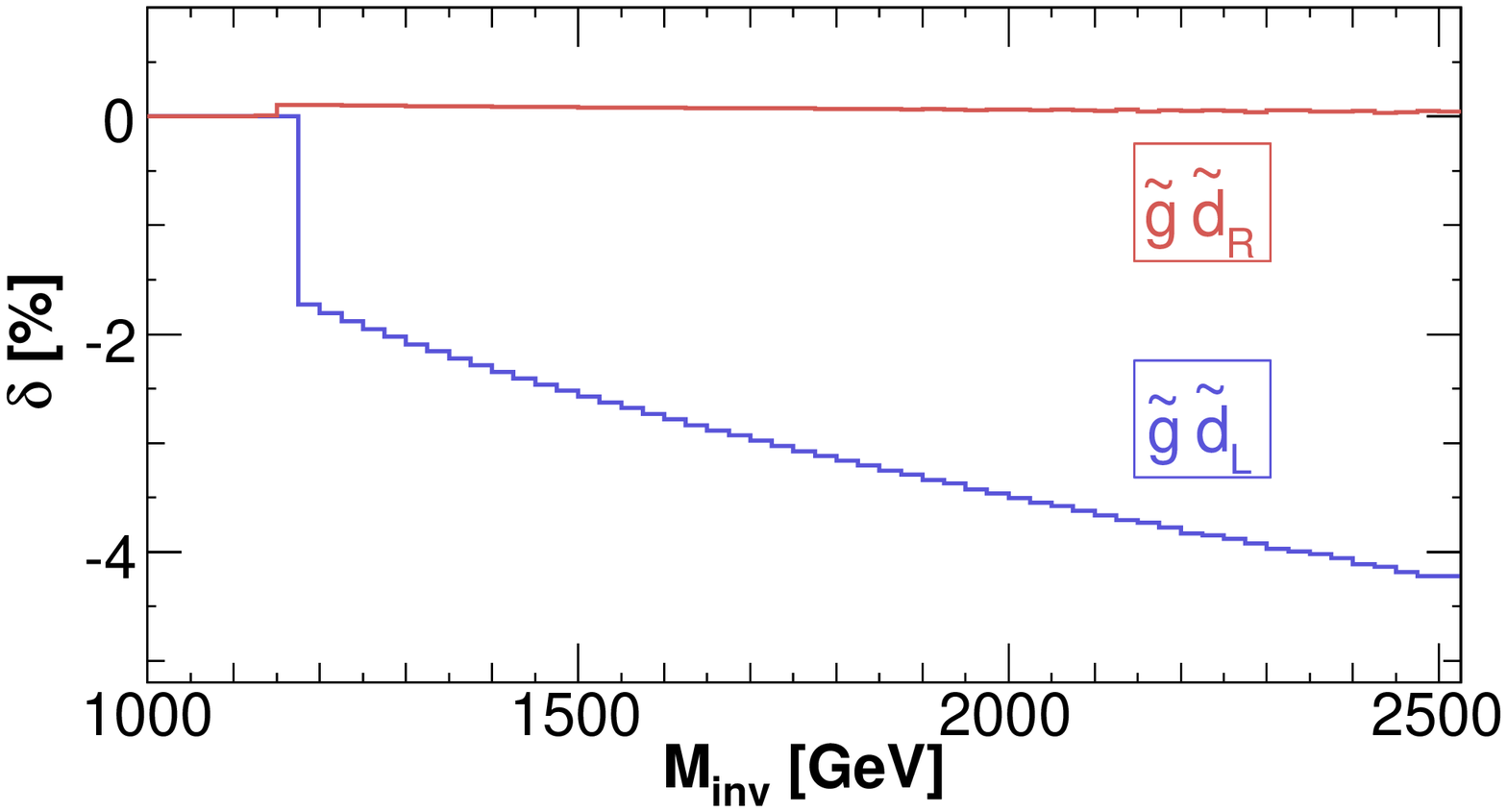}%
\\[-.1ex]
	\hspace*{0ex}%
        \image{7.31cm}{}{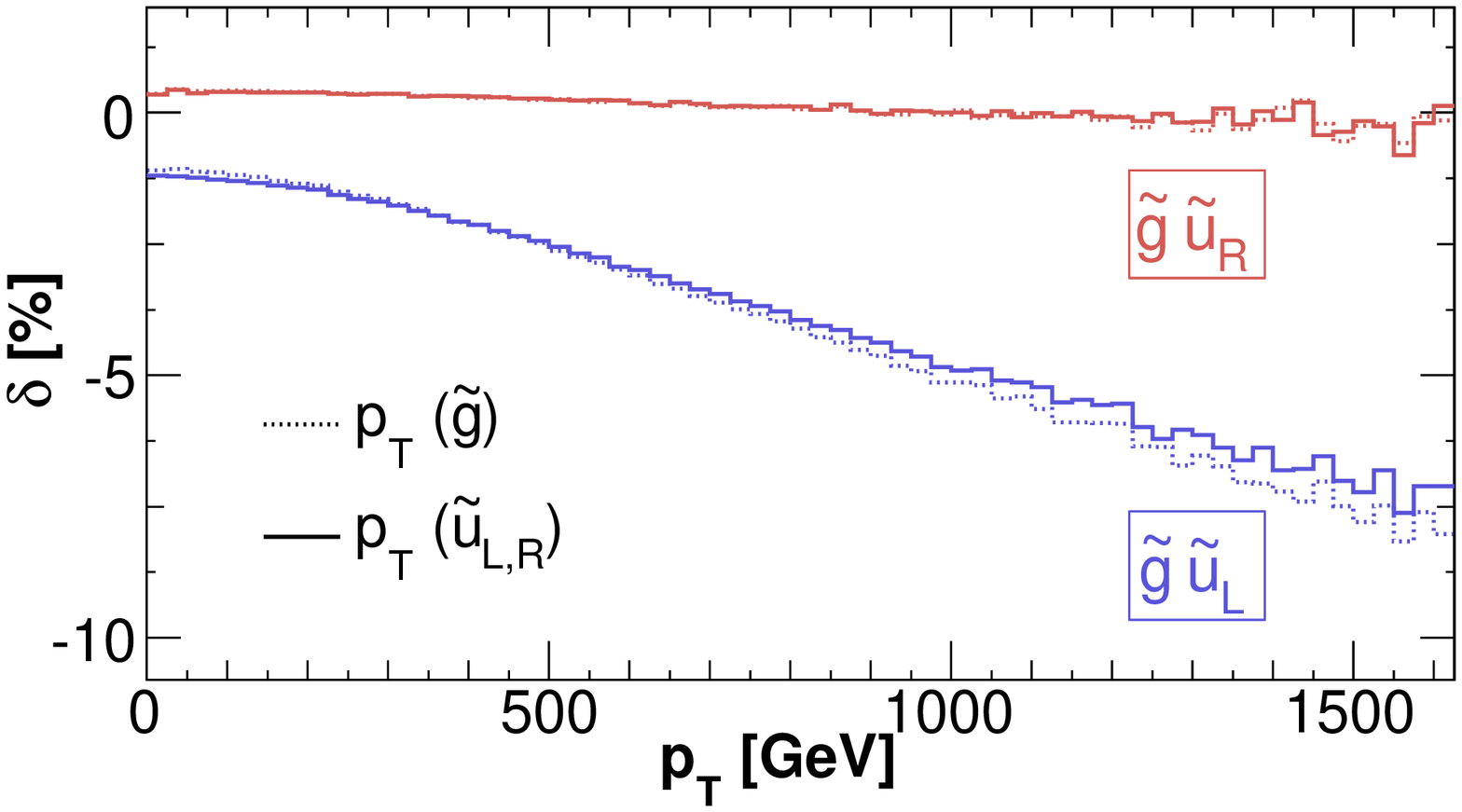}%
        \image{7.31cm}{}{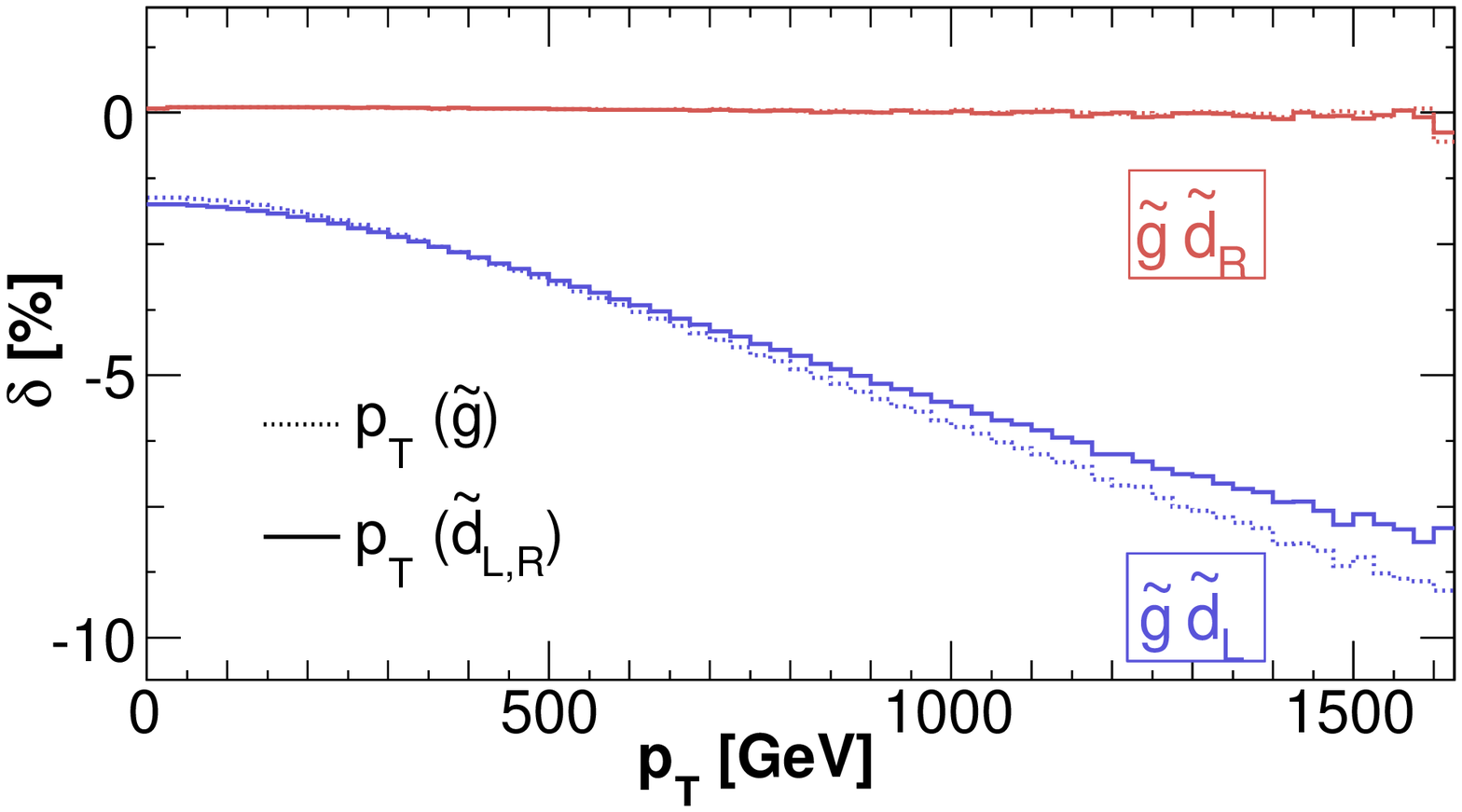}%
\\[-.1ex]
	\hspace*{0ex}%
        \image{7.31cm}{trim=0 25 0 0}{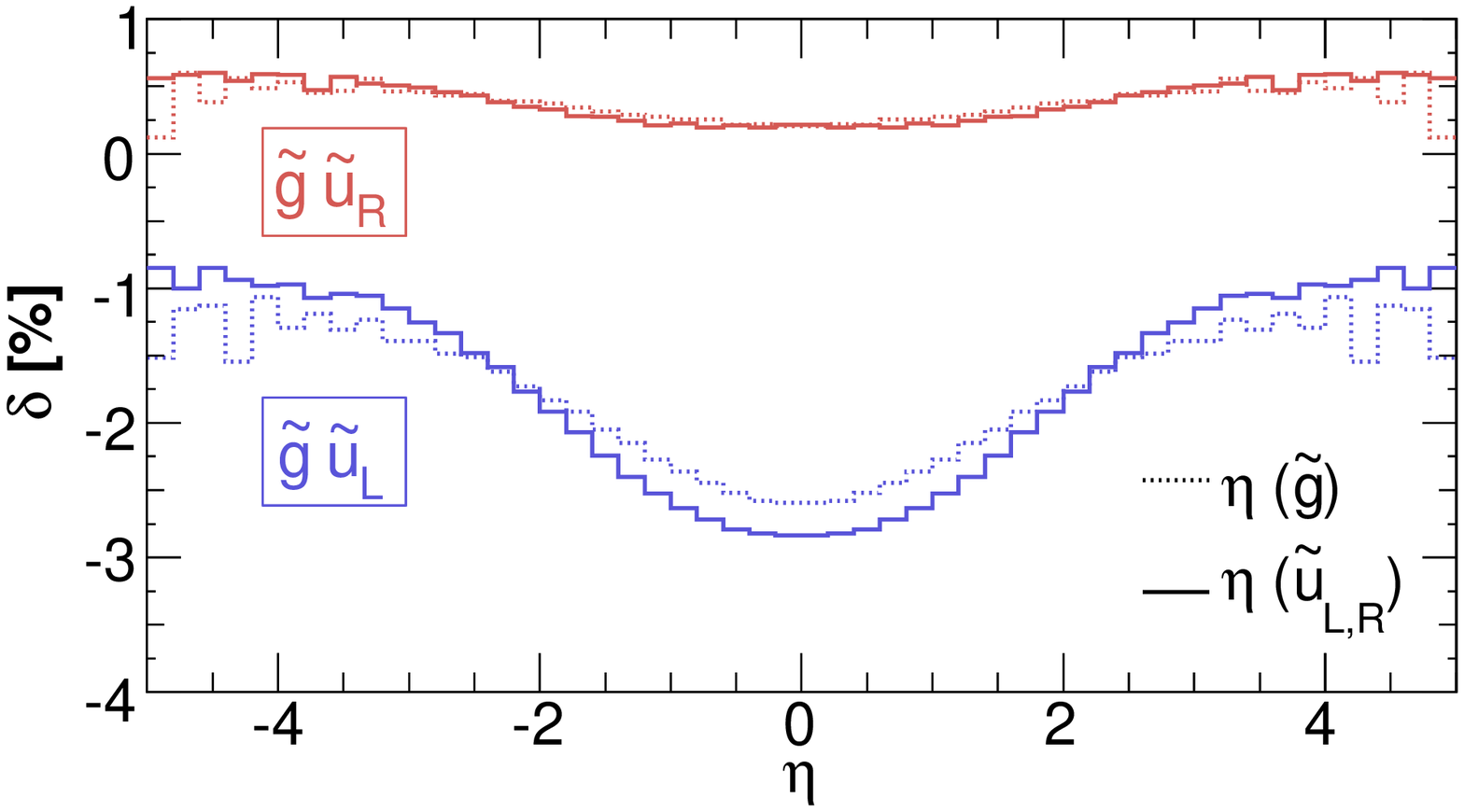}%
        \image{7.31cm}{trim=0 25 0 0}{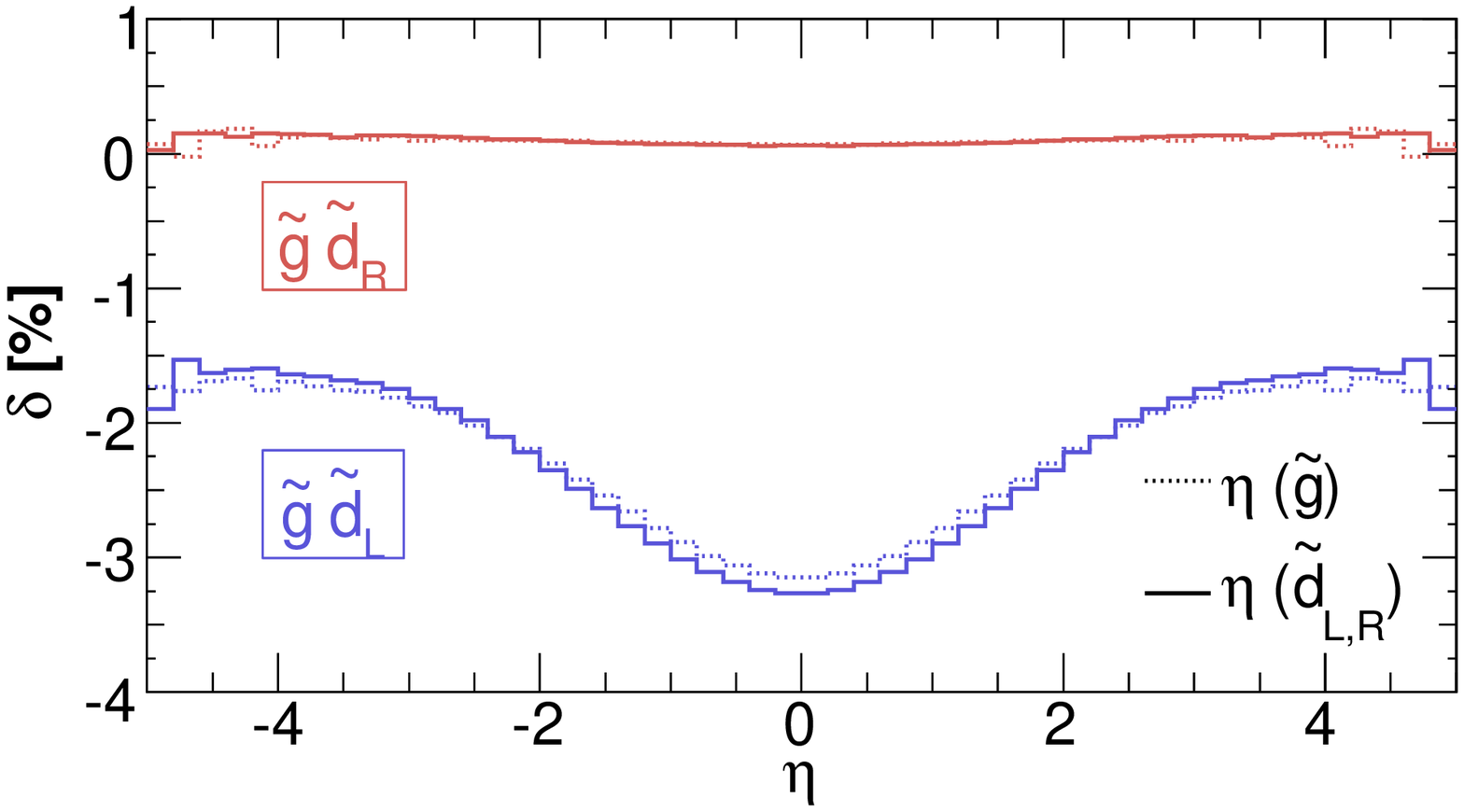}%
       \caption{Relative EW contribution to squark--gluino production at 
	the LHC within the SPS1a$'$ scenario. The left plots refer to $\gluino\tilde{u}_L$ 
	and $\gluino\tilde{u}_R$ production, the right plots to  $\gluino\tilde{d}_L$ 
	and $\gluino\tilde{d}_R$ production. 
	Shown are the invariant mass distribution (top panels), 
	the distributions with respect to the transverse momentum (middle panels)
	 of the gluino (dashed lines) and of the squark (solid lines), 
	and the distributions with respect to the pseudo rapidity (bottom panels) 
	of the gluino (dashed) and the 	squark (solid).}
        \label{fig_relcorrs}
}

Next, we consider the complete EW contribution relative to the LO result, $\delta$. 
In Fig.~\ref{fig_relcorrs}, the distributions with respect to $M_{\rm inv}$, and to
$p_T$ and $\eta$ of both the squark and the gluino
are given, for all four $\gluino\tilde{u}_L$, $\gluino\tilde{d}_L$, $\gluino\tilde{u}_R$, 
$\gluino\tilde{d}_R$ production processes.
As expected, the shape of the relative corrections is similar for up- and down-type squarks 
of the same chirality, and also the size is comparable. 
For right-handed squark production, the distributions are almost flat and contribute negligibly.

For left-handed squarks, 
the EW contribution in the $M_{\rm inv}$ distribution amounts $-2\%$ near threshold 
and increases up to $-4\%$ in the considered $M_{\rm inv}$ range 
($M_{\rm inv} < 2500$~GeV). 
Larger corrections arise in the $p_T$ distribution, where
the EW contributions reach the $-10\%$ level for $p_T > 1500$~GeV. 
The distributions with respect to $p_T(\gluino)$ and $p_T (\squark)$ differ slightly because of 
the different contributions they receive from real photon and real quark radiation processes. 
In particular the $qq$~channels affect the $p_T$ of the 
squark more, reducing (in absolute size) the EW contribution in the high $p_T$ range.

With respect to $\eta$, the EW contribution is largest in the central region ($-3\%$ for 
left-handed squarks). Differences between $\eta(\gluino)$ and $\eta(\squark)$
are related to the real emission processes, and also to the different masses 
of the two final particles which affect the definition of $\eta$ already at the lowest order.
\medskip

%
\FIGURE{ 
        \image{7.2cm}{trim=0 25 0 0}{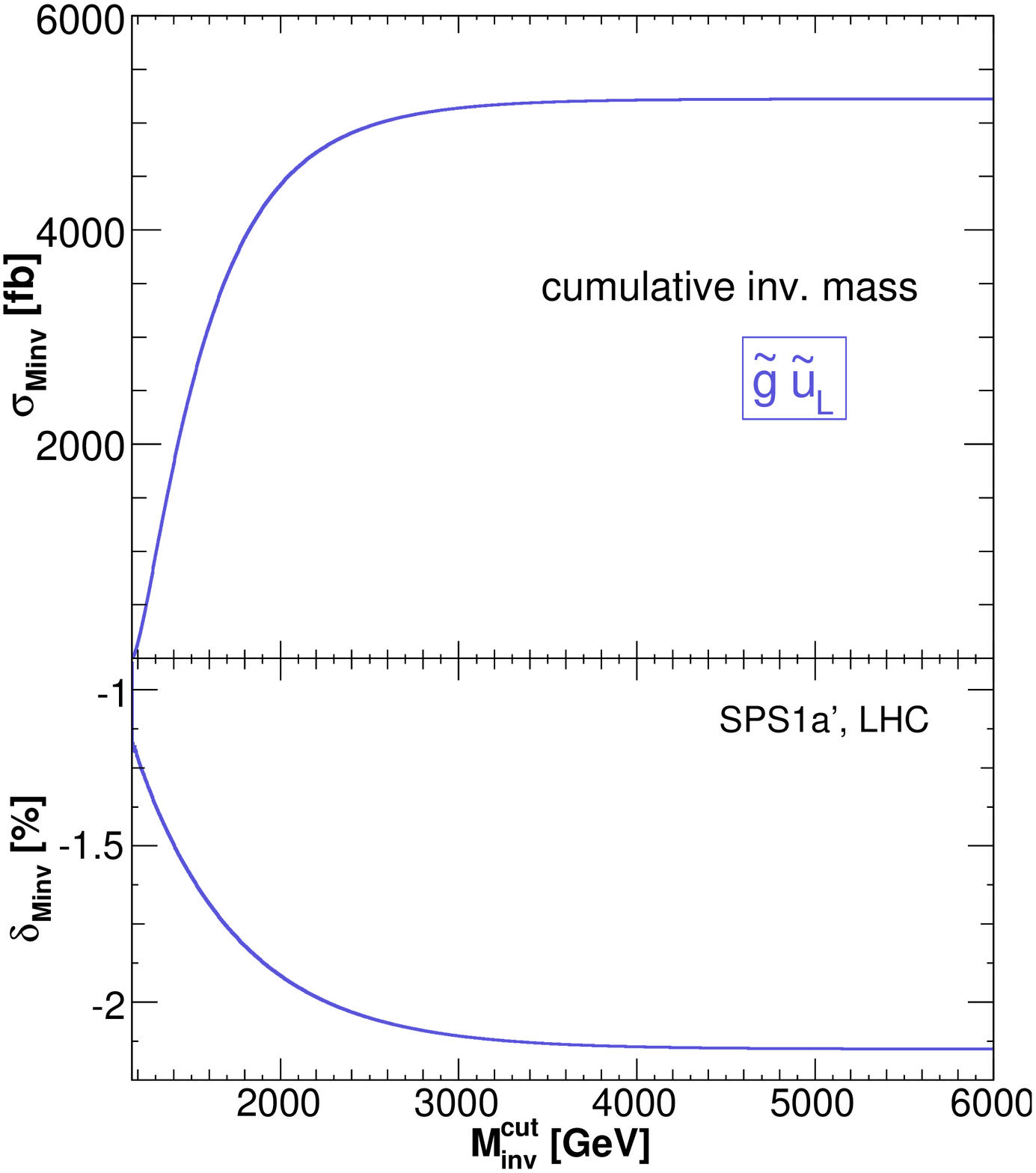}%
        \image{7.2cm}{trim=0 25 0 0}{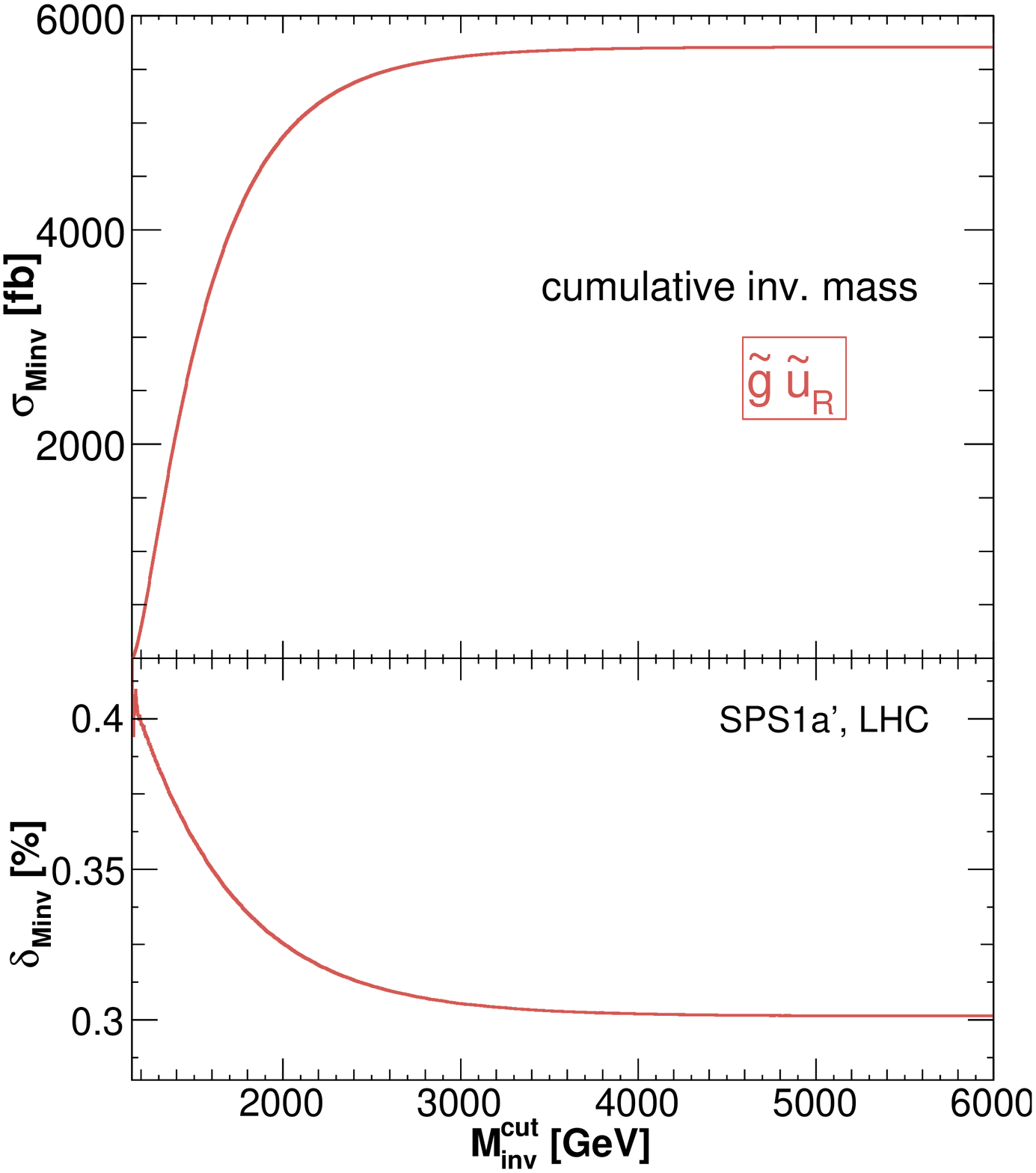}%
       \caption{NLO cumulative invariant mass and relative EW contribution to the same observable, \cf \eqref{eq_cuminvMass},
	for left- and right-handed up-type squark production in association with a gluino.}
        \label{fig_invMcumulative}
}

\FIGURE[t]{ 
	\image{7.2cm}{trim=0 25 0 0}{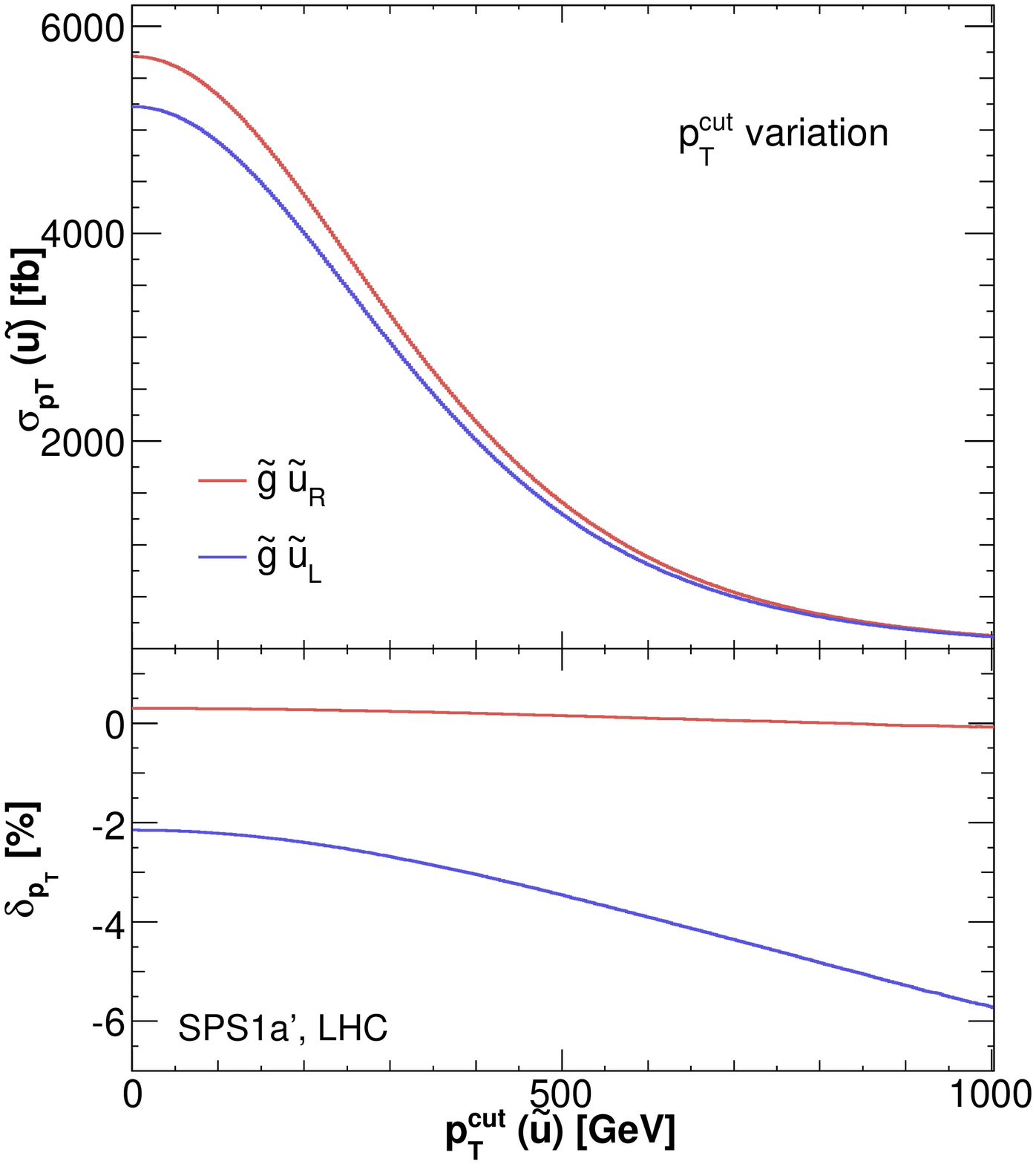}%
	\image{7.2cm}{trim=0 25 0 0}{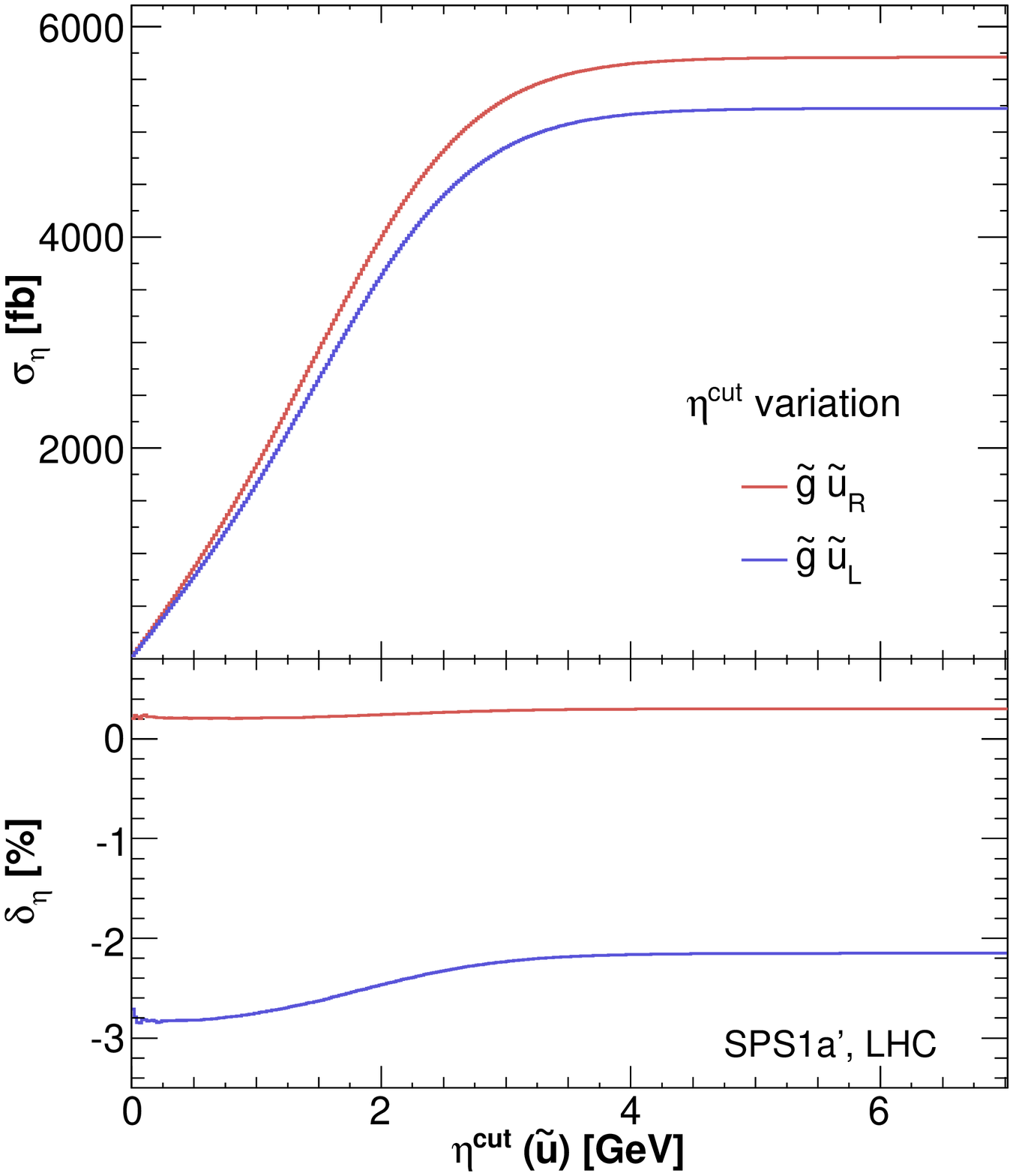}%
        \caption{Hadronic cross sections and relative corrections
	as a function of $p_T^{\rm cut}$ (left panels) and $\eta^{\rm cut}$ (right panels), 
	\cf \eqref{eq_sigmacut}, 
	for up-type squark production in association with a gluino. 
	The cuts refer to $p_T$  and $\eta$ of the produced 
	squark.}
        \label{fig_pTcut-etacut}
}

In order to study the behavior of the EW contribution close to the threshold  we consider the distribution
of the ``cumulative invariant mass'', defined as
\begin{align}
	\sigma(M_{\rm inv}^{\rm cut}) = \int_{m_{\tilde{g}}+m_{\tilde{q}_a}}^{M_{\rm inv}^{\rm cut}} \, \frac{d\sigma}{dM_{\rm inv}} \, dM_{\rm inv}.
 \label{eq_cuminvMass}
\end{align}
In Fig.~\ref{fig_invMcumulative} the cumulative invariant mass including the EW contribution 
and the relative yield of the EW contribution is depicted for the case of 
$\gluino \ul$ (left panel) and $\gluino \ur$ (right panel) production. 
For left-handed squarks, the relative EW contribution increases in absolute size
as $M_{\rm inv}^{\rm cut}$ increases. 
This is a clear signal that the relative yield of the EW corrections increases 
in high $M_{\rm inv}$ region, a general feature that can also be seen in Fig.~\ref{fig_relcorrs}. 
Interestingly, the situation is reversed for right-handed squarks.
In absolute numbers, the relative EW contribution to the cumulative invariant mass decreases 
for increasing $M_{\rm inv}^{\rm cut}$: In the high invariant mass range 
the virtual corrections to the $gq$ channel 
receive negative contributions from Sudakov-like double and single logarithms 
and the positive, non-logarithmically enhanced part of the amplitude is suppressed.
\medskip


In experimental analyses, usually cuts on the kinematically allowed phase space of 
the final state particles are applied. These include lower cuts $p_{T}^{\rm cut}$ 
on the transverse momenta, to focus on high-$p_T$ jets, 
and cuts on the pseudo rapidity $\eta^{\rm cut}$ to restrict the scattering angles 
to the central region in the detector.
For illustration, we give
in Fig.~\ref{fig_pTcut-etacut} the hadronic cross sections as a function of 
these cuts,
\begin{align}
	\sigma(p_T^{\rm cut}) = \int_{p_T^{\rm cut}}^{\infty} \, \frac{d\sigma}{dp_T} \, dp_T,
	\qquad
	\sigma(\eta^{\rm cut}) = \int_{-\eta^{\rm cut}}^{\eta^{\rm cut}} \, \frac{d\sigma}{d\eta} \, d\eta, 
\label{eq_sigmacut}
\end{align} 
together with the corresponding relative corrections.
Since the difference of LO and NLO results are small, only the NLO hadronic
cross sections are plotted. We refer to cuts on $p_T$ and $\eta$ of the (up-type) squark. 
As argued above, results are similar for down-type squarks and 
for cuts on $p_T (\gluino)$ or $\eta(\gluino)$.
As we can see from the left panel of Fig.~\ref{fig_pTcut-etacut}, a cut on $p_T$ enlarges 
the relative EW contribution. The total cross section is about halved for $p_T^{\rm cut} =300$~GeV.  A cut on $\eta$, see 
right panel of Fig.~\ref{fig_pTcut-etacut},  affects the EW contribution only weakly. 
The cross section however, falls rapidly for  $\eta^{\rm cut} < 3$. 
\medskip

\FIGURE[ht]{ 
        \image{7.2cm}{trim=0 25 0 0}{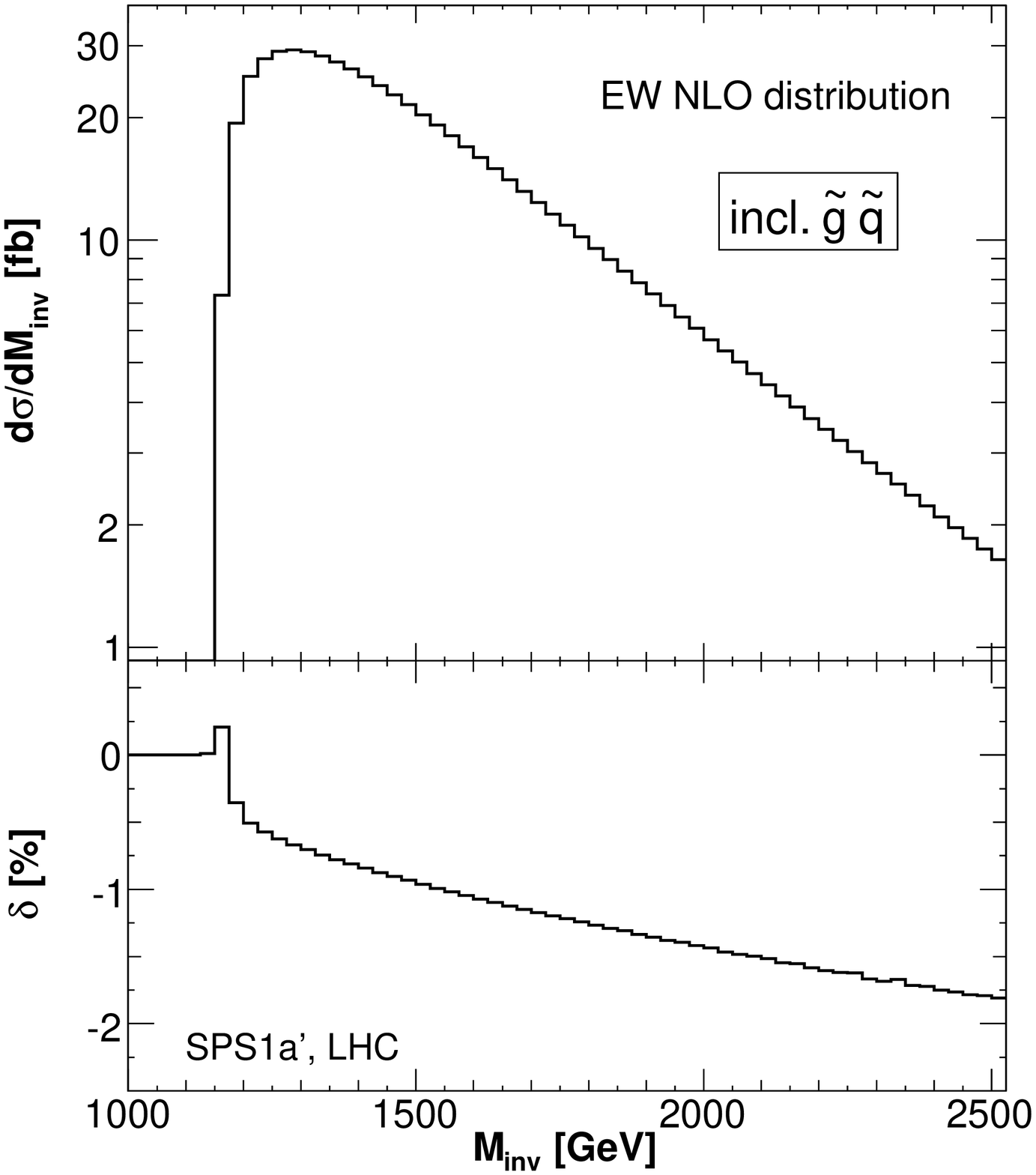}%
        \image{7.2cm}{trim=0 25 0 0}{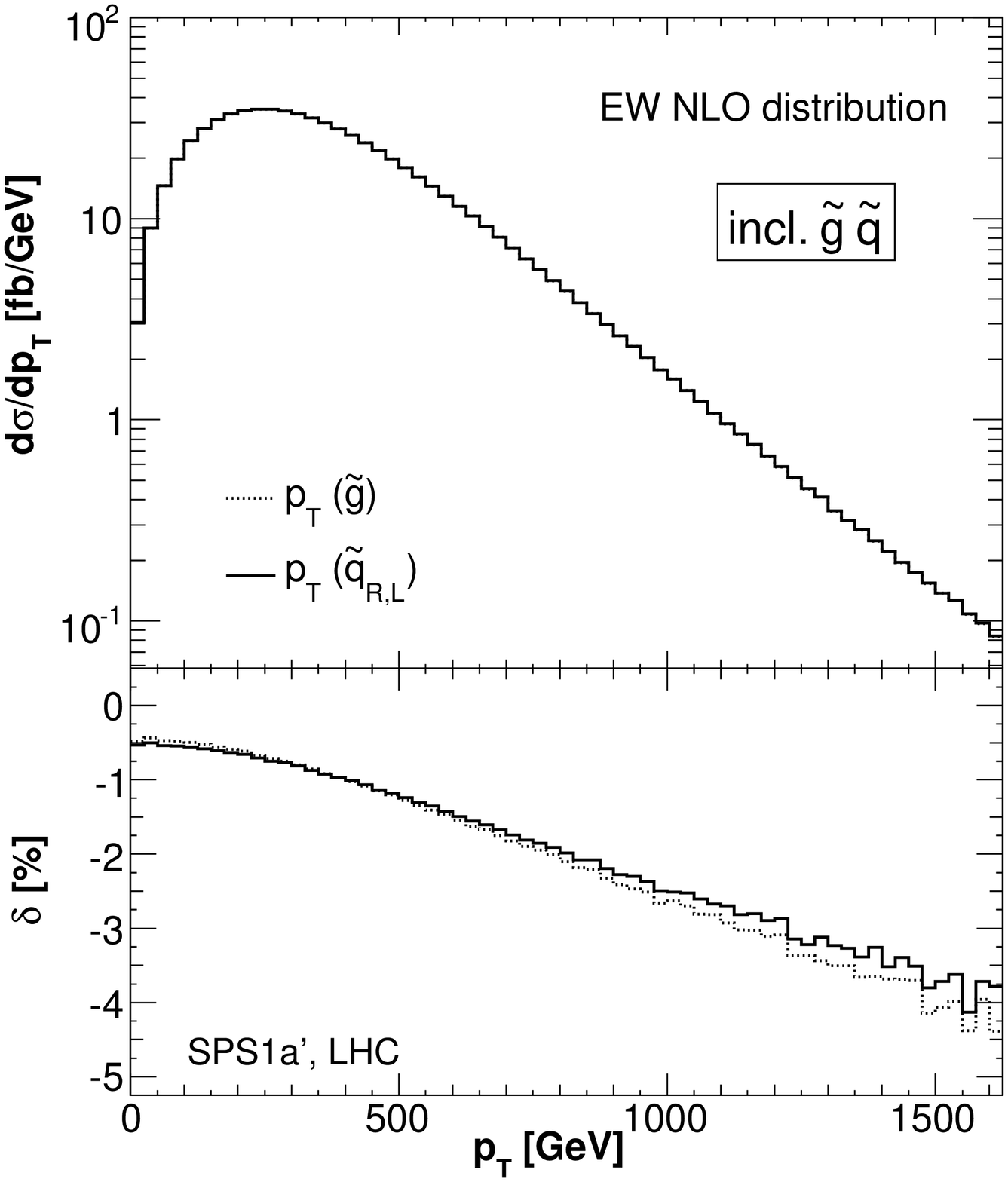}%
        \caption{Hadronic cross sections including the EW contribution
	 (upper panels) and relative EW contribution (lower panels) 
	for inclusive $\squark \gluino$ production.
	Left panel: differential distribution with respect to the 
	invariant mass of the squark and the gluino.
	Right panel: differential distribution with respect to the 
	transverse momentum of the produced squark (solid) or gluino (dashed).
}
        \label{fig_invMpT_inclusive}
}

Finally, we consider inclusive squark--gluino production and show
in Fig.~\ref{fig_invMpT_inclusive} the differential 
hadronic cross sections at EW NLO (\ie LO plus EW contribution), together with the relative
corrections $\delta$, with respect to $M_{\rm inv}$ and to $p_T(\gluino)$ and $p_T(\squark)$.
The relative EW contribution grows in the high-$M_{\rm inv}$ and high-$p_T$ range , but 
owing to the small corrections for right-handed squarks, it remains at the percent 
level only.  
\medskip


\subsection{Dependence on squark and gluino mass}
\label{subsec_scan}

\FIGURE[t]{ 
        \image{7.2cm}{trim=0 25 0 0}{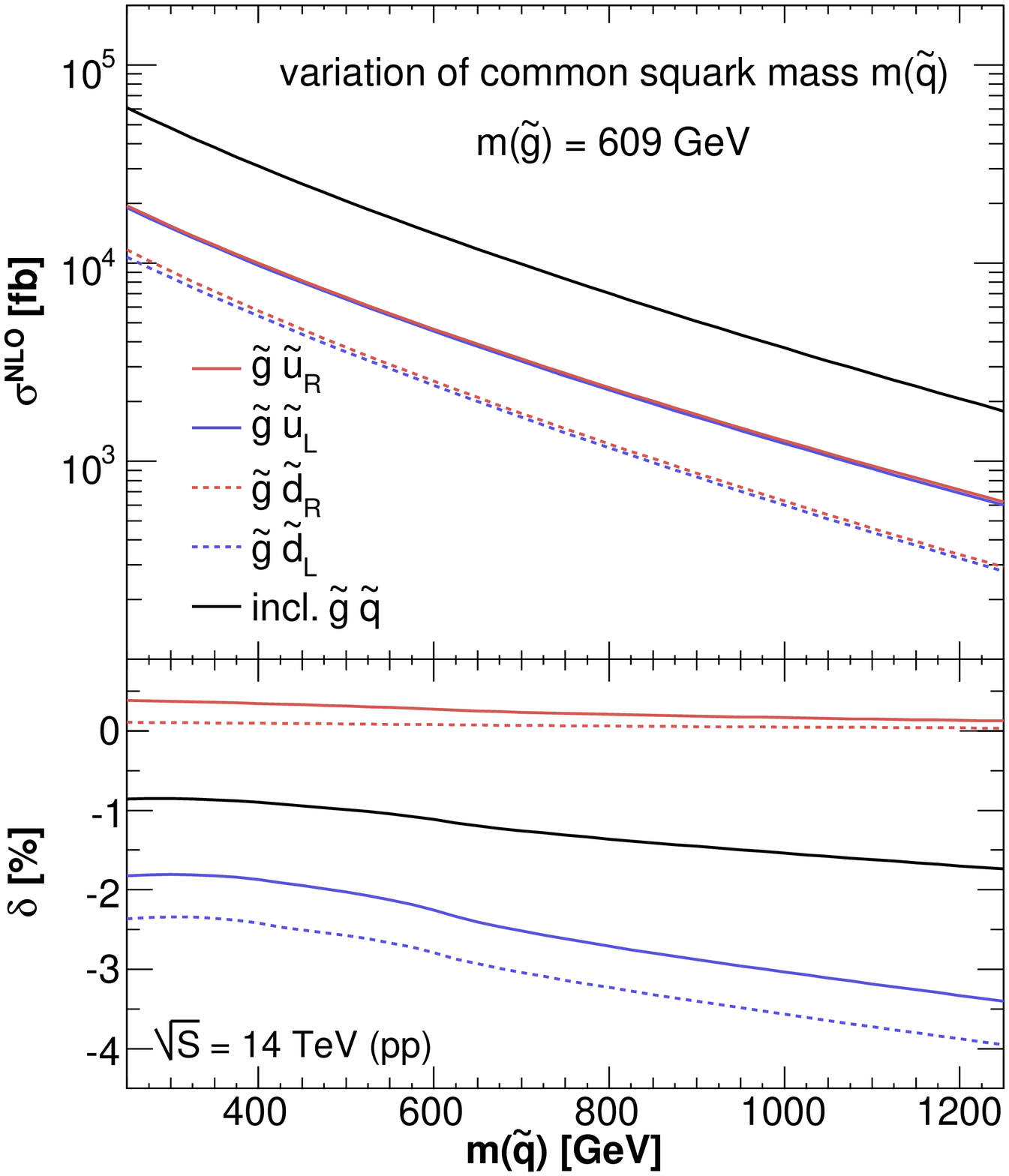}%
        \image{7.2cm}{trim=0 25 0 0}{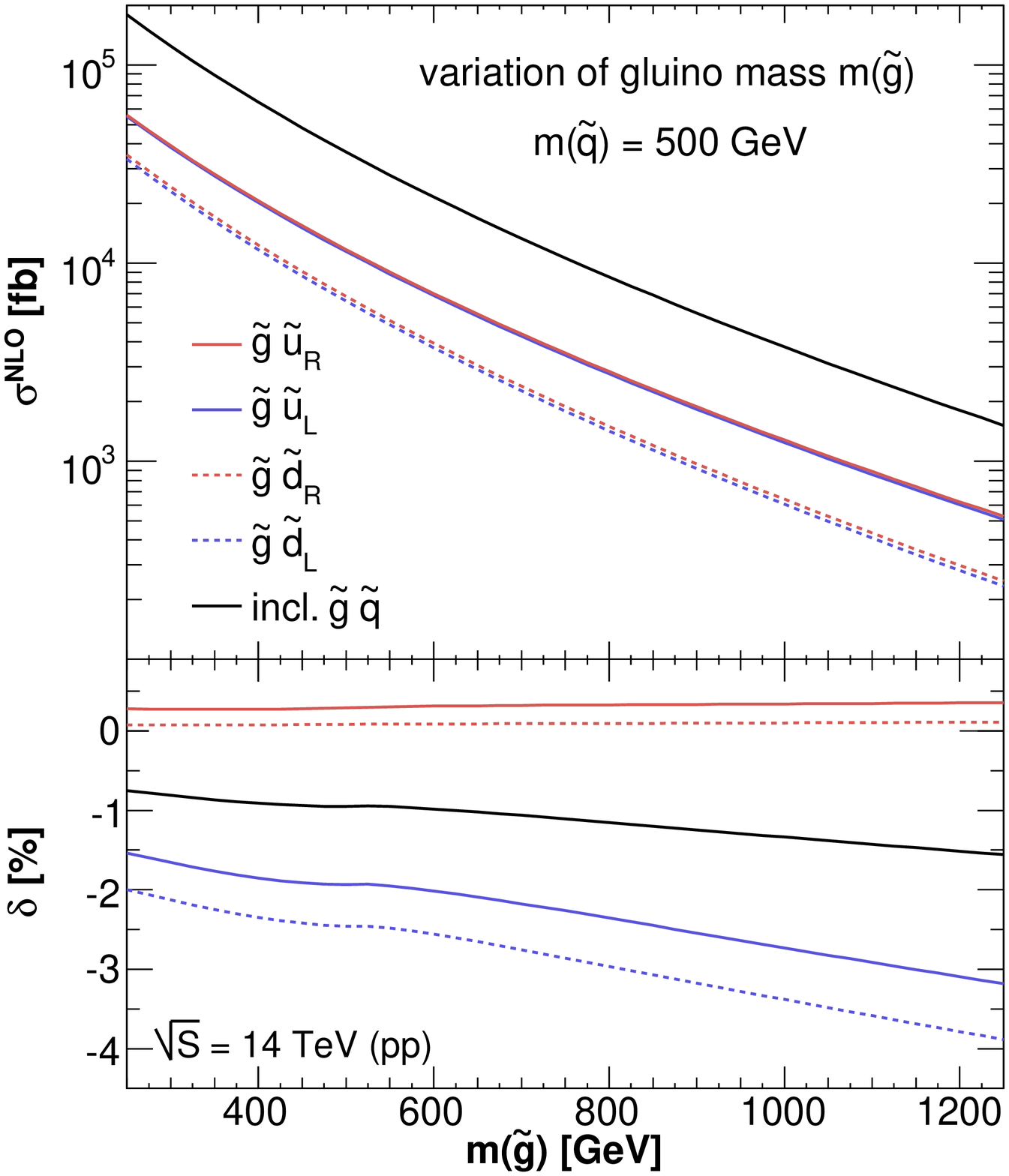}%
        \caption{Hadronic cross sections as a function of a common squark mass 
	 (left panel) and of the gluino mass (right panel). Masses of squarks of the first and 
	second generation are set equal to $m(\tilde{q})$. 
	All other parameters are fixed to their SPS1a$'$ values.
	Shown are the hadronic cross sections at EW NLO and the relative EW 
	contribution for $\gluino \ur$, $\gluino \ul$, $\gluino \dr$,	 $\gluino \dl$ 
	production and the $\gluino\squark$ production.}
        \label{fig_mass-scans}
}

At LO, the only SUSY parameters that enter the production cross section are the masses 
of the final state particles. These parameters are thus 
crucial for the total size of the cross section and it is worth to investigate the dependence 
of the cross section and the EW contribution on the squark and gluino masses.
To this aim, we set the independent squark masses of the first and second generation to a 
common value $m(\squark)$, which is varied for the 'squark mass variation' and 
fixed (to 500~GeV) for the 'gluino mass variation'. 
The fourth, dependent squark mass is computed at each SUSY point 
according to \eqref{eq_app_4thmass}. 
All other SUSY parameters are kept at their SPS1a$'$ values.
We give the results in Fig.~\ref{fig_mass-scans} for the variation of the common squark mass 
$m(\squark)$ (left) and the variation of the gluino mass (right). 
In the upper panels, the total cross sections including the EW contribution, 
and in the lower panels, the relative EW contribution are shown. 
Up-type squark production contributes twice as large as down-type squark production 
to the inclusive result. Again, this is due to the respective parton densities.
The relative EW contribution to right-handed squark production can be neglected ($< 0.5\%$) 
for the considered parameter points. 
For left-handed squarks the corrections vary around $-2\%$ for light masses ($m< 600$~GeV) 
and grow up to $-4\%$ for squark and gluino masses at the TeV range. 
One observes a change in the slope of the relative corrections at the point $m(\gluino) = m(\squark)$ 
since the cross section depends also on the difference of the masses. 
If squarks are heavier than gluinos, the resonance contributions from the $qq$ channels 
have been subtracted as described in Sect.~\ref{subsec_qqchannels} 
and the final contributions from these channels are tiny. 

As a consequence, the relative EW contribution to inclusive gluino--squark production depends 
only weekly on the final state masses and is rather small ($\approx -1\%$).

\section{Conclusions}

We have computed the complete EW contribution to squark--gluino production at 
hadron colliders. At $\mathcal{O}(\alpha_s^2 \alpha)$, the EW contributions are 
of NLO, including EW one-loop corrections together with real photon and real quark radiation processes.
Furthermore,  there are 
tree level contributions arising from photon induced channels at $\mathcal{O}(\alpha_s \alpha)$. 

We discussed in detail the EW contribution to each case of producing a 
left- or right-handed, up- or down-type squark in association with a gluino. Experimentally
distinguishable is $\tilde{b}\gluino$~production, which has not been considered here.
A numerical analysis is presented for squark--gluino production at the LHC within the 
SPS1a$'$ scenario. The EW contribution can be sizable in distributions, in particular for 
left-handed squarks where the virtual $\mathcal{O}(\alpha_s^2 \alpha)$ 
and real photon corrections dominate. 
We also investigated the dependence on the masses of the 
final state squark and gluino, which are crucial for the absolute size of the cross section. 
However the relative EW contribution to inclusive squark--gluino production 
depends only weekly on the masses and ranges at  the $-1\%$~level.

Compared to $\squark_a \squark_a^*$ production~\cite{Hollik:2008}
and to $\tilde{t}_1 \tilde{t}_1^*$ production~\cite{Hollik:2007wf},
the EW contribution to squark--gluino production is small.
Squark pair production profits from additional tree level EW processes that give also 
non-zero interference contributions with the LO QCD diagrams. 
These $\mathcal{O}(\alpha_s \alpha + \alpha^2)$ channels add up to 
the tree level quark radiation processes of $\mathcal{O}(\alpha_s^2 \alpha)$ 
and enhance the EW contribution.
For squarks of the third generation, L--R-mixing has to be taken into account. 
As a consequence, both top-squark mass eigenstates are partially left-handed and 
the EW contribution to the (mainly right-handed) $\tilde{t}_1$~production 
is less suppressed than for  $\squark_R$~production.

\acknowledgments

We thank Stefano Pozzorini and  Jianhui Zhang for valuable discussions. 

\section*{Appendix}
\appendix

\section{SPS1a$'$ input parameters}
\label{app_input}
For the numerical analysis, we consider the mSUGRA scenario SPS1a$'$ that has been proposed by 
the SPA convention \cite{AguilarSaavedra:2005pw}.  
We use {\tt Softsusy 2.0.17} \cite{Allanach:2001kg}, to evolve the GUT parameters down to 
the scale $\mu = 1$~TeV. The mass of the top is fixed
as $m_t = 170.9$~GeV~\cite{CDFtopmass}, while the other SM parameters are chosen in accordance  
with~\cite{AguilarSaavedra:2005pw}.
\medskip

In the renormalization scheme we are using the input parameters are the on-shell ($\os$) masses
of the right- and left-handed up and charm squarks and of the right-handed down and strange
squark. Since  the input parameters for the SPS1a$'$ scenario are defined in $\drbar$ scheme 
a translation of the squark masses into the $\os$ scheme is needed. 
This can be achieved by exploiting the one-loop relation between masses renormalized 
in different schemes:
\begin{align}
	m^2_{\drbar} + \delta m^2_{\drbar} = m^2_{\os} + \delta m^2_{\os},
\end{align}
where $m^2$ is the (squark) mass squared and $\delta m^2$ the corresponding (one-loop)
counter term.  
%
\TABULAR[t!]{c| c c c c c c c c c}{
\hline\hline 
Particle        & $m_{\tilde{u}_L}$ & $m_{\tilde{u}_R}$ & $m_{\tilde{d}_L}$ & $m_{\tilde{d}_R}$ 
                & $m_{\tilde{c}_L}$ & $m_{\tilde{c}_R}$ & $m_{\tilde{s}_L}$ & $m_{\tilde{s}_R}$ &  $m_{\tilde{g}}$  \\
\hline 
 $\drbar$  mass & 523.3   &  506.0  &   529.2 & 501.7 & 523.3  &  506.0  &  529.2  & 501.7   & --  \\
 $\os$     mass & 560.7   &  543.4  &   566.4 & 539.4 & 560.7  &  543.4  &   566.4 & 539.4   & 609.0   \\         
\hline\hline 
}
{ $\drbar$ and $\os$ masses of squarks and gluino (expressed in GeV) within the SPS1a$'$~scenario~\cite{AguilarSaavedra:2005pw}.
\label{tab_masses}
}
%
Owing to the $SU(2)$ invariance, the masses of the left-handed 
down and strange squark are  dependent parameters and are not longer the $\os$ ones.    
At one loop the $\os$ masses can be obtained exploiting the relation:
\begin{align}
\begin{split}
	m_{\tilde{q}_L,~\os}^{2}&=  m_{\tilde{q}_L,~\rm dep.}
		+ \delta m_{\tilde{q}_L}^2 - \Re \left \{ \Sigma_{\tilde{q}_{L}}(m_{\tilde{q}_L}^2)\right \}, \qquad  q=d,s.
\label{eq_app_4thmass}
\end{split}
\end{align}
$\delta m_{\tilde{q}_L}^2$ is the (dependent) counter term for $m_{\tilde{q}_L}^{2}$
whose explicit expression can be found in Appendix B of Ref.~\cite{Hollik:2008}, while 
$\Sigma_{\tilde{q}_L}$ is the self energy of the squark $\tilde{q}_L$. 
\medskip

The $\drbar$ and $\os$ masses of the squarks are collected in Table~\ref{tab_masses}. 
For completeness, we also quote the $\os$  mass of the gluino.

\section{Feynman diagrams}
\label{app_Fey}

We show all Feynman diagrams at the parton level for the example process 
$g\, u \rightarrow \tilde{g} \, \tilde{u}_L$.
The diagrams for (s)quarks of different flavor, charge, and chirality can be obtained in complete analogy.  
The index $i$ runs over all six quark flavors, and $a (b)$ over the chirality
eigenstates $L,\,R$.  We use a common label $V$ to denote the three gauge bosons $\gamma$, $Z$, and $W$.
The label $S^0$ refers to the neutral Higgs (and Goldstone) bosons $h^0,\,H^0,\,A^0,\,G^0$, 
and the label $S^{\pm}$ to the charged Higgs (and Goldstone) bosons $H^{\pm},\,G^{\pm}$.
For neutralinos and charginos, we use a common index $n$ to number the eigenstates,
i.\,e. $\neut{n}$ = $\neut{1,2,3,4}$ and \mbox{$\tilde{\chi}_{n}^{\pm}$ = $\tilde{\chi}_{1,2}^{\pm}$}.

\unitlength=12mm 
\small

\FIGURE{

\vspace*{-.7cm}
\scalebox{.9}{
\begin{feynartspicture}(9,3)(3,1)

\FADiagram{}
\FAProp(0.,15.)(6.,10.)(0.,){/Cycles}{0}
\FALabel(2.48771,11.7893)[tr]{$g$}
\FAProp(0.,5.)(6.,10.)(0.,){/Straight}{1}
\FALabel(3.51229,6.78926)[tl]{$u$}
\FAProp(20.,15.)(14.,10.)(0.,){/Straight}{0}
\FALabel(16.6478,13.0187)[br]{$\tilde g$}
\FAProp(20.,5.)(14.,10.)(0.,){/ScalarDash}{-1}
\FALabel(17.5123,8.21074)[bl]{$\tilde u_L$}
\FAProp(6.,10.)(14.,10.)(0.,){/Straight}{1}
\FALabel(10.,8.93)[t]{$u$}
\FAVert(6.,10.){0}
\FAVert(14.,10.){0}

\FADiagram{}
\FAProp(0.,15.)(10.,14.)(0.,){/Cycles}{0}
\FALabel(4.84577,13.4377)[t]{$g$}
\FAProp(0.,5.)(10.,6.)(0.,){/Straight}{1}
\FALabel(5.15423,4.43769)[t]{$u$}
\FAProp(20.,15.)(10.,14.)(0.,){/Straight}{0}
\FALabel(14.8706,15.3135)[b]{$\tilde g$}
\FAProp(20.,5.)(10.,6.)(0.,){/ScalarDash}{-1}
\FALabel(15.1542,6.56231)[b]{$\tilde u_L$}
\FAProp(10.,14.)(10.,6.)(0.,){/Straight}{0}
\FALabel(9.18,10.)[r]{$\tilde g$}
\FAVert(10.,14.){0}
\FAVert(10.,6.){0}

\FADiagram{}
\FAProp(0.,15.)(10.,14.)(0.,){/Cycles}{0}
\FALabel(4.84577,13.4377)[t]{$g$}
\FAProp(0.,5.)(10.,6.)(0.,){/Straight}{1}
\FALabel(5.15423,4.43769)[t]{$u$}
\FAProp(20.,15.)(10.,6.)(0.,){/Straight}{0}
\FALabel(16.98,13.02)[br]{$\tilde g$}
\FAProp(20.,5.)(10.,14.)(0.,){/ScalarDash}{-1}
\FALabel(17.6872,8.20582)[bl]{$\tilde u_L$}
\FAProp(10.,14.)(10.,6.)(0.,){/ScalarDash}{-1}
\FALabel(9.03,10.)[r]{$\tilde u_L$}
\FAVert(10.,14.){0}
\FAVert(10.,6.){0}

\end{feynartspicture}
}

\caption{LO Feynman diagrams for the process $g\, u \rightarrow \tilde{g} \, \tilde{u}_L$.}
\label{fig_borndiags}
}

\FIGURE{

\vspace*{-.7cm}
\scalebox{.9}{
\begin{feynartspicture}(12,3)(4,1)

\FADiagram{}

\FADiagram{}
\FAProp(0.,15.)(6.,10.)(0.,){/Sine}{0}
\FALabel(2.48771,11.7893)[tr]{$\gamma$}
\FAProp(0.,5.)(6.,10.)(0.,){/Straight}{1}
\FALabel(3.51229,6.78926)[tl]{$u$}
\FAProp(20.,15.)(14.,10.)(0.,){/Straight}{0}
\FALabel(16.6478,13.0187)[br]{$\tilde g$}
\FAProp(20.,5.)(14.,10.)(0.,){/ScalarDash}{-1}
\FALabel(17.5123,8.21074)[bl]{$\tilde u_L$}
\FAProp(6.,10.)(14.,10.)(0.,){/Straight}{1}
\FALabel(10.,8.93)[t]{$u$}
\FAVert(6.,10.){0}
\FAVert(14.,10.){0}

\FADiagram{}
\FAProp(0.,15.)(10.,14.)(0.,){/Sine}{0}
\FALabel(4.84577,13.4377)[t]{$\gamma$}
\FAProp(0.,5.)(10.,6.)(0.,){/Straight}{1}
\FALabel(5.15423,4.43769)[t]{$u$}
\FAProp(20.,15.)(10.,6.)(0.,){/Straight}{0}
\FALabel(16.98,13.02)[br]{$\tilde g$}
\FAProp(20.,5.)(10.,14.)(0.,){/ScalarDash}{-1}
\FALabel(17.6872,8.20582)[bl]{$\tilde u_L$}
\FAProp(10.,14.)(10.,6.)(0.,){/ScalarDash}{-1}
\FALabel(9.03,10.)[r]{$\tilde u_L$}
\FAVert(10.,14.){0}
\FAVert(10.,6.){0}

\FADiagram{}

\end{feynartspicture}}

\caption{Feynman diagrams for photon--quark fusion at lowest order.}
\label{fig_quarkphoton}
}

%

\FIGURE{
\vspace*{-.3cm}
\hspace*{-.3cm}
\scalebox{.85}{
\begin{feynartspicture}(9,3)(3,1)

\FADiagram{}
\FAProp(0.,15.)(10.,13.)(0.,){/Cycles}{0}
\FALabel(5.44126,15.7263)[b]{$g$}
\FAProp(0.,5.)(6.5,7.)(0.,){/Straight}{1}
\FALabel(3.70583,4.99854)[t]{$u$}
\FAProp(20.,15.)(13.5,7.)(0.,){/Straight}{0}
\FALabel(17.53,13.12)[br]{$\tilde g$}
\FAProp(20.,5.)(10.,13.)(0.,){/ScalarDash}{-1}
\FALabel(17.6738,5.68478)[tr]{$\tilde u_L$}
\FAProp(6.5,7.)(13.5,7.)(0.,){/Straight}{1}
\FALabel(10.,5.93)[t]{$q$}
\FAProp(6.5,7.)(10.,13.)(0.,){/Sine}{0}
\FALabel(7.47752,9.5)[br]{$V$}
\FAProp(13.5,7.)(10.,13.)(0.,){/ScalarDash}{1}
\FALabel(11,10)[tr]{$\tilde q_a$}
\FAVert(6.5,7.){0}
\FAVert(13.5,7.){0}
\FAVert(10.,13.){0}
  
\FADiagram{}
\FAProp(0.,15.)(6.5,13.5)(0.,){/Cycles}{0}
\FALabel(3.75593,15.9624)[b]{$g$}
\FAProp(0.,5.)(6.5,6.5)(0.,){/Straight}{1}
\FALabel(3.59853,4.71969)[t]{$u$}
\FAProp(20.,15.)(13.5,13.5)(0.,){/Straight}{0}
\FALabel(16.4577,15.0367)[b]{$\tilde g$}
\FAProp(20.,5.)(13.5,6.5)(0.,){/ScalarDash}{-1}
\FALabel(16.4015,4.71969)[t]{$\tilde u_L$}
\FAProp(6.5,13.5)(6.5,6.5)(0.,){/Straight}{-1}
\FALabel(5.43,10.)[r]{$q$}
\FAProp(6.5,13.5)(13.5,13.5)(0.,){/Straight}{1}
\FALabel(10.,14.57)[b]{$q$}
\FAProp(6.5,6.5)(13.5,6.5)(0.,){/Sine}{0}
\FALabel(10.,5.43)[t]{$V$}
\FAProp(13.5,13.5)(13.5,6.5)(0.,){/ScalarDash}{1}
\FALabel(14.57,10.)[l]{$\tilde q_a$}
\FAVert(6.5,13.5){0}
\FAVert(6.5,6.5){0}
\FAVert(13.5,13.5){0}
\FAVert(13.5,6.5){0}

\FADiagram{}
\FAProp(0.,15.)(13.5,13.)(0.,){/Cycles}{0}
\FALabel(3.51281,16.2204)[b]{$g$}
\FAProp(0.,5.)(6.5,6.)(0.,){/Straight}{1}
\FALabel(3.48569,4.44802)[t]{$u$}
\FAProp(20.,15.)(6.5,13.)(0.,){/Straight}{0}
\FALabel(16.6264,15.2806)[b]{$\tilde g$}
\FAProp(20.,5.)(13.5,6.)(0.,){/ScalarDash}{-1}
\FALabel(16.5143,4.44802)[t]{$\tilde u_L$}
\FAProp(13.5,13.)(6.5,13.)(0.,){/ScalarDash}{-1}
\FALabel(10.,11.93)[t]{$\tilde{q}_a$}
\FAProp(13.5,13.)(13.5,6.)(0.,){/ScalarDash}{1}
\FALabel(14.57,9.5)[l]{$\tilde{q}_a$}
\FAProp(6.5,6.)(6.5,13.)(0.,){/Straight}{1}
\FALabel(5.43,9.5)[r]{$q$}
\FAProp(6.5,6.)(13.5,6.)(0.,){/Sine}{0}
\FALabel(10.,4.93)[t]{$V$}
\FAVert(13.5,13.){0}
\FAVert(6.5,6.){0}
\FAVert(6.5,13.){0}
\FAVert(13.5,6.){0}
  
\end{feynartspicture}

\begin{feynartspicture}(6,3)(2,1)

\FADiagram{}
\FAProp(0.,15.)(13.5,13.)(0.,){/Cycles}{0}
\FALabel(3.51281,16.2204)[b]{$g$}
\FAProp(0.,5.)(6.5,6.)(0.,){/Straight}{1}
\FALabel(3.48569,4.44802)[t]{$u$}
\FAProp(20.,15.)(6.5,13.)(0.,){/Straight}{0}
\FALabel(16.6264,15.2806)[b]{$\tilde g$}
\FAProp(20.,5.)(13.5,6.)(0.,){/ScalarDash}{-1}
\FALabel(16.5143,4.44802)[t]{$\tilde u_L$}
\FAProp(13.5,13.)(6.5,13.)(0.,){/Straight}{-1}
\FALabel(10.,11.93)[t]{$q$}
\FAProp(13.5,13.)(13.5,6.)(0.,){/Straight}{1}
\FALabel(14.57,9.5)[l]{$q$}
\FAProp(6.5,6.)(6.5,13.)(0.,){/ScalarDash}{1}
\FALabel(5.43,9.5)[r]{$\tilde{q}_a$}
\FAProp(6.5,6.)(13.5,6.)(0.,){/Straight}{0}
\FALabel(10.,5.18)[t]{$\tilde{\chi}_n^{0/\pm}$}
\FAVert(13.5,13.){0}
\FAVert(6.5,6.){0}
\FAVert(6.5,13.){0}
\FAVert(13.5,6.){0}

\FADiagram{}
\FAProp(0.,15.)(6.5,13.5)(0.,){/Cycles}{0}
\FALabel(3.75593,15.9624)[b]{$g$}
\FAProp(0.,5.)(6.5,6.5)(0.,){/Straight}{1}
\FALabel(3.59853,4.71969)[t]{$u$}
\FAProp(20.,15.)(13.5,13.5)(0.,){/Straight}{0}
\FALabel(16.4577,15.0367)[b]{$\tilde g$}
\FAProp(20.,5.)(13.5,6.5)(0.,){/ScalarDash}{-1}
\FALabel(16.4015,4.71969)[t]{$\tilde u_L$}
\FAProp(6.5,13.5)(6.5,6.5)(0.,){/ScalarDash}{-1}
\FALabel(5.43,10.)[r]{$\tilde q_a$}
\FAProp(6.5,13.5)(13.5,13.5)(0.,){/ScalarDash}{1}
\FALabel(10.,14.57)[b]{$\tilde q_a$}
\FAProp(6.5,6.5)(13.5,6.5)(0.,){/Straight}{0}
\FALabel(10.,5.68)[t]{$\tilde \chi_n^{0/ \pm}$}
\FAProp(13.5,13.5)(13.5,6.5)(0.,){/Straight}{1}
\FALabel(14.57,10.)[l]{$q$}
\FAVert(6.5,13.5){0}
\FAVert(6.5,6.5){0}
\FAVert(13.5,13.5){0}
\FAVert(13.5,6.5){0}

\end{feynartspicture}}

\\
\put(2,.2){(a)}
\\[-.3ex]
\vspace*{-.3cm}
\hspace*{-.3cm}
\scalebox{.84}{

\begin{feynartspicture}(9,3)(3,1)

\FADiagram{}
\FAProp(0.,17.5)(10.,17)(0.,){/Cycles}{0}
\FALabel(5.11236,18.5)[b]{$g$}
\FAProp(0.,5.)(10.,5.5)(0.,){/Straight}{1}
\FALabel(5.0774,4.3)[t]{$u$}
\FAProp(20.,17.)(10.,5.5)(0.,){/Straight}{0}
\FALabel(17.,15.)[b]{$\tilde g$}
\FAProp(20.,5.)(10.,17)(0.,){/ScalarDash}{-1}
\FALabel(17.,6.86964)[t]{$\tilde u_L$}
\FAProp(10.,17)(10.,11.25)(0.,){/ScalarDash}{-1}
\FALabel(9.5,15.1)[r]{$\tilde u_L$}
\FAProp(10.,5.5)(10.,11.25)(0.,){/ScalarDash}{1}
\FALabel(9.5,7.)[r]{$\tilde u_a$}
\FAProp(10.,11.25)(10.,11.25)(5.,11.25){/ScalarDash}{0}
\FALabel(4.8,11.6)[r]{$S^{0,\pm}$}
\FAVert(10.,17){0}
\FAVert(10.,5.5){0}
\FAVert(10.,11.25){0}
 
\FADiagram{}
\FAProp(0.,17.5)(10.,17)(0.,){/Cycles}{0}
\FALabel(5.11236,18.5)[b]{$g$}
\FAProp(0.,5.)(10.,5.5)(0.,){/Straight}{1}
\FALabel(5.0774,4.3)[t]{$u$}
\FAProp(20.,17.)(10.,5.5)(0.,){/Straight}{0}
\FALabel(17.,15.)[b]{$\tilde g$}
\FAProp(20.,5.)(10.,17)(0.,){/ScalarDash}{-1}
\FALabel(17.,6.86964)[t]{$\tilde u_L$}
\FAProp(10.,17)(10.,11.25)(0.,){/ScalarDash}{-1}
\FALabel(9.5,15.1)[r]{$\tilde u_L$}
\FAProp(10.,5.5)(10.,11.25)(0.,){/ScalarDash}{1}
\FALabel(9.5,7.)[r]{$\tilde u_a$}
\FAProp(10.,11.25)(10.,11.25)(5.,11.25){/ScalarDash}{0}
\FALabel(4.18,11.25)[r]{$\tilde \ell_{ib}, \tilde q_{ib}$}
\FAVert(10.,17){0}
\FAVert(10.,5.5){0}
\FAVert(10.,11.25){0}
 
\FADiagram{}
\FAProp(0.,17.5)(10.,17)(0.,){/Cycles}{0}
\FALabel(5.11236,18.5)[b]{$g$}
\FAProp(0.,5.)(10.,5.5)(0.,){/Straight}{1}
\FALabel(5.0774,4.3)[t]{$u$}
\FAProp(20.,17.)(10.,5.5)(0.,){/Straight}{0}
\FALabel(17.,15.)[b]{$\tilde g$}
\FAProp(20.,5.)(10.,17)(0.,){/ScalarDash}{-1}
\FALabel(17.,6.86964)[t]{$\tilde u_L$}
\FAProp(10.,17)(10.,11.25)(0.,){/ScalarDash}{-1}
\FALabel(9.5,15.1)[r]{$\tilde u_L$}
\FAProp(10.,5.5)(10.,11.25)(0.,){/ScalarDash}{1}
\FALabel(9.5,7.)[r]{$\tilde u_L$}
\FAProp(10.,11.25)(10.,11.25)(5.,11.25){/Sine}{0}
\FALabel(4.18,11.25)[r]{$V$}
\FAVert(10.,17){0}
\FAVert(10.,5.5){0}
\FAVert(10.,11.25){0}
 
\end{feynartspicture}

\begin{feynartspicture}(6,3)(2,1)

\FADiagram{}
\FAProp(0.,17.5)(3.,11.25)(0.,){/Cycles}{0}
\FALabel(0.650886,14.1825)[tr]{$g$}
\FAProp(0.,5.)(3.,11.25)(0.,){/Straight}{1}
\FALabel(2.,6.8)[tl]{$u$}
\FAProp(20.,17.5)(17.,11.25)(0.,){/Straight}{0}
\FALabel(17.8653,14.6888)[br]{$\tilde g$}
\FAProp(20.,5.)(17.,11.25)(0.,){/ScalarDash}{-1}
\FALabel(19.,7.7)[bl]{$\tilde u_L$}
\FAProp(3.,11.25)(7.,11.25)(0.,){/Straight}{1}
\FALabel(5.,12.37)[b]{$u$}
\FAProp(17.,11.25)(13.,11.25)(0.,){/Straight}{-1}
\FALabel(15.,10)[t]{$u$}
\FAProp(7.,11.25)(13.,11.25)(0.8,){/Straight}{0}
\FALabel(10.2,8)[t]{$\tilde \chi_n^{0/ \pm}$}
\FAProp(7.,11.25)(13.,11.25)(-0.8,){/ScalarDash}{1}
\FALabel(10.,14.5)[b]{$\tilde q_a$}
\FAVert(3.,11.25){0}
\FAVert(17.,11.25){0}
\FAVert(7.,11.25){0}
\FAVert(13.,11.25){0}
 
\FADiagram{}
\FAProp(0.,17.5)(3.,11.25)(0.,){/Cycles}{0}
\FALabel(.650886,15.1825)[tr]{$g$}
\FAProp(0.,5.)(3.,11.25)(0.,){/Straight}{1}
\FALabel(2.,6.8)[tl]{$u$}
\FAProp(20.,17.5)(17.,11.25)(0.,){/Straight}{0}
\FALabel(17.8653,15.6888)[br]{$\tilde g$}
\FAProp(20.,5.)(17.,11.25)(0.,){/ScalarDash}{-1}
\FALabel(19.,7.7)[bl]{$\tilde u_L$}
\FAProp(3.,11.25)(7.,11.25)(0.,){/Straight}{1}
\FALabel(5.,12.37)[b]{$u$}
\FAProp(17.,11.25)(13.,11.25)(0.,){/Straight}{-1}
\FALabel(15.,10)[t]{$u$}
\FAProp(7.,11.25)(13.,11.25)(0.8,){/Straight}{1}
\FALabel(10.2,8)[t]{$q$}
\FAProp(7.,11.25)(13.,11.25)(-0.8,){/Sine}{0}
\FALabel(10.,14.5)[b]{$V$}
\FAVert(3.,11.25){0}
\FAVert(17.,11.25){0}
\FAVert(7.,11.25){0}
\FAVert(13.,11.25){0}
 
\end{feynartspicture}
}

\hspace*{1.2cm}
\scalebox{.84}{

\begin{feynartspicture}(9,3)(3,1)


\FADiagram{}
\FAProp(0.,17.5)(10.,17)(0.,){/Cycles}{0}
\FALabel(5.11236,18.5)[b]{$g$}
\FAProp(0.,5.)(10.,5.5)(0.,){/Straight}{1}
\FALabel(5.0774,4.18)[t]{$u$}
\FAProp(20.,17.)(10.,5.5)(0.,){/Straight}{0}
\FALabel(17.,15.)[b]{$\tilde g$}
\FAProp(20.,5.)(10.,17)(0.,){/ScalarDash}{-1}
\FALabel(17.,6.86964)[t]{$\tilde u_L$}
\FAProp(10.,17)(10.,13.5)(0.,){/ScalarDash}{-1}
\FALabel(9.5,15.)[r]{$\tilde u_L$}
\FAProp(10.,5.5)(10.,9.)(0.,){/ScalarDash}{1}
\FALabel(9.5,7.4)[r]{$\tilde u_a$}
\FAProp(10.,13.5)(10.,9.)(.85,){/Straight}{0}
\FALabel(7.4,11.5)[r]{$\tilde \chi_n^{0/ \pm}$}
\FAProp(10.,13.5)(10.,9.)(-.75,){/Straight}{-1}
\FALabel(12.5,11.1)[l]{$q$}
\FAVert(10.,17){0}
\FAVert(10.,5.5){0}
\FAVert(10.,13.5){0}
\FAVert(10.,9.){0}

\FADiagram{}
\FAProp(0.,17.5)(10.,17)(0.,){/Cycles}{0}
\FALabel(5.11236,18.5)[b]{$g$}
\FAProp(0.,5.)(10.,5.5)(0.,){/Straight}{1}
\FALabel(5.0774,4.18)[t]{$u$}
\FAProp(20.,16.7)(10.,5.5)(0.,){/Straight}{0}
\FALabel(17.2,15.1)[b]{$\tilde g$}
\FAProp(20.,5.5)(10.,17)(0.,){/ScalarDash}{-1}
\FALabel(17.2,6.6)[t]{$\tilde u_L$}
\FAProp(10.,17)(10.,13.5)(0.,){/ScalarDash}{-1}
\FALabel(9.5,15.)[r]{$\tilde u_L$}
\FAProp(10.,5.5)(10.,9.)(0.,){/ScalarDash}{1}
\FALabel(9.5,7.4)[r]{$\tilde u_a$}
\FAProp(10.,13.5)(10.,9.)(.85,){/ScalarDash}{0}
\FALabel(7.4,11.5)[r]{$S^{0/\pm}$}
\FAProp(10.,13.5)(10.,9.)(-.67,){/ScalarDash}{-1}
\FALabel(12.2,11.2)[l]{$\tilde q_b$}
\FAVert(10.,17){0}
\FAVert(10.,5.5){0}
\FAVert(10.,13.5){0}
\FAVert(10.,9.){0}

\FADiagram{}
\FAProp(0.,17.5)(10.,17)(0.,){/Cycles}{0}
\FALabel(5.11236,18.5)[b]{$g$}
\FAProp(0.,5.)(10.,5.5)(0.,){/Straight}{1}
\FALabel(5.0774,4.18)[t]{$u$}
\FAProp(20.,16.7)(10.,5.5)(0.,){/Straight}{0}
\FALabel(17.2,15.1)[b]{$\tilde g$}
\FAProp(20.,5.5)(10.,17)(0.,){/ScalarDash}{-1}
\FALabel(17.2,6.6)[t]{$\tilde u_L$}
\FAProp(10.,17)(10.,13.5)(0.,){/ScalarDash}{-1}
\FALabel(9.5,15.)[r]{$\tilde u_L$}
\FAProp(10.,5.5)(10.,9.)(0.,){/ScalarDash}{1}
\FALabel(9.5,7.4)[r]{$\tilde u_{L/a}$}
\FAProp(10.,13.5)(10.,9.)(.9,){/Sine}{0}
\FALabel(7.,11.5)[r]{$V$}
\FAProp(10.,13.5)(10.,9.)(-.67,){/ScalarDash}{-1}
\FALabel(12.2,11.2)[l]{$\tilde q_b$}
\FAVert(10.,17){0}
\FAVert(10.,5.5){0}
\FAVert(10.,13.5){0}
\FAVert(10.,9.){0}

\end{feynartspicture}
}
\vspace*{-.3cm}\\
\put(2,.2){(b)}
\\[-.3ex]

\hspace*{-.3cm}
\vspace*{-.3cm}
\scalebox{.85}{
\begin{feynartspicture}(12,3)(4,1)

\FADiagram{}
\FAProp(0.,15.)(10.,14.5)(0.,){/Cycles}{0}
\FALabel(5.,16.51)[b]{$g$}
\FAProp(0.,3.)(6.5,3.5)(0.,){/Straight}{1}
\FALabel(3.4,5.5)[t]{$u$}
\FAProp(20.,15.)(10.,14.5)(0.,){/Straight}{0}
\FALabel(17.3,15.3)[br]{$\tilde g$}
\FAProp(20.,3.)(13.5,3.5)(0.,){/ScalarDash}{-1}
\FALabel(16.6311,4.5)[b]{$\tilde u_L$}
\FAProp(10.,14.5)(10.,9.5)(0.,){/Straight}{0}
\FALabel(10.82,11.75)[l]{$\tilde g$}
\FAProp(6.5,3.5)(13.5,3.5)(0.,){/Sine}{0}
\FALabel(10.,2.5)[t]{$V$}
\FAProp(6.5,3.5)(10.,9.5)(0.,){/Straight}{1}
\FALabel(7.42232,7.2)[br]{$q$}
\FAProp(13.5,3.5)(10.,9.5)(0.,){/ScalarDash}{-1}
\FALabel(12.5777,7.2)[bl]{$\tilde q_a$}
\FAVert(10.,14.5){0}
\FAVert(6.5,3.5){0}
\FAVert(13.5,3.5){0}
\FAVert(10.,9.5){0}

\FADiagram{}
\FAProp(0.,15.)(10.,14.5)(0.,){/Cycles}{0}
\FALabel(5,16.51)[b]{$g$}
\FAProp(0.,3.)(6.5,3.5)(0.,){/Straight}{1}
\FALabel(3.4,5.5)[t]{$u$}
\FAProp(20.,15.)(10.,14.5)(0.,){/Straight}{0}
\FALabel(17.3,15.3)[br]{$\tilde g$}
\FAProp(20.,3.)(13.5,3.5)(0.,){/ScalarDash}{-1}
\FALabel(16.6311,4.5)[b]{$\tilde u_L$}
\FAProp(10.,14.5)(10.,9.5)(0.,){/Straight}{0}
\FALabel(10.82,11.75)[l]{$\tilde g$}
\FAProp(6.5,3.5)(13.5,3.5)(0.,){/Straight}{0}
\FALabel(10.,3.5)[t]{$\tilde \chi_n^{0/\pm}$}
\FAProp(6.5,3.5)(10.,9.5)(0.,){/ScalarDash}{1}
\FALabel(7.42232,7.2)[br]{$\tilde q_a$}
\FAProp(13.5,3.5)(10.,9.5)(0.,){/Straight}{-1}
\FALabel(12.5777,7.2)[bl]{$q$}
\FAVert(10.,14.5){0}
\FAVert(6.5,3.5){0}
\FAVert(13.5,3.5){0}
\FAVert(10.,9.5){0}

\FADiagram{ }
\FAProp(0.,15.)(10.,14.5)(0.,){/Cycles}{0}
\FALabel(5.,16.51)[b]{$g$}
\FAProp(0.,3.)(6.5,3.5)(0.,){/Straight}{1}
\FALabel(3.2,5.5)[t]{$u$}
\FAProp(20.,15.)(13.5,3.5)(0.,){/Straight}{0}
\FALabel(18.3,14.5)[br]{$\tilde g$}
\FAProp(20.,3.)(10,14.5)(0.,){/ScalarDash}{-1}
\FALabel(18.3,5.)[rt]{$\tilde u_L$}
\FAProp(10.,14.5)(10.,9.5)(0.,){/ScalarDash}{-1}
\FALabel(9.03,12.)[r]{$\tilde u_L$}
\FAProp(6.5,3.5)(13.5,3.5)(0.,){/Straight}{1}
\FALabel(10.,3)[t]{$u$}
\FAProp(6.5,3.5)(10.,9.5)(0.,){/Sine}{0}
\FALabel(8,6.5)[br]{$V$}
\FAProp(13.5,3.5)(10.,9.5)(0.,){/ScalarDash}{1}
\FALabel(12.,7.2)[bl]{$\tilde q_a$}
\FAVert(10.,14.5){0}
\FAVert(6.5,3.5){0}
\FAVert(13.5,3.5){0}
\FAVert(10.,9.5){0}

\FADiagram{ }
\FAProp(0.,15.)(10.,14.5)(0.,){/Cycles}{0}
\FALabel(5.,16.51)[b]{$g$}
\FAProp(0.,3.)(6.5,3.5)(0.,){/Straight}{1}
\FALabel(3.2,5.5)[t]{$u$}
\FAProp(20.,15.)(13.5,3.5)(0.,){/Straight}{0}
\FALabel(18.3,14.5)[br]{$\tilde g$}
\FAProp(20.,3.)(10,14.5)(0.,){/ScalarDash}{-1}
\FALabel(18.3,5.)[rt]{$\tilde u_L$}
\FAProp(10.,14.5)(10.,9.5)(0.,){/ScalarDash}{-1}
\FALabel(9.03,12.)[r]{$\tilde u_L$}
\FAProp(6.5,3.5)(13.5,3.5)(0.,){/ScalarDash}{1}
\FALabel(10.,3)[t]{$\tilde q_a$}
\FAProp(6.5,3.5)(10.,9.5)(0.,){/Straight}{0}
\FALabel(9,7.)[br]{$\tilde \chi_n^{0/\pm}$}
\FAProp(13.5,3.5)(10.,9.5)(0.,){/Straight}{1}
\FALabel(12.,7.2)[bl]{$q$}
\FAVert(10.,14.5){0}
\FAVert(6.5,3.5){0}
\FAVert(13.5,3.5){0}
\FAVert(10.,9.5){0}

\end{feynartspicture}
}

\hspace*{-.3cm}
\vspace*{-.5cm}
\scalebox{.89}{
\begin{feynartspicture}(12,3)(4,1)

\FADiagram{ }
\FAProp(1.,15.)(7.5,14.5)(0.,){/Cycles}{0}
\FALabel(4.4,16.4)[b]{$g$}
\FAProp(1.,3.)(11.,3.5)(0.,){/Straight}{1}
\FALabel(4.4,5.5)[t]{$u$}
\FAProp(20.,3.)(11.,3.5)(0.,){/ScalarDash}{-1}
\FALabel(18.3,4.5)[rb]{$\tilde u_L$}
\FAProp(20.,15.)(13.5,14.5)(0.,){/Straight}{0}
\FALabel(18.,15.5)[br]{$\tilde g$}
\FAProp(11.,3.5)(11.,8.5)(0.,){/Straight}{0}
\FALabel(10.18,6.5)[r]{$\tilde \chi_n^0$}
\FAProp(7.5,14.5)(13.5,14.5)(0.,){/Straight}{-1}
\FALabel(11.,15.57)[b]{$q_i$}
\FAProp(7.5,14.5)(11.,8.5)(0.,){/Straight}{1}
\FALabel(8.39114,11.199)[tr]{$q_i$}
\FAProp(13.5,14.5)(11.,8.5)(0.,){/ScalarDash}{-1}
\FALabel(12.6,11.199)[tl]{$\tilde{q}_{ia}$}
\FAVert(7.5,14.5){0}
\FAVert(11.,3.5){0}
\FAVert(13.5,14.5){0}
\FAVert(11.,8.5){0}

\FADiagram{ }
\FAProp(1.,15.)(7.5,14.5)(0.,){/Cycles}{0}
\FALabel(4.4,16.5134)[b]{$g$}
\FAProp(1.,3.)(11.,3.5)(0.,){/Straight}{1}
\FALabel(4.4,5.5)[t]{$u$}
\FAProp(20.,3.)(11.,3.5)(0.,){/ScalarDash}{-1}
\FALabel(18.3,4.5)[rb]{$\tilde u_L$}
\FAProp(20.,15.)(13.5,14.5)(0.,){/Straight}{0}
\FALabel(18.,15.5)[br]{$\tilde g$}
\FAProp(11.,3.5)(11.,8.5)(0.,){/Straight}{0}
\FALabel(10.18,6.5)[r]{$\tilde \chi_n^0$}
\FAProp(7.5,14.5)(13.5,14.5)(0.,){/Straight}{1}
\FALabel(11.,15.57)[b]{$q_i$}
\FAProp(7.5,14.5)(11.,8.5)(0.,){/Straight}{-1}
\FALabel(8.39114,11.199)[tr]{$q_i$}
\FAProp(13.5,14.5)(11.,8.5)(0.,){/ScalarDash}{1}
\FALabel(12.6,11.199)[tl]{$\tilde q_{ia}$}
\FAVert(7.5,14.5){0}
\FAVert(11.,3.5){0}
\FAVert(13.5,14.5){0}
\FAVert(11.,8.5){0}

\FADiagram{}
\FAProp(1.,15.)(7.5,14.5)(0.,){/Cycles}{0}
\FALabel(4.4,16.5134)[b]{$g$}
\FAProp(1.,3.)(11.,3.5)(0.,){/Straight}{1}
\FALabel(4.4,5.5)[t]{$u$}
\FAProp(20.,3.)(11.,3.5)(0.,){/ScalarDash}{-1}
\FALabel(18.3,4.5)[rb]{$\tilde u_L$}
\FAProp(20.,15.)(13.5,14.5)(0.,){/Straight}{0}
\FALabel(18.,15.5)[br]{$\tilde g$}
\FAProp(11.,3.5)(11.,8.5)(0.,){/Straight}{0}
\FALabel(10.18,6.5)[r]{$\tilde \chi_n^0$}
\FAProp(7.5,14.5)(13.5,14.5)(0.,){/ScalarDash}{-1}
\FALabel(11.,15.57)[b]{$\tilde{q}_{ia}$}
\FAProp(7.5,14.5)(11.,8.5)(0.,){/ScalarDash}{1}
\FALabel(8.39114,11.199)[tr]{$\tilde{q}_{ia}$}
\FAProp(13.5,14.5)(11.,8.5)(0.,){/Straight}{-1}
\FALabel(12.8,11.199)[tl]{$q_i$}
\FAVert(7.5,14.5){0}
\FAVert(11.,3.5){0}
\FAVert(13.5,14.5){0}
\FAVert(11.,8.5){0}

\FADiagram{}
\FAProp(1.,15.)(7.5,14.5)(0.,){/Cycles}{0}
\FALabel(4.4,16.5134)[b]{$g$}
\FAProp(1.,3.)(11.,3.5)(0.,){/Straight}{1}
\FALabel(4.4,5.5)[t]{$u$}
\FAProp(20.,3.)(11.,3.5)(0.,){/ScalarDash}{-1}
\FALabel(18.3,4.5)[rb]{$\tilde u_L$}
\FAProp(20.,15.)(13.5,14.5)(0.,){/Straight}{0}
\FALabel(18.,15.5)[br]{$\tilde g$}
\FAProp(11.,3.5)(11.,8.5)(0.,){/Straight}{0}
\FALabel(10.18,6.5)[r]{$\tilde \chi_n^0$}
\FAProp(7.5,14.5)(13.5,14.5)(0.,){/ScalarDash}{1}
\FALabel(11.,15.57)[b]{$\tilde{q}_{ia}$}
\FAProp(7.5,14.5)(11.,8.5)(0.,){/ScalarDash}{-1}
\FALabel(8.39114,11.199)[tr]{$\tilde{q}_{ia}$}
\FAProp(13.5,14.5)(11.,8.5)(0.,){/Straight}{1}
\FALabel(12.8,11.199)[tl]{$q_i$}
\FAVert(7.5,14.5){0}
\FAVert(11.,3.5){0}
\FAVert(13.5,14.5){0}
\FAVert(11.,8.5){0}

\end{feynartspicture}
}

\vspace*{-.3cm}
\hspace*{-.3cm}
\scalebox{.85}{
\begin{feynartspicture}(15,3)(5,1)

\FADiagram{ }
\FAProp(0.,15.)(6.5,14.5)(0.,){/Cycles}{0}
\FALabel(3.4,16.5134)[b]{$g$}
\FAProp(0.,3.)(10.,3.5)(0.,){/Straight}{1}
\FALabel(3.2,5.5)[t]{$u$}
\FAProp(20.,15.)(10.,3.5)(0.,){/Straight}{0}
\FALabel(18.3,14.5)[br]{$\tilde g$}
\FAProp(20.,4.)(12.5,14.5)(0.,){/ScalarDash}{-1}
\FALabel(18.3,5.5)[rt]{$\tilde u_L$}
\FAProp(10.,3.5)(10.,9)(0.,){/ScalarDash}{1}
\FALabel(9.3,7.1)[r]{$\tilde{u}_{L/a}$}
\FAProp(6.5,14.5)(12.5,14.5)(0.,){/ScalarDash}{1}
\FALabel(10.,15.4)[b]{$\tilde q_b$}
\FAProp(6.5,14.5)(10.,9)(0.,){/ScalarDash}{-1}
\FALabel(7.39114,11.5)[tr]{$\tilde q_b$}
\FAProp(12.5,14.5)(10.,9)(0.,){/Sine}{0}
\FALabel(11.5,10.9)[tl]{$V$}
\FAVert(6.5,14.5){0}
\FAVert(10.,3.5){0}
\FAVert(12.5,14.5){0}
\FAVert(10.,9){0}

\FADiagram{ }
\FAProp(0.,15.)(6.5,14.5)(0.,){/Cycles}{0}
\FALabel(3.4,16.5134)[b]{$g$}
\FAProp(0.,3.)(10.,3.5)(0.,){/Straight}{1}
\FALabel(3.4,5.5)[t]{$u$}
\FAProp(20.,15.)(10.,3.5)(0.,){/Straight}{0}
\FALabel(18.3,14.5)[br]{$\tilde g$}
\FAProp(20.,4.)(12.5,14.5)(0.,){/ScalarDash}{-1}
\FALabel(18.3,5.5)[rt]{$\tilde u_L$}
\FAProp(10.,3.5)(10.,9)(0.,){/ScalarDash}{1}
\FALabel(8.93,7.)[r]{$\tilde u_a$}
\FAProp(6.5,14.5)(12.5,14.5)(0.,){/ScalarDash}{1}
\FALabel(10.,15.4)[b]{$\tilde q_b$}
\FAProp(6.5,14.5)(10.,9)(0.,){/ScalarDash}{-1}
\FALabel(7.39114,11.5)[tr]{$\tilde q_b$}
\FAProp(12.5,14.5)(10.,9)(0.,){/ScalarDash}{0}
\FALabel(10.8,11.)[tl]{$S^{0/\pm}$}
\FAVert(6.5,14.5){0}
\FAVert(10.,3.5){0}
\FAVert(12.5,14.5){0}
\FAVert(10.,9){0}

\FADiagram{ }
\FAProp(0.,15.)(6.5,14.5)(0.,){/Cycles}{0}
\FALabel(3.4,16.5134)[b]{$g$}
\FAProp(0.,3.)(10.,3.5)(0.,){/Straight}{1}
\FALabel(3.4,5.5)[t]{$u$}
\FAProp(20.,15.)(10.,3.5)(0.,){/Straight}{0}
\FALabel(18.3,14.5)[br]{$\tilde g$}
\FAProp(20.,4.)(12.5,14.5)(0.,){/ScalarDash}{-1}
\FALabel(18.3,5.5)[rt]{$\tilde u_L$}
\FAProp(10.,3.5)(10.,9.)(0.,){/ScalarDash}{1}
\FALabel(8.93,7.)[r]{$\tilde u_a$}
\FAProp(6.5,14.5)(12.5,14.5)(0.,){/Straight}{1}
\FALabel(10.,15.4)[b]{$q$}
\FAProp(6.5,14.5)(10.,9.)(0.,){/Straight}{-1}
\FALabel(7.39114,11.5)[tr]{$q$}
\FAProp(12.5,14.5)(10.,9.)(0.,){/Straight}{0}
\FALabel(10.3,11.)[tl]{$\tilde \chi_n^{0/\pm}$}
\FAVert(6.5,14.5){0}
\FAVert(10.,5.5){0}
\FAVert(12.5,14.5){0}
\FAVert(10.,9.){0}
 
\FADiagram{}
\FAProp(0.,15.)(4.,9.)(0.,){/Cycles}{0}
\FALabel(2,11.0117)[tr]{$g$}
\FAProp(0.,3.)(4.,9.)(0.,){/Straight}{1}
\FALabel(2.73035,6.)[tl]{$u$}
\FAProp(20.,15.)(16.,12.5)(0.,){/Straight}{0}
\FALabel(17.5435,14.2)[b]{$\tilde g$}
\FAProp(20.,3.)(16.,5.5)(0.,){/ScalarDash}{-1}
\FALabel(17.4558,3.77869)[t]{$\tilde u_L$}
\FAProp(4.,9.)(10.,9.)(0.,){/Straight}{1}
\FALabel(7.,10.07)[b]{$u$}
\FAProp(16.,12.5)(16.,5.5)(0.,){/ScalarDash}{1}
\FALabel(17.07,9.)[l]{$\tilde q_a$}
\FAProp(16.,12.5)(10.,9.)(0.,){/Straight}{-1}
\FALabel(12.699,11.6089)[br]{$q$}
\FAProp(16.,5.5)(10.,9.)(0.,){/Sine}{0}
\FALabel(13.2,6.1)[tr]{$V$}
\FAVert(4.,9.){0}
\FAVert(16.,12.5){0}
\FAVert(16.,5.5){0}
\FAVert(10.,9.){0}

\FADiagram{}
\FAProp(0.,15.)(4.,9.)(0.,){/Cycles}{0}
\FALabel(2,11.0117)[tr]{$g$}
\FAProp(0.,3.)(4.,9.)(0.,){/Straight}{1}
\FALabel(2.73035,5.)[tl]{$u$}
\FAProp(20.,15.)(16.,12.5)(0.,){/Straight}{0}
\FALabel(17.5435,14.2)[b]{$\tilde g$}
\FAProp(20.,3.)(16.,5.5)(0.,){/ScalarDash}{-1}
\FALabel(17.4558,3.77869)[t]{$\tilde u_L$}
\FAProp(4.,9.)(10.,9.)(0.,){/Straight}{1}
\FALabel(7.,10.07)[b]{$u$}
\FAProp(16.,12.5)(16.,5.5)(0.,){/Straight}{1}
\FALabel(17.07,9.)[l]{$q$}
\FAProp(16.,12.5)(10.,9.)(0.,){/ScalarDash}{-1}
\FALabel(12.699,11.6089)[br]{$\tilde q_a$}
\FAProp(16.,5.5)(10.,9.)(0.,){/Straight}{0}
\FALabel(13,6.60709)[tr]{$\tilde \chi_n^{0/ \pm}$}
\FAVert(4.,9.){0}
\FAVert(16.,12.5){0}
\FAVert(16.,5.5){0}
\FAVert(10.,9.){0}

\end{feynartspicture}
}

\scalebox{.85}{
\begin{feynartspicture}(15,3)(5,1)

\FADiagram{}
\FAProp(0.,15.)(4.5,14.5)(0.,){/Cycles}{0}
\FALabel(2.49847,16.5062)[b]{$g$}
\FAProp(0.,3.)(10.,3.5)(0.,){/Straight}{1}
\FALabel(3.4,5.5)[t]{$u$}
\FAProp(20.,15.)(10.,3.5)(0.,){/Straight}{0}
\FALabel(18,14.)[br]{$\tilde g$}
\FAProp(20.,5.)(10.,14.)(0.,){/ScalarDash}{-1}
\FALabel(18.3,5.5)[rt]{$\tilde u_L$}
\FAProp(10.,3.5)(10.,14.)(0.,){/ScalarDash}{1}
\FALabel(11.07,9.)[l]{$\tilde u_a$}
\FAProp(4.5,14.5)(10.,14.)(0.8,){/ScalarDash}{-1}
\FALabel(6.90967,10.9864)[t]{$\tilde{q}_{ib}$}
\FAProp(4.5,14.5)(10.,14.)(-0.8,){/ScalarDash}{1}
\FALabel(7.59033,17.5136)[b]{$\tilde{q}_{ib}$}
\FAVert(4.5,14.5){0}
\FAVert(10.,3.5){0}
\FAVert(10.,14.){0}

\FADiagram{}
\FAProp(0.,15.)(10.,4.)(0.,){/Cycles}{0}
\FALabel(3.65552,13.7261)[bl]{$g$}
\FAProp(0.,3.)(10.,14.)(0.,){/Straight}{1}
\FALabel(2.5,4)[tl]{$u$}
\FAProp(20.,15.)(10.,14.)(0.,){/Straight}{0}
\FALabel(17,15.3135)[b]{$\tilde g$}
\FAProp(20.,3.)(15.55,3.4)(0.,){/ScalarDash}{-1}
\FALabel(17.9138,5)[b]{$\tilde u_L$}
\FAProp(10.,14.)(10.,4.)(0.,){/ScalarDash}{1}
\FALabel(10.92,11.)[l]{$\tilde u_L$}
\FAProp(15.55,3.4)(10.,4.)(0.8,){/ScalarDash}{-1}
\FALabel(13.1816,7.)[b]{$\tilde q_b$}
\FAProp(15.55,3.4)(10.,4.)(-0.8,){/Sine}{0}
\FALabel(12.3684,1.)[t]{$V$}
\FAVert(10.,14.){0}
\FAVert(15.55,3.4){0}
\FAVert(10.,4.){0}

\FADiagram{}
\FAProp(0.,15.)(9.5,14.5)(0.,){/Cycles}{0}
\FALabel(4.86826,16.5169)[b]{$g$}
\FAProp(0.,3.)(10.,3.5)(0.,){/Straight}{1}
\FALabel(5.,2.8)[t]{$u$}
\FAProp(20.,15.)(10.,3.5)(0.,){/Straight}{0}
\FALabel(17.3443,13.3164)[br]{$\tilde g$}
\FAProp(20.,3.)(9.5,14.5)(0.,){/ScalarDash}{-1}
\FALabel(17.5,5.5)[tr]{$\tilde u_L$}
\FAProp(10.,3.5)(7.5,9.)(0.,){/ScalarDash}{1}
\FALabel(7.42657,7.53469)[tr]{$\tilde u_{L/a}$}
\FAProp(7.5,9.)(9.5,14.5)(0.8,){/ScalarDash}{1}
\FALabel(10.511,9.3382)[l]{$\tilde q_b$}
\FAProp(7.5,9.)(9.5,14.5)(-0.8,){/Sine}{0}
\FALabel(5,10.986)[r]{$V$}
\FAVert(10.,3.5){0}
\FAVert(7.5,9.){0}
\FAVert(9.5,14.5){0}

\FADiagram{}
\FAProp(0.,15.)(4.,12.5)(0.,){/Cycles}{0}
\FALabel(3,14.8767)[bl]{$g$}
\FAProp(0.,3.)(4.,5.5)(0.,){/Straight}{1}
\FALabel(3,3.77869)[t]{$u$}
\FAProp(20.,15.)(16.,9.)(0.,){/Straight}{0}
\FALabel(17.5,12.5)[br]{$\tilde g$}
\FAProp(20.,3.)(16.,9.)(0.,){/ScalarDash}{-1}
\FALabel(17.8,3.2)[br]{$\tilde u_L$}
\FAProp(16.,9.)(10.,9.)(0.,){/Straight}{-1}
\FALabel(13.,10)[b]{$u$}
\FAProp(4.,12.5)(4.,5.5)(0.,){/Straight}{-1}
\FALabel(2.93,9.)[r]{$q$}
\FAProp(4.,12.5)(10.,9.)(0.,){/Straight}{1}
\FALabel(7.301,11.6089)[bl]{$q$}
\FAProp(4.,5.5)(10.,9.)(0.,){/Sine}{0}
\FALabel(7,6.5)[tl]{$V$}
\FAVert(4.,12.5){0}
\FAVert(4.,5.5){0}
\FAVert(16.,9.){0}
\FAVert(10.,9.){0}

\FADiagram{}
\FAProp(0.,15.)(4.,12.5)(0.,){/Cycles}{0}
\FALabel(3,14.8767)[bl]{$g$}
\FAProp(0.,3.)(4.,5.5)(0.,){/Straight}{1}
\FALabel(3,3.77869)[t]{$u$}
\FAProp(20.,15.)(16.,9.)(0.,){/Straight}{0}
\FALabel(17.5,12.5)[br]{$\tilde g$}
\FAProp(20.,3.)(16.,9.)(0.,){/ScalarDash}{-1}
\FALabel(17.8,3.2)[br]{$\tilde u_L$}
\FAProp(16.,9.)(10.,9.)(0.,){/Straight}{-1}
\FALabel(13.,10)[b]{$u$}
\FAProp(4.,12.5)(4.,5.5)(0.,){/ScalarDash}{-1}
\FALabel(2.93,9.)[r]{$\tilde q_a$}
\FAProp(4.,12.5)(10.,9.)(0.,){/ScalarDash}{1}
\FALabel(7.301,11.6089)[bl]{$\tilde q_a$}
\FAProp(4.,5.5)(10.,9.)(0.,){/Straight}{0}
\FALabel(7.,6.9)[tl]{$\tilde \chi_n^{0/\pm}$}
\FAVert(4.,12.5){0}
\FAVert(4.,5.5){0}
\FAVert(16.,9.){0}
\FAVert(10.,9.){0}

\end{feynartspicture}
}
\\
\put(0,0){(c)}
\caption{Feynman diagrams for (a) box, (b) self energy, and (c) vertex correction contributions. 
In case of $\gamma$ exchange, $q$ denotes an $u$ quark, and $\tilde{q}_a \equiv \tilde{u}_L$.
For $Z/W$ boson, $\tilde{\chi}^0_n/\, \tilde{\chi}_n^{\pm}$, and $S^0/\,S^{\pm}$
 exchange, it is $q \equiv u/d$ and $\tilde{q}_a \equiv \tilde{u}_a/\tilde{d}_a$.}
\label{fig_vertexdiags}
}

\FIGURE{

\vspace*{-.5cm}
\hspace*{-.3cm}
\scalebox{.85}{
\begin{feynartspicture}(15,3)(5,1)

\FADiagram{}
\FAProp(0.,15.)(5.5,10.)(0.,){/Cycles}{0}
\FALabel(2.18736,11.8331)[tr]{$g$}
\FAProp(0.,5.)(5.5,10.)(0.,){/Straight}{1}
\FALabel(3.31264,6.83309)[tl]{$u$}
\FAProp(20.,17.)(11.5,10.)(0.,){/Straight}{0}
\FALabel(15.4036,14.0235)[br]{$\tilde g$}
\FAProp(20.,10.)(15.5,6.5)(0.,){/ScalarDash}{-1}
\FALabel(17.2784,8.9935)[br]{$\tilde u_L$}
\FAProp(20.,3.)(15.5,6.5)(0.,){/Sine}{0}
\FALabel(18.2216,5.4935)[bl]{$\gamma$}
\FAProp(5.5,10.)(11.5,10.)(0.,){/Straight}{1}
\FALabel(8.5,11.07)[b]{$u$}
\FAProp(11.5,10.)(15.5,6.5)(0.,){/ScalarDash}{1}
\FALabel(12.9593,7.56351)[tr]{$\tilde u_L$}
\FAVert(5.5,10.){0}
\FAVert(11.5,10.){0}
\FAVert(15.5,6.5){0}

\FADiagram{}
\FAProp(0.,15.)(10.,14.5)(0.,){/Cycles}{0}
\FALabel(5.11236,16.5172)[b]{$g$}
\FAProp(0.,5.)(10.,7.)(0.,){/Straight}{1}
\FALabel(5.30398,4.9601)[t]{$u$}
\FAProp(20.,17.)(10.,14.5)(0.,){/Straight}{0}
\FALabel(14.6847,16.5312)[b]{$\tilde g$}
\FAProp(20.,10.)(15.5,6.5)(0.,){/ScalarDash}{-1}
\FALabel(17.2784,8.9935)[br]{$\tilde u_L$}
\FAProp(20.,3.)(15.5,6.5)(0.,){/Sine}{0}
\FALabel(17.2784,4.0065)[tr]{$\gamma$}
\FAProp(10.,14.5)(10.,7.)(0.,){/Straight}{0}
\FALabel(9.18,10.75)[r]{$\tilde g$}
\FAProp(10.,7.)(15.5,6.5)(0.,){/ScalarDash}{1}
\FALabel(12.6097,5.68637)[t]{$\tilde u_L$}
\FAVert(10.,14.5){0}
\FAVert(10.,7.){0}
\FAVert(15.5,6.5){0}

\FADiagram{}
\FAProp(0.,15.)(10.,7.)(0.,){/Cycles}{0}
\FALabel(3.26346,14.4618)[bl]{$g$}
\FAProp(0.,5.)(10.,14.5)(0.,){/Straight}{1}
\FALabel(3.04219,6.54875)[tl]{$u$}
\FAProp(20.,17.)(10.,14.5)(0.,){/Straight}{0}
\FALabel(14.6847,16.5312)[b]{$\tilde g$}
\FAProp(20.,10.)(15.5,6.5)(0.,){/ScalarDash}{-1}
\FALabel(17.2784,8.9935)[br]{$\tilde u_L$}
\FAProp(20.,3.)(15.5,6.5)(0.,){/Sine}{0}
\FALabel(17.2784,4.0065)[tr]{$\gamma$}
\FAProp(10.,7.)(10.,14.5)(0.,){/ScalarDash}{-1}
\FALabel(11.07,10.75)[l]{$\tilde u_L$}
\FAProp(10.,7.)(15.5,6.5)(0.,){/ScalarDash}{1}
\FALabel(12.6097,5.68637)[t]{$\tilde u_L$}
\FAVert(10.,7.){0}
\FAVert(10.,14.5){0}
\FAVert(15.5,6.5){0}

\FADiagram{}
\FAProp(0.,15.)(10.,14.5)(0.,){/Cycles}{0}
\FALabel(5.11236,16.5172)[b]{$g$}
\FAProp(0.,5.)(10.,5.5)(0.,){/Straight}{1}
\FALabel(5.0774,4.18193)[t]{$u$}
\FAProp(20.,17.)(10.,14.5)(0.,){/Straight}{0}
\FALabel(14.6847,16.5312)[b]{$\tilde g$}
\FAProp(20.,10.)(10.,10.)(0.,){/ScalarDash}{-1}
\FALabel(15.,11.07)[b]{$\tilde u_L$}
\FAProp(20.,3.)(10.,5.5)(0.,){/Sine}{0}
\FALabel(15.3759,5.27372)[b]{$\gamma$}
\FAProp(10.,14.5)(10.,10.)(0.,){/Straight}{0}
\FALabel(9.18,12.25)[r]{$\tilde g$}
\FAProp(10.,5.5)(10.,10.)(0.,){/Straight}{1}
\FALabel(8.93,7.75)[r]{$u$}
\FAVert(10.,14.5){0}
\FAVert(10.,5.5){0}
\FAVert(10.,10.){0}

\FADiagram{}
\FAProp(0.,15.)(10.,14.5)(0.,){/Cycles}{0}
\FALabel(5.11236,16.5172)[b]{$g$}
\FAProp(0.,5.)(10.,5.5)(0.,){/Straight}{1}
\FALabel(5.0774,4.18193)[t]{$u$}
\FAProp(20.,17.)(10.,10.)(0.,){/Straight}{0}
\FALabel(17.5821,16.1028)[br]{$\tilde g$}
\FAProp(20.,10.)(10.,14.5)(0.,){/ScalarDash}{-1}
\FALabel(17.2774,10.302)[tr]{$\tilde u_L$}
\FAProp(20.,3.)(10.,5.5)(0.,){/Sine}{0}
\FALabel(14.6241,3.22628)[t]{$\gamma$}
\FAProp(10.,14.5)(10.,10.)(0.,){/ScalarDash}{-1}
\FALabel(8.93,12.25)[r]{$\tilde u_L$}
\FAProp(10.,5.5)(10.,10.)(0.,){/Straight}{1}
\FALabel(8.93,7.75)[r]{$u$}
\FAVert(10.,14.5){0}
\FAVert(10.,5.5){0}
\FAVert(10.,10.){0}
\end{feynartspicture}
}

\hspace*{-.3cm}
\vspace*{-.2cm}
\scalebox{.85}{
\begin{feynartspicture}(12,3)(4,1)

\FADiagram{}
\FAProp(0.,15.)(10.,13.)(0.,){/Cycles}{0}
\FALabel(5.44126,15.7263)[b]{$g$}
\FAProp(0.,5.)(10.,5.5)(0.,){/Straight}{1}
\FALabel(5.0774,4.18193)[t]{$u$}
\FAProp(20.,17.)(15.5,13.5)(0.,){/Straight}{0}
\FALabel(17.4319,15.7962)[br]{$\tilde g$}
\FAProp(20.,10.)(15.5,13.5)(0.,){/ScalarDash}{-1}
\FALabel(18.2216,12.4935)[bl]{$\tilde u_L$}
\FAProp(20.,3.)(10.,5.5)(0.,){/Sine}{0}
\FALabel(15.3759,5.27372)[b]{$\gamma$}
\FAProp(10.,13.)(10.,5.5)(0.,){/Straight}{-1}
\FALabel(8.93,9.25)[r]{$u$}
\FAProp(10.,13.)(15.5,13.5)(0.,){/Straight}{1}
\FALabel(12.8903,12.1864)[t]{$u$}
\FAVert(10.,13.){0}
\FAVert(10.,5.5){0}
\FAVert(15.5,13.5){0}

\FADiagram{}
\FAProp(0.,15.)(10.,6.)(0.,){/Cycles}{0}
\FALabel(3.75552,13.9761)[bl]{$g$}
\FAProp(0.,5.)(10.,14.5)(0.,){/Straight}{1}
\FALabel(2.92329,6.67666)[tl]{$u$}
\FAProp(20.,17.)(10.,14.5)(0.,){/Straight}{0}
\FALabel(14.6847,16.5312)[b]{$\tilde g$}
\FAProp(20.,10.)(10.,6.)(0.,){/ScalarDash}{-1}
\FALabel(17.5929,7.74766)[t]{$\tilde u_L$}
\FAProp(20.,3.)(10.,6.)(0.,){/Sine}{0}
\FALabel(14.5546,3.49537)[t]{$\gamma$}
\FAProp(10.,14.5)(10.,6.)(0.,){/ScalarDash}{1}
\FALabel(11.07,10.25)[l]{$\tilde u_L$}
\FAVert(10.,14.5){0}
\FAVert(10.,6.){0}

\FADiagram{}
\FAProp(0.,15.)(5.5,10.)(0.,){/Cycles}{0}
\FALabel(2.18736,11.8331)[tr]{$g$}
\FAProp(0.,5.)(5.5,10.)(0.,){/Straight}{1}
\FALabel(3.31264,6.83309)[tl]{$u$}
\FAProp(20.,17.)(15.5,13.5)(0.,){/Straight}{0}
\FALabel(17.4319,15.7962)[br]{$\tilde g$}
\FAProp(20.,10.)(15.5,13.5)(0.,){/ScalarDash}{-1}
\FALabel(18.2216,12.4935)[bl]{$\tilde u_L$}
\FAProp(20.,3.)(12.,10.)(0.,){/Sine}{0}
\FALabel(15.4593,5.81351)[tr]{$\gamma$}
\FAProp(5.5,10.)(12.,10.)(0.,){/Straight}{1}
\FALabel(8.75,8.93)[t]{$u$}
\FAProp(15.5,13.5)(12.,10.)(0.,){/Straight}{-1}
\FALabel(13.134,12.366)[br]{$u$}
\FAVert(5.5,10.){0}
\FAVert(15.5,13.5){0}
\FAVert(12.,10.){0}

\FADiagram{}
\FAProp(0.,15.)(10.,14.5)(0.,){/Cycles}{0}
\FALabel(5.11236,16.5172)[b]{$g$}
\FAProp(0.,5.)(10.,5.5)(0.,){/Straight}{1}
\FALabel(5.0774,4.18193)[t]{$u$}
\FAProp(20.,17.)(10.,5.5)(0.,){/Straight}{0}
\FALabel(17.48,15.02)[br]{$\tilde g$}
\FAProp(20.,10.)(10.,14.5)(0.,){/ScalarDash}{-1}
\FALabel(17.8274,10.002)[tr]{$\tilde u_L$}
\FAProp(20.,3.)(10.,10.)(0.,){/Sine}{0}
\FALabel(14.5911,5.71019)[tr]{$\gamma$}
\FAProp(10.,14.5)(10.,10.)(0.,){/ScalarDash}{-1}
\FALabel(8.93,12.25)[r]{$\tilde u_L$}
\FAProp(10.,5.5)(10.,10.)(0.,){/ScalarDash}{1}
\FALabel(8.93,7.75)[r]{$\tilde u_L$}
\FAVert(10.,14.5){0}
\FAVert(10.,5.5){0}
\FAVert(10.,10.){0}

\end{feynartspicture}
}

\caption{Feynman diagrams for real photon radiation.
The first six diagrams are IR divergent, the last three are IR finite.}
\label{fig_realcorrs}
}

\FIGURE{

\hspace*{-.4cm}
\vspace*{-.6cm}
\scalebox{.845}{
\begin{feynartspicture}(12,3)(4,1)

\FADiagram{}
\FAProp(0.,15.)(5.5,10.)(0.,){/Straight}{1}
\FALabel(2.18736,11.8331)[tr]{$q_i$}
\FAProp(0.,5.)(5.5,10.)(0.,){/Straight}{-1}
\FALabel(3.31264,6.83309)[tl]{$\bar{q}_i$}
\FAProp(20.,17.)(15.5,13.5)(0.,){/Straight}{0}
\FALabel(17.4319,15.7962)[br]{$\tilde g$}
\FAProp(20.,10.)(15.5,13.5)(0.,){/ScalarDash}{-1}
\FALabel(18.2216,12.4935)[bl]{$\tilde u_L$}
\FAProp(20.,3.)(12.,10.)(0.,){/Straight}{1}
\FALabel(15.4593,5.81351)[tr]{$\bar{u}$}
\FAProp(5.5,10.)(12.,10.)(0.,){/Sine}{-1}
\FALabel(8.75,8.93)[t]{$V^0$}
\FAProp(15.5,13.5)(12.,10.)(0.,){/Straight}{-1}
\FALabel(13.134,12.366)[br]{$u$}
\FAVert(5.5,10.){0}
\FAVert(15.5,13.5){0}
\FAVert(12.,10.){0}

\FADiagram{}
\FAProp(0.,15.)(4.5,10.)(0.,){/Straight}{1}
\FALabel(1.57789,11.9431)[tr]{$q_i$}
\FAProp(0.,5.)(4.5,10.)(0.,){/Straight}{-1}
\FALabel(2.92211,6.9431)[tl]{$\bar{q}_i$}
\FAProp(20.,17.)(13.,14.5)(0.,){/Straight}{0}
\FALabel(16.0628,16.4943)[b]{$\tilde g$}
\FAProp(20.,10.)(10.85,8.4)(0.,){/ScalarDash}{-1}
\FALabel(18.4569,10.7663)[b]{$\tilde u_L$}
\FAProp(20.,3.)(13.,14.5)(0.,){/Straight}{1}
\FALabel(17.7665,4.80001)[tr]{$\bar{u}$}
\FAProp(4.5,10.)(10.85,8.4)(0.,){/Sine}{-1}
\FALabel(7.29629,8.17698)[t]{$V^0$}
\FAProp(13.,14.5)(10.85,8.4)(0.,){/ScalarDash}{1}
\FALabel(10.9431,11.9652)[r]{$\tilde u_a$}
\FAVert(4.5,10.){0}
\FAVert(13.,14.5){0}
\FAVert(10.85,8.4){0}

\FADiagram{}
\FAProp(0.,15.)(10.,7.)(0.,){/Straight}{1}
\FALabel(2.82617,13.9152)[bl]{$q_i$}
\FAProp(0.,5.)(10.,14.5)(0.,){/Straight}{-1}
\FALabel(3.04219,6.54875)[tl]{$\bar{q}_i$}
\FAProp(20.,17.)(10.,14.5)(0.,){/Straight}{0}
\FALabel(14.6847,16.5312)[b]{$\tilde g$}
\FAProp(20.,10.)(15.5,6.5)(0.,){/ScalarDash}{-1}
\FALabel(17.2784,8.9935)[br]{$\tilde u_L$}
\FAProp(20.,3.)(15.5,6.5)(0.,){/Straight}{1}
\FALabel(17.2784,4.0065)[tr]{$\bar{u}$}
\FAProp(10.,7.)(10.,14.5)(0.,){/ScalarDash}{1}
\FALabel(11.07,10.75)[l]{$\tilde q_{ia}$}
\FAProp(10.,7.)(15.5,6.5)(0.,){/Straight}{-1}
\FALabel(12.6097,5.68637)[t]{$\tilde \chi_n^{0}$}
\FAVert(10.,7.){0}
\FAVert(10.,14.5){0}
\FAVert(15.5,6.5){0}

\FADiagram{}
\FAProp(0.,15.)(10.,14.5)(0.,){/Straight}{1}
\FALabel(5.0774,15.8181)[b]{$q_i$}
\FAProp(0.,5.)(10.,7.)(0.,){/Straight}{-1}
\FALabel(5.30398,4.9601)[t]{$\bar{q}_i$}
\FAProp(20.,17.)(10.,14.5)(0.,){/Straight}{0}
\FALabel(14.6847,16.5312)[b]{$\tilde g$}
\FAProp(20.,10.)(15.5,6.5)(0.,){/ScalarDash}{-1}
\FALabel(17.2784,8.9935)[br]{$\tilde u_L$}
\FAProp(20.,3.)(15.5,6.5)(0.,){/Straight}{1}
\FALabel(17.2784,4.0065)[tr]{$\bar{u}$}
\FAProp(10.,14.5)(10.,7.)(0.,){/ScalarDash}{1}
\FALabel(8.93,10.75)[r]{$\tilde q_{ia}$}
\FAProp(10.,7.)(15.5,6.5)(0.,){/Straight}{-1}
\FALabel(12.6097,5.68637)[t]{$\tilde \chi_n^{0}$}
\FAVert(10.,14.5){0}
\FAVert(10.,7.){0}
\FAVert(15.5,6.5){0}

\end{feynartspicture}
}

\hspace*{.4cm}
\scalebox{.845}{
\begin{feynartspicture}(12,3)(4,1)
 
\FADiagram{}
\FAProp(0.,15.)(10.,14.5)(0.,){/Straight}{1}
\FALabel(5.0774,15.8181)[b]{$q$}
\FAProp(0.,5.)(10.,5.5)(0.,){/Straight}{-1}
\FALabel(5.0774,4.18193)[t]{$\bar{q}$}
\FAProp(20.,17.)(10.,14.5)(0.,){/Straight}{0}
\FALabel(14.6847,16.5312)[b]{$\tilde g$}
\FAProp(20.,10.)(10.,10.)(0.,){/ScalarDash}{-1}
\FALabel(15.,11.07)[b]{$\tilde u_L$}
\FAProp(20.,3.)(10.,5.5)(0.,){/Straight}{1}
\FALabel(15.3759,5.27372)[b]{$\bar{u}$}
\FAProp(10.,14.5)(10.,10.)(0.,){/ScalarDash}{1}
\FALabel(8.93,12.25)[r]{$\tilde q_a$}
\FAProp(10.,5.5)(10.,10.)(0.,){/Sine}{0}
\FALabel(8.93,7.75)[r]{$V$}
\FAVert(10.,14.5){0}
\FAVert(10.,5.5){0}
\FAVert(10.,10.){0}

\FADiagram{}
\FAProp(0.,15.)(10.,13.)(0.,){/Straight}{1}
\FALabel(5.30398,15.0399)[b]{$q$}
\FAProp(0.,5.)(10.,5.5)(0.,){/Straight}{1}
\FALabel(5.0774,4.18193)[t]{$\bar{q}$}
\FAProp(20.,17.)(15.5,13.5)(0.,){/Straight}{0}
\FALabel(17.4319,15.7962)[br]{$\tilde g$}
\FAProp(20.,10.)(15.5,13.5)(0.,){/ScalarDash}{-1}
\FALabel(18.2216,12.4935)[bl]{$\tilde u_L$}
\FAProp(20.,3.)(10.,5.5)(0.,){/Straight}{-1}
\FALabel(15.3759,5.27372)[b]{$\bar{u}$}
\FAProp(10.,13.)(10.,5.5)(0.,){/Sine}{0}
\FALabel(8.93,9.25)[r]{$V$}
\FAProp(10.,13.)(15.5,13.5)(0.,){/Straight}{1}
\FALabel(12.8903,12.1864)[t]{$u$}
\FAVert(10.,13.){0}
\FAVert(10.,5.5){0}
\FAVert(15.5,13.5){0}

\FADiagram{}
\FAProp(0.,15.)(10.,14.5)(0.,){/Straight}{1}
\FALabel(5.0774,15.8181)[b]{$q$}
\FAProp(0.,5.)(10.,5.5)(0.,){/Straight}{-1}
\FALabel(5.0774,4.18193)[t]{$\bar{q}$}
\FAProp(20.,17.)(10.,5.5)(0.,){/Straight}{0}
\FALabel(17.48,15.02)[br]{$\tilde g$}
\FAProp(20.,10.)(10.,14.5)(0.,){/ScalarDash}{-1}
\FALabel(17.8274,10.002)[tr]{$\tilde u_L$}
\FAProp(20.,3.)(10.,10.)(0.,){/Straight}{1}
\FALabel(14.5911,5.71019)[tr]{$\bar{u}$}
\FAProp(10.,14.5)(10.,10.)(0.,){/Straight}{0}
\FALabel(9.7,11.7)[r]{$\tilde \chi_n^{0/\pm}$}
\FAProp(10.,5.5)(10.,10.)(0.,){/ScalarDash}{-1}
\FALabel(8.93,7.75)[r]{$\tilde q_{a}$}
\FAVert(10.,14.5){0}
\FAVert(10.,5.5){0}
\FAVert(10.,10.){0}

\FADiagram{}
\FAProp(0.,15.)(10.,14.5)(0.,){/Straight}{1}
\FALabel(5.0774,15.8181)[b]{$q$}
\FAProp(0.,5.)(10.,5.5)(0.,){/Straight}{-1}
\FALabel(5.0774,4.18193)[t]{$\bar{q}$}
\FAProp(20.,17.)(15.5,8.)(0.,){/Straight}{0}
\FALabel(18.28,14.97)[br]{$\tilde g$}
\FAProp(20.,10.)(10.,14.5)(0.,){/ScalarDash}{-1}
\FALabel(12.6226,14.248)[bl]{$\tilde u_L$}
\FAProp(20.,3.)(15.5,8.)(0.,){/Straight}{1}
\FALabel(18.4221,6.0569)[bl]{$\bar{u}$}
\FAProp(10.,14.5)(10.,5.5)(0.,){/Straight}{0}
\FALabel(9.18,10.)[r]{$\tilde \chi_n^{0/\pm}$}
\FAProp(10.,5.5)(15.5,8.)(0.,){/ScalarDash}{-1}
\FALabel(12.9114,5.81893)[tl]{$\tilde u_{a}$}
\FAVert(10.,14.5){0}
\FAVert(10.,5.5){0}
\FAVert(15.5,8.){0}

\end{feynartspicture}
}
\\
\put(2,.2){(a)}
\\[-.3ex]
\hspace*{-.35cm}
\vspace*{-.6cm}
\scalebox{.845}{
\begin{feynartspicture}(15,3)(5,1)

\FADiagram{}
\FAProp(0.,15.)(5.5,10.)(0.,){/Straight}{1}
\FALabel(2.18736,11.8331)[tr]{$q_i$}
\FAProp(0.,5.)(5.5,10.)(0.,){/Straight}{-1}
\FALabel(3.31264,6.83309)[tl]{$\bar{q}_i$}
\FAProp(20.,17.)(15.5,13.5)(0.,){/Straight}{0}
\FALabel(17.4319,15.7962)[br]{$\tilde g$}
\FAProp(20.,10.)(15.5,13.5)(0.,){/ScalarDash}{-1}
\FALabel(18.2216,12.4935)[bl]{$\tilde u_L$}
\FAProp(20.,3.)(12.,10.)(0.,){/Straight}{1}
\FALabel(15.4593,5.81351)[tr]{$\bar{u}$}
\FAProp(5.5,10.)(12.,10.)(0.,){/Cycles}{0}
\FALabel(8.75,8.93)[t]{$g$}
\FAProp(15.5,13.5)(12.,10.)(0.,){/Straight}{-1}
\FALabel(13.134,12.366)[br]{$u$}
\FAVert(5.5,10.){0}
\FAVert(15.5,13.5){0}
\FAVert(12.,10.){0}

\FADiagram{}
\FAProp(0.,15.)(4.5,10.)(0.,){/Straight}{1}
\FALabel(1.57789,11.9431)[tr]{$q_i$}
\FAProp(0.,5.)(4.5,10.)(0.,){/Straight}{-1}
\FALabel(2.92211,6.9431)[tl]{$\bar{q}_i$}
\FAProp(20.,17.)(13.,14.5)(0.,){/Straight}{0}
\FALabel(16.0628,16.4943)[b]{$\tilde g$}
\FAProp(20.,10.)(10.85,8.4)(0.,){/ScalarDash}{-1}
\FALabel(18.4569,10.7663)[b]{$\tilde u_L$}
\FAProp(20.,3.)(13.,14.5)(0.,){/Straight}{1}
\FALabel(17.7665,4.80001)[tr]{$\bar{u}$}
\FAProp(4.5,10.)(10.85,8.4)(0.,){/Cycles}{0}
\FALabel(7.29629,8.17698)[t]{$g$}
\FAProp(13.,14.5)(10.85,8.4)(0.,){/ScalarDash}{1}
\FALabel(10.9431,11.9652)[r]{$\tilde u_L$}
\FAVert(4.5,10.){0}
\FAVert(13.,14.5){0}
\FAVert(10.85,8.4){0}

\FADiagram{}
\FAProp(0.,15.)(5.5,10.)(0.,){/Straight}{1}
\FALabel(2.18736,11.8331)[tr]{$q_i$}
\FAProp(0.,5.)(5.5,10.)(0.,){/Straight}{-1}
\FALabel(3.31264,6.83309)[tl]{$\bar{q}_i$}
\FAProp(20.,17.)(11.5,10.)(0.,){/Straight}{0}
\FALabel(15.4036,14.0235)[br]{$\tilde g$}
\FAProp(20.,10.)(15.5,6.5)(0.,){/ScalarDash}{-1}
\FALabel(17.2784,8.9935)[br]{$\tilde u_L$}
\FAProp(20.,3.)(15.5,6.5)(0.,){/Straight}{1}
\FALabel(18.2216,5.4935)[bl]{$\bar{u}$}
\FAProp(5.5,10.)(11.5,10.)(0.,){/Cycles}{0}
\FALabel(8.5,11.77)[b]{$g$}
\FAProp(11.5,10.)(15.5,6.5)(0.,){/Straight}{0}
\FALabel(13.1239,7.75165)[tr]{$\tilde g$}
\FAVert(5.5,10.){0}
\FAVert(11.5,10.){0}
\FAVert(15.5,6.5){0}

\FADiagram{}
\FAProp(0.,15.)(10.,14.5)(0.,){/Straight}{1}
\FALabel(5.0774,15.8181)[b]{$q_i$}
\FAProp(0.,5.)(10.,7.)(0.,){/Straight}{-1}
\FALabel(5.30398,4.9601)[t]{$\bar{q}_i$}
\FAProp(20.,17.)(10.,14.5)(0.,){/Straight}{0}
\FALabel(14.6847,16.5312)[b]{$\tilde g$}
\FAProp(20.,10.)(15.5,6.5)(0.,){/ScalarDash}{-1}
\FALabel(17.2784,8.9935)[br]{$\tilde u_L$}
\FAProp(20.,3.)(15.5,6.5)(0.,){/Straight}{1}
\FALabel(17.2784,4.0065)[tr]{$\bar{u}$}
\FAProp(10.,14.5)(10.,7.)(0.,){/ScalarDash}{1}
\FALabel(8.93,10.75)[r]{$\tilde q_{ia}$}
\FAProp(10.,7.)(15.5,6.5)(0.,){/Straight}{0}
\FALabel(12.6323,5.93534)[t]{$\tilde g$}
\FAVert(10.,14.5){0}
\FAVert(10.,7.){0}
\FAVert(15.5,6.5){0}

\FADiagram{}
\FAProp(0.,15.)(10.,7.)(0.,){/Straight}{1}
\FALabel(2.82617,13.9152)[bl]{$q_i$}
\FAProp(0.,5.)(10.,14.5)(0.,){/Straight}{-1}
\FALabel(3.04219,6.54875)[tl]{$\bar{q}_i$}
\FAProp(20.,17.)(10.,14.5)(0.,){/Straight}{0}
\FALabel(14.6847,16.5312)[b]{$\tilde g$}
\FAProp(20.,10.)(15.5,6.5)(0.,){/ScalarDash}{-1}
\FALabel(17.2784,8.9935)[br]{$\tilde u_L$}
\FAProp(20.,3.)(15.5,6.5)(0.,){/Straight}{1}
\FALabel(17.2784,4.0065)[tr]{$\bar{u}$}
\FAProp(10.,7.)(10.,14.5)(0.,){/ScalarDash}{1}
\FALabel(11.07,10.75)[l]{$\tilde q_{ia}$}
\FAProp(10.,7.)(15.5,6.5)(0.,){/Straight}{0}
\FALabel(12.6323,5.93534)[t]{$\tilde g$}
\FAVert(10.,7.){0}
\FAVert(10.,14.5){0}
\FAVert(15.5,6.5){0}

\end{feynartspicture}
}

\hspace*{-.35cm}
\vspace*{-.2cm}
\scalebox{.845}{
\begin{feynartspicture}(15,3)(5,1)
 
\FADiagram{}
\FAProp(0.,15.)(10.,13.)(0.,){/Straight}{1}
\FALabel(5.30398,15.0399)[b]{$u$}
\FAProp(0.,5.)(10.,5.5)(0.,){/Straight}{-1}
\FALabel(5.0774,4.18193)[t]{$\bar{u}$}
\FAProp(20.,17.)(15.5,13.5)(0.,){/Straight}{0}
\FALabel(17.4319,15.7962)[br]{$\tilde g$}
\FAProp(20.,10.)(15.5,13.5)(0.,){/ScalarDash}{-1}
\FALabel(18.2216,12.4935)[bl]{$\tilde u_L$}
\FAProp(20.,3.)(10.,5.5)(0.,){/Straight}{1}
\FALabel(15.3759,5.27372)[b]{$\bar{u}$}
\FAProp(10.,13.)(10.,5.5)(0.,){/Cycles}{0}
\FALabel(8.93,9.25)[r]{$g$}
\FAProp(10.,13.)(15.5,13.5)(0.,){/Straight}{1}
\FALabel(12.8903,12.1864)[t]{$u$}
\FAVert(10.,13.){0}
\FAVert(10.,5.5){0}
\FAVert(15.5,13.5){0}

\FADiagram{}
\FAProp(0.,15.)(10.,14.5)(0.,){/Straight}{1}
\FALabel(5.0774,15.8181)[b]{$u$}
\FAProp(0.,5.)(10.,5.5)(0.,){/Straight}{-1}
\FALabel(5.0774,4.18193)[t]{$\bar{u}$}
\FAProp(20.,17.)(10.,14.5)(0.,){/Straight}{0}
\FALabel(14.6847,16.5312)[b]{$\tilde g$}
\FAProp(20.,10.)(10.,10.)(0.,){/ScalarDash}{-1}
\FALabel(15.,11.07)[b]{$\tilde u_L$}
\FAProp(20.,3.)(10.,5.5)(0.,){/Straight}{1}
\FALabel(15.3759,5.27372)[b]{$\bar{u}$}
\FAProp(10.,14.5)(10.,10.)(0.,){/ScalarDash}{1}
\FALabel(8.93,12.25)[r]{$\tilde u_L$}
\FAProp(10.,5.5)(10.,10.)(0.,){/Cycles}{0}
\FALabel(8.23,7.75)[r]{$g$}
\FAVert(10.,14.5){0}
\FAVert(10.,5.5){0}
\FAVert(10.,10.){0}

\FADiagram{}
\FAProp(0.,15.)(10.,14.5)(0.,){/Straight}{1}
\FALabel(5.0774,15.8181)[b]{$u$}
\FAProp(0.,5.)(10.,5.5)(0.,){/Straight}{-1}
\FALabel(5.0774,4.18193)[t]{$\bar{u}$}
\FAProp(20.,17.)(10.,10.)(0.,){/Straight}{0}
\FALabel(17.5821,16.1028)[br]{$\tilde g$}
\FAProp(20.,10.)(10.,14.5)(0.,){/ScalarDash}{-1}
\FALabel(17.2774,10.302)[tr]{$\tilde u_L$}
\FAProp(20.,3.)(10.,5.5)(0.,){/Straight}{1}
\FALabel(14.6241,3.22628)[t]{$\bar{u}$}
\FAProp(10.,14.5)(10.,10.)(0.,){/Straight}{0}
\FALabel(9.18,12.25)[r]{$\tilde g$}
\FAProp(10.,5.5)(10.,10.)(0.,){/Cycles}{0}
\FALabel(8.23,7.75)[r]{$g$}
\FAVert(10.,14.5){0}
\FAVert(10.,5.5){0}
\FAVert(10.,10.){0}

\FADiagram{}
\FAProp(0.,15.)(10.,14.5)(0.,){/Straight}{1}
\FALabel(5.0774,15.8181)[b]{$u$}
\FAProp(0.,5.)(10.,5.5)(0.,){/Straight}{-1}
\FALabel(5.0774,4.18193)[t]{$\bar{u}$}
\FAProp(20.,17.)(10.,5.5)(0.,){/Straight}{0}
\FALabel(17.48,15.02)[br]{$\tilde g$}
\FAProp(20.,10.)(10.,14.5)(0.,){/ScalarDash}{-1}
\FALabel(17.8274,10.002)[tr]{$\tilde u_L$}
\FAProp(20.,3.)(10.,10.)(0.,){/Straight}{1}
\FALabel(14.5911,5.71019)[tr]{$\bar{u}$}
\FAProp(10.,14.5)(10.,10.)(0.,){/Straight}{0}
\FALabel(9.18,12.25)[r]{$\tilde g$}
\FAProp(10.,5.5)(10.,10.)(0.,){/ScalarDash}{-1}
\FALabel(8.93,7.75)[r]{$\tilde u_{a}$}
\FAVert(10.,14.5){0}
\FAVert(10.,5.5){0}
\FAVert(10.,10.){0}

\FADiagram{}
\FAProp(0.,15.)(10.,14.5)(0.,){/Straight}{1}
\FALabel(5.0774,15.8181)[b]{$u$}
\FAProp(0.,5.)(10.,5.5)(0.,){/Straight}{-1}
\FALabel(5.0774,4.18193)[t]{$\bar{u}$}
\FAProp(20.,17.)(15.5,8.)(0.,){/Straight}{0}
\FALabel(18.28,14.97)[br]{$\tilde g$}
\FAProp(20.,10.)(10.,14.5)(0.,){/ScalarDash}{-1}
\FALabel(12.6226,14.248)[bl]{$\tilde u_L$}
\FAProp(20.,3.)(15.5,8.)(0.,){/Straight}{1}
\FALabel(18.4221,6.0569)[bl]{$\bar{u}$}
\FAProp(10.,14.5)(10.,5.5)(0.,){/Straight}{0}
\FALabel(9.18,10.)[r]{$\tilde g$}
\FAProp(10.,5.5)(15.5,8.)(0.,){/ScalarDash}{-1}
\FALabel(12.6,6.3)[tl]{$\tilde u_{a}$}
\FAVert(10.,14.5){0}
\FAVert(10.,5.5){0}
\FAVert(15.5,8.){0}

\end{feynartspicture}
}

\\
\put(0,0){(b)}
\caption{Feynman diagrams for quark radiation via $q_i \bar{q}_i \rightarrow \tilde{g} \tilde{u}_L \bar{u}$, with 
$q_i = u,\,d,\,c,\,s$. Only interference terms from EW (a) and QCD (b) diagrams contribute at $\mathcal{O}(\alpha_s^2\alpha)$. In panel (a), the diagrams of the second row contribute only for $q_i=u,d$. In panel (b), the diagrams of the second row contribute only for $q_i = u$.}
\label{fig_qqbarchannels}
}

\FIGURE{
\hspace*{-.4cm}
\vspace*{-.5cm}
\scalebox{.845}{
\begin{feynartspicture}(12,3)(4,1)
 
\FADiagram{}
\FAProp(0.,15.)(10.,14.5)(0.,){/Straight}{1}
\FALabel(5.0774,15.8181)[b]{$u$}
\FAProp(0.,5.)(10.,5.5)(0.,){/Straight}{1}
\FALabel(5.0774,4.18193)[t]{$q_i$}
\FAProp(20.,17.)(10.,14.5)(0.,){/Straight}{0}
\FALabel(14.6847,16.5312)[b]{$\tilde g$}
\FAProp(20.,10.)(10.,10.)(0.,){/ScalarDash}{-1}
\FALabel(15.,11.07)[b]{$\tilde u_L$}
\FAProp(20.,3.)(10.,5.5)(0.,){/Straight}{-1}
\FALabel(15.3759,5.27372)[b]{$q_i$}
\FAProp(10.,14.5)(10.,10.)(0.,){/ScalarDash}{1}
\FALabel(8.93,12.25)[r]{$\tilde u_L$}
\FAProp(10.,5.5)(10.,10.)(0.,){/Sine}{0}
\FALabel(8.93,7.75)[r]{$V^0$}
\FAVert(10.,14.5){0}
\FAVert(10.,5.5){0}
\FAVert(10.,10.){0}

\FADiagram{}
\FAProp(0.,15.)(10.,13.)(0.,){/Straight}{1}
\FALabel(5.30398,15.0399)[b]{$u$}
\FAProp(0.,5.)(10.,5.5)(0.,){/Straight}{1}
\FALabel(5.0774,4.18193)[t]{$q_i$}
\FAProp(20.,17.)(15.5,13.5)(0.,){/Straight}{0}
\FALabel(17.4319,15.7962)[br]{$\tilde g$}
\FAProp(20.,10.)(15.5,13.5)(0.,){/ScalarDash}{-1}
\FALabel(18.2216,12.4935)[bl]{$\tilde u_L$}
\FAProp(20.,3.)(10.,5.5)(0.,){/Straight}{-1}
\FALabel(15.3759,5.27372)[b]{$q_i$}
\FAProp(10.,13.)(10.,5.5)(0.,){/Sine}{0}
\FALabel(8.93,9.25)[r]{$V^0$}
\FAProp(10.,13.)(15.5,13.5)(0.,){/Straight}{1}
\FALabel(12.8903,12.1864)[t]{$u$}
\FAVert(10.,13.){0}
\FAVert(10.,5.5){0}
\FAVert(15.5,13.5){0}

\FADiagram{}
\FAProp(0.,15.)(10.,14.5)(0.,){/Straight}{1}
\FALabel(5.0774,15.8181)[b]{$u$}
\FAProp(0.,5.)(10.,5.5)(0.,){/Straight}{1}
\FALabel(5.0774,4.18193)[t]{$q_i$}
\FAProp(20.,17.)(10.,5.5)(0.,){/Straight}{0}
\FALabel(17.48,15.02)[br]{$\tilde g$}
\FAProp(20.,10.)(10.,14.5)(0.,){/ScalarDash}{-1}
\FALabel(17.8274,10.002)[tr]{$\tilde u_L$}
\FAProp(20.,3.)(10.,10.)(0.,){/Straight}{-1}
\FALabel(14.5911,5.71019)[tr]{$q_i$}
\FAProp(10.,14.5)(10.,10.)(0.,){/Straight}{0}
\FALabel(9.18,12.25)[r]{$\tilde \chi_n^0$}
\FAProp(10.,5.5)(10.,10.)(0.,){/ScalarDash}{1}
\FALabel(8.93,7.75)[r]{$\tilde q_{ia}$}
\FAVert(10.,14.5){0}
\FAVert(10.,5.5){0}
\FAVert(10.,10.){0}

\FADiagram{}
\FAProp(0.,15.)(10.,14.5)(0.,){/Straight}{1}
\FALabel(5.0774,15.8181)[b]{$u$}
\FAProp(0.,5.)(10.,5.5)(0.,){/Straight}{1}
\FALabel(5.0774,4.18193)[t]{$q_i$}
\FAProp(20.,17.)(15.5,8.)(0.,){/Straight}{0}
\FALabel(18.28,14.97)[br]{$\tilde g$}
\FAProp(20.,10.)(10.,14.5)(0.,){/ScalarDash}{-1}
\FALabel(12.6226,14.248)[bl]{$\tilde u_L$}
\FAProp(20.,3.)(15.5,8.)(0.,){/Straight}{-1}
\FALabel(18.4221,6.0569)[bl]{$q_i$}
\FAProp(10.,14.5)(10.,5.5)(0.,){/Straight}{0}
\FALabel(9.18,10.)[r]{$\tilde \chi_n^0$}
\FAProp(10.,5.5)(15.5,8.)(0.,){/ScalarDash}{1}
\FALabel(12.9114,5.81893)[tl]{$\tilde q_{ia}$}
\FAVert(10.,14.5){0}
\FAVert(10.,5.5){0}
\FAVert(15.5,8.){0}

\end{feynartspicture}
}

\hspace*{-.4cm}
\vspace*{-.5cm}
\scalebox{.85}{
\begin{feynartspicture}(12,3)(4,1)

\FADiagram{}
\FAProp(0.,15.)(10.,5.5)(0.,){/Straight}{1}
\FALabel(3.19219,13.2012)[bl]{$u$}
\FAProp(0.,5.)(10.,14.5)(0.,){/Straight}{1}
\FALabel(3.10297,6.58835)[tl]{$q$}
\FAProp(20.,17.)(10.,14.5)(0.,){/Straight}{0}
\FALabel(14.6847,16.5312)[b]{$\tilde g$}
\FAProp(20.,10.)(10.,10.)(0.,){/ScalarDash}{-1}
\FALabel(15.95,11.07)[b]{$\tilde u_L$}
\FAProp(20.,3.)(10.,5.5)(0.,){/Straight}{-1}
\FALabel(14.6241,3.22628)[t]{$q$}
\FAProp(10.,5.5)(10.,10.)(0.,){/Sine}{0}
\FALabel(11.07,7.75)[l]{$V$}
\FAProp(10.,14.5)(10.,10.)(0.,){/ScalarDash}{1}
\FALabel(11.07,12.25)[l]{$\tilde q_a$}
\FAVert(10.,5.5){0}
\FAVert(10.,14.5){0}
\FAVert(10.,10.){0}

\FADiagram{}
\FAProp(0.,15.)(10.,5.5)(0.,){/Straight}{1}
\FALabel(3.19219,13.2012)[bl]{$u$}
\FAProp(0.,5.)(10.,13.)(0.,){/Straight}{1}
\FALabel(3.17617,6.28478)[tl]{$q$}
\FAProp(20.,17.)(16.,13.5)(0.,){/Straight}{0}
\FALabel(17.6239,15.7483)[br]{$\tilde g$}
\FAProp(20.,10.)(16.,13.5)(0.,){/ScalarDash}{-1}
\FALabel(17.4593,11.0635)[tr]{$\tilde u_L$}
\FAProp(20.,3.)(10.,5.5)(0.,){/Straight}{-1}
\FALabel(14.6241,3.22628)[t]{$q$}
\FAProp(10.,5.5)(10.,13.)(0.,){/Sine}{0}
\FALabel(11.07,9.25)[l]{$V$}
\FAProp(10.,13.)(16.,13.5)(0.,){/Straight}{1}
\FALabel(12.8713,14.3146)[b]{$u$}
\FAVert(10.,5.5){0}
\FAVert(10.,13.){0}
\FAVert(16.,13.5){0}

\FADiagram{}
\FAProp(0.,15.)(10.,14.5)(0.,){/Straight}{1}
\FALabel(5.0774,15.8181)[b]{$u$}
\FAProp(0.,5.)(10.,5.5)(0.,){/Straight}{1}
\FALabel(5.0774,4.18193)[t]{$q$}
\FAProp(20.,17.)(10.,14.5)(0.,){/Straight}{0}
\FALabel(14.6847,16.5312)[b]{$\tilde g$}
\FAProp(20.,10.)(10.,5.5)(0.,){/ScalarDash}{-1}
\FALabel(17.2909,9.58086)[br]{$\tilde u_L$}
\FAProp(20.,3.)(10.,10.)(0.,){/Straight}{-1}
\FALabel(17.3366,3.77519)[tr]{$q$}
\FAProp(10.,14.5)(10.,10.)(0.,){/ScalarDash}{1}
\FALabel(8.93,12.25)[r]{$\tilde u_a$}
\FAProp(10.,5.5)(10.,10.)(0.,){/Straight}{0}
\FALabel(9.5,8.2)[r]{$\tilde \chi_n^{0/\pm}$}
\FAVert(10.,14.5){0}
\FAVert(10.,5.5){0}
\FAVert(10.,10.){0}

\FADiagram{}
\FAProp(0.,15.)(10.,14.5)(0.,){/Straight}{1}
\FALabel(5.0774,15.8181)[b]{$u$}
\FAProp(0.,5.)(10.,5.5)(0.,){/Straight}{1}
\FALabel(5.0774,4.18193)[t]{$q$}
\FAProp(20.,17.)(15.5,14.)(0.,){/Straight}{0}
\FALabel(17.5089,16.1017)[br]{$\tilde g$}
\FAProp(20.,10.)(10.,5.5)(0.,){/ScalarDash}{-1}
\FALabel(13.8364,8.43358)[br]{$\tilde u_L$}
\FAProp(20.,3.)(15.5,14.)(0.,){/Straight}{-1}
\FALabel(18.,4.3)[r]{$q$}
\FAProp(10.,14.5)(10.,5.5)(0.,){/Straight}{0}
\FALabel(9.5,10.)[r]{$\tilde \chi_n^{0/\pm}$}
\FAProp(10.,14.5)(15.5,14.)(0.,){/ScalarDash}{1}
\FALabel(12.8903,15.3136)[b]{$\tilde q_a$}
\FAVert(10.,14.5){0}
\FAVert(10.,5.5){0}
\FAVert(15.5,14.){0}

\end{feynartspicture}
}

\hspace*{.4cm}
\vspace*{-.2cm}
\scalebox{.845}{
\begin{feynartspicture}(12,3)(4,1)

\FADiagram{}
\FAProp(0.,15.)(5.5,10.)(0.,){/Straight}{1}
\FALabel(2.18736,11.8331)[tr]{$u$}
\FAProp(0.,5.)(5.5,10.)(0.,){/Straight}{-1}
\FALabel(3.31264,6.83309)[tl]{$\bar{d}$}
\FAProp(20.,17.)(15.5,13.5)(0.,){/Straight}{0}
\FALabel(17.4319,15.7962)[br]{$\tilde g$}
\FAProp(20.,10.)(15.5,13.5)(0.,){/ScalarDash}{-1}
\FALabel(18.2216,12.4935)[bl]{$\tilde u_L$}
\FAProp(20.,3.)(12.,10.)(0.,){/Straight}{1}
\FALabel(15.4593,5.81351)[tr]{$\bar{d}$}
\FAProp(5.5,10.)(12.,10.)(0.,){/Sine}{-1}
\FALabel(8.75,8.93)[t]{$W$}
\FAProp(15.5,13.5)(12.,10.)(0.,){/Straight}{-1}
\FALabel(13.134,12.366)[br]{$u$}
\FAVert(5.5,10.){0}
\FAVert(15.5,13.5){0}
\FAVert(12.,10.){0}

\FADiagram{}
\FAProp(0.,15.)(4.5,10.)(0.,){/Straight}{1}
\FALabel(1.57789,11.9431)[tr]{$u$}
\FAProp(0.,5.)(4.5,10.)(0.,){/Straight}{-1}
\FALabel(2.92211,6.9431)[tl]{$\bar{d}$}
\FAProp(20.,17.)(13.,14.5)(0.,){/Straight}{0}
\FALabel(16.0628,16.4943)[b]{$\tilde g$}
\FAProp(20.,10.)(10.85,8.4)(0.,){/ScalarDash}{-1}
\FALabel(18.4569,10.7663)[b]{$\tilde u_L$}
\FAProp(20.,3.)(13.,14.5)(0.,){/Straight}{1}
\FALabel(17.7665,4.80001)[tr]{$\bar{d}$}
\FAProp(4.5,10.)(10.85,8.4)(0.,){/Sine}{-1}
\FALabel(7.29629,8.17698)[t]{$W$}
\FAProp(13.,14.5)(10.85,8.4)(0.,){/ScalarDash}{1}
\FALabel(10.9431,11.9652)[r]{$\tilde d_a$}
\FAVert(4.5,10.){0}
\FAVert(13.,14.5){0}
\FAVert(10.85,8.4){0}

\FADiagram{}
\FAProp(0.,15.)(10.,7.)(0.,){/Straight}{1}
\FALabel(2.82617,13.9152)[bl]{$u$}
\FAProp(0.,5.)(10.,14.5)(0.,){/Straight}{-1}
\FALabel(3.04219,6.54875)[tl]{$\bar{d}$}
\FAProp(20.,17.)(10.,14.5)(0.,){/Straight}{0}
\FALabel(14.6847,16.5312)[b]{$\tilde g$}
\FAProp(20.,10.)(15.5,6.5)(0.,){/ScalarDash}{-1}
\FALabel(17.2784,8.9935)[br]{$\tilde u_L$}
\FAProp(20.,3.)(15.5,6.5)(0.,){/Straight}{1}
\FALabel(17.2784,4.0065)[tr]{$\bar{d}$}
\FAProp(10.,7.)(10.,14.5)(0.,){/ScalarDash}{1}
\FALabel(11.07,10.75)[l]{$\tilde d_a$}
\FAProp(10.,7.)(15.5,6.5)(0.,){/Straight}{-1}
\FALabel(12.6097,5.68637)[t]{$\tilde \chi_n^{\pm}$}
\FAVert(10.,7.){0}
\FAVert(10.,14.5){0}
\FAVert(15.5,6.5){0}

\FADiagram{}
\FAProp(0.,15.)(10.,14.5)(0.,){/Straight}{1}
\FALabel(5.0774,15.8181)[b]{$u$}
\FAProp(0.,5.)(10.,7.)(0.,){/Straight}{-1}
\FALabel(5.30398,4.9601)[t]{$\bar{d}$}
\FAProp(20.,17.)(10.,14.5)(0.,){/Straight}{0}
\FALabel(14.6847,16.5312)[b]{$\tilde g$}
\FAProp(20.,10.)(15.5,6.5)(0.,){/ScalarDash}{-1}
\FALabel(17.2784,8.9935)[br]{$\tilde u_L$}
\FAProp(20.,3.)(15.5,6.5)(0.,){/Straight}{1}
\FALabel(17.2784,4.0065)[tr]{$\bar{d}$}
\FAProp(10.,14.5)(10.,7.)(0.,){/ScalarDash}{1}
\FALabel(8.93,10.75)[r]{$\tilde u_a$}
\FAProp(10.,7.)(15.5,6.5)(0.,){/Straight}{-1}
\FALabel(12.6097,5.68637)[t]{$\tilde \chi_n^{\pm}$}
\FAVert(10.,14.5){0}
\FAVert(10.,7.){0}
\FAVert(15.5,6.5){0}

\end{feynartspicture}
}

\\
\put(2,.2){(a)}
\\[-.3ex]

\hspace*{-.4cm}
\vspace*{-.5cm}
\scalebox{.845}{
\begin{feynartspicture}(15,3)(5,1)
 
\FADiagram{}
\FAProp(0.,15.)(10.,13.)(0.,){/Straight}{1}
\FALabel(5.30398,15.0399)[b]{$u$}
\FAProp(0.,5.)(10.,5.5)(0.,){/Straight}{1}
\FALabel(5.0774,4.18193)[t]{$q_i$}
\FAProp(20.,17.)(15.5,13.5)(0.,){/Straight}{0}
\FALabel(17.4319,15.7962)[br]{$\tilde g$}
\FAProp(20.,10.)(15.5,13.5)(0.,){/ScalarDash}{-1}
\FALabel(18.2216,12.4935)[bl]{$\tilde u_L$}
\FAProp(20.,3.)(10.,5.5)(0.,){/Straight}{-1}
\FALabel(15.3759,5.27372)[b]{$q_i$}
\FAProp(10.,13.)(10.,5.5)(0.,){/Cycles}{0}
\FALabel(8.93,9.25)[r]{$g$}
\FAProp(10.,13.)(15.5,13.5)(0.,){/Straight}{1}
\FALabel(12.8903,12.1864)[t]{$u$}
\FAVert(10.,13.){0}
\FAVert(10.,5.5){0}
\FAVert(15.5,13.5){0}

\FADiagram{}
\FAProp(0.,15.)(10.,14.5)(0.,){/Straight}{1}
\FALabel(5.0774,15.8181)[b]{$u$}
\FAProp(0.,5.)(10.,5.5)(0.,){/Straight}{1}
\FALabel(5.0774,4.18193)[t]{$q_i$}
\FAProp(20.,17.)(10.,14.5)(0.,){/Straight}{0}
\FALabel(14.6847,16.5312)[b]{$\tilde g$}
\FAProp(20.,10.)(10.,10.)(0.,){/ScalarDash}{-1}
\FALabel(15.,11.07)[b]{$\tilde u_L$}
\FAProp(20.,3.)(10.,5.5)(0.,){/Straight}{-1}
\FALabel(15.3759,5.27372)[b]{$q_i$}
\FAProp(10.,14.5)(10.,10.)(0.,){/ScalarDash}{1}
\FALabel(8.93,12.25)[r]{$\tilde u_L$}
\FAProp(10.,5.5)(10.,10.)(0.,){/Cycles}{0}
\FALabel(8.23,7.75)[r]{$g$}
\FAVert(10.,14.5){0}
\FAVert(10.,5.5){0}
\FAVert(10.,10.){0}

\FADiagram{}
\FAProp(0.,15.)(10.,14.5)(0.,){/Straight}{1}
\FALabel(5.0774,15.8181)[b]{$u$}
\FAProp(0.,5.)(10.,5.5)(0.,){/Straight}{1}
\FALabel(5.0774,4.18193)[t]{$q_i$}
\FAProp(20.,17.)(10.,10.)(0.,){/Straight}{0}
\FALabel(17.5821,16.1028)[br]{$\tilde g$}
\FAProp(20.,10.)(10.,14.5)(0.,){/ScalarDash}{-1}
\FALabel(17.2774,10.302)[tr]{$\tilde u_L$}
\FAProp(20.,3.)(10.,5.5)(0.,){/Straight}{-1}
\FALabel(14.6241,3.22628)[t]{$q_i$}
\FAProp(10.,14.5)(10.,10.)(0.,){/Straight}{0}
\FALabel(9.18,12.25)[r]{$\tilde g$}
\FAProp(10.,5.5)(10.,10.)(0.,){/Cycles}{0}
\FALabel(8.23,7.75)[r]{$g$}
\FAVert(10.,14.5){0}
\FAVert(10.,5.5){0}
\FAVert(10.,10.){0}

\FADiagram{}
\FAProp(0.,15.)(10.,14.5)(0.,){/Straight}{1}
\FALabel(5.0774,15.8181)[b]{$u$}
\FAProp(0.,5.)(10.,5.5)(0.,){/Straight}{1}
\FALabel(5.0774,4.18193)[t]{$q_i$}
\FAProp(20.,17.)(10.,5.5)(0.,){/Straight}{0}
\FALabel(17.48,15.02)[br]{$\tilde g$}
\FAProp(20.,10.)(10.,14.5)(0.,){/ScalarDash}{-1}
\FALabel(17.8274,10.002)[tr]{$\tilde u_L$}
\FAProp(20.,3.)(10.,10.)(0.,){/Straight}{-1}
\FALabel(14.5911,5.71019)[tr]{$q_i$}
\FAProp(10.,14.5)(10.,10.)(0.,){/Straight}{0}
\FALabel(9.18,12.25)[r]{$\tilde g$}
\FAProp(10.,5.5)(10.,10.)(0.,){/ScalarDash}{1}
\FALabel(8.93,7.75)[r]{$\tilde q_{ia}$}
\FAVert(10.,14.5){0}
\FAVert(10.,5.5){0}
\FAVert(10.,10.){0}

\FADiagram{}
\FAProp(0.,15.)(10.,14.5)(0.,){/Straight}{1}
\FALabel(5.0774,15.8181)[b]{$u$}
\FAProp(0.,5.)(10.,5.5)(0.,){/Straight}{1}
\FALabel(5.0774,4.18193)[t]{$q_i$}
\FAProp(20.,17.)(15.5,8.)(0.,){/Straight}{0}
\FALabel(18.28,14.97)[br]{$\tilde g$}
\FAProp(20.,10.)(10.,14.5)(0.,){/ScalarDash}{-1}
\FALabel(12.6226,14.248)[bl]{$\tilde u_L$}
\FAProp(20.,3.)(15.5,8.)(0.,){/Straight}{-1}
\FALabel(18.4221,6.0569)[bl]{$q_i$}
\FAProp(10.,14.5)(10.,5.5)(0.,){/Straight}{0}
\FALabel(9.18,10.)[r]{$\tilde g$}
\FAProp(10.,5.5)(15.5,8.)(0.,){/ScalarDash}{1}
\FALabel(12.6,6.3)[tl]{$\tilde q_{ia}$}
\FAVert(10.,14.5){0}
\FAVert(10.,5.5){0}
\FAVert(15.5,8.){0}

\end{feynartspicture}
}

\hspace*{-.4cm}
\vspace*{-.5cm}
\scalebox{.845}{
\begin{feynartspicture}(15,3)(5,1)

\FADiagram{}
\FAProp(0.,15.)(10.,5.5)(0.,){/Straight}{1}
\FALabel(3.19219,13.2012)[bl]{$u$}
\FAProp(0.,5.)(10.,13.)(0.,){/Straight}{1}
\FALabel(3.17617,6.28478)[tl]{$u$}
\FAProp(20.,17.)(16.,13.5)(0.,){/Straight}{0}
\FALabel(17.6239,15.7483)[br]{$\tilde g$}
\FAProp(20.,10.)(16.,13.5)(0.,){/ScalarDash}{-1}
\FALabel(17.4593,11.0635)[tr]{$\tilde u_L$}
\FAProp(20.,3.)(10.,5.5)(0.,){/Straight}{-1}
\FALabel(14.6241,3.22628)[t]{$u$}
\FAProp(10.,5.5)(10.,13.)(0.,){/Cycles}{0}
\FALabel(11.07,9.25)[l]{$g$}
\FAProp(10.,13.)(16.,13.5)(0.,){/Straight}{1}
\FALabel(12.8713,14.3146)[b]{$u$}
\FAVert(10.,5.5){0}
\FAVert(10.,13.){0}
\FAVert(16.,13.5){0}

\FADiagram{}
\FAProp(0.,15.)(10.,5.5)(0.,){/Straight}{1}
\FALabel(3.19219,13.2012)[bl]{$u$}
\FAProp(0.,5.)(10.,14.5)(0.,){/Straight}{1}
\FALabel(3.10297,6.58835)[tl]{$u$}
\FAProp(20.,17.)(10.,14.5)(0.,){/Straight}{0}
\FALabel(14.6847,16.5312)[b]{$\tilde g$}
\FAProp(20.,10.)(10.,10.)(0.,){/ScalarDash}{-1}
\FALabel(15.95,11.07)[b]{$\tilde u_L$}
\FAProp(20.,3.)(10.,5.5)(0.,){/Straight}{-1}
\FALabel(14.6241,3.22628)[t]{$u$}
\FAProp(10.,5.5)(10.,10.)(0.,){/Cycles}{0}
\FALabel(11.07,7.75)[l]{$g$}
\FAProp(10.,14.5)(10.,10.)(0.,){/ScalarDash}{1}
\FALabel(11.07,12.25)[l]{$\tilde u_L$}
\FAVert(10.,5.5){0}
\FAVert(10.,14.5){0}
\FAVert(10.,10.){0}

\FADiagram{}
\FAProp(0.,15.)(10.,14.5)(0.,){/Straight}{1}
\FALabel(5.0774,15.8181)[b]{$u$}
\FAProp(0.,5.)(10.,5.5)(0.,){/Straight}{1}
\FALabel(5.0774,4.18193)[t]{$u$}
\FAProp(20.,17.)(10.,10.)(0.,){/Straight}{0}
\FALabel(15.98,15.02)[br]{$\tilde g$}
\FAProp(20.,10.)(10.,5.5)(0.,){/ScalarDash}{-1}
\FALabel(17.7693,10.1016)[br]{$\tilde u_L$}
\FAProp(20.,3.)(10.,14.5)(0.,){/Straight}{-1}
\FALabel(16.7913,4.81596)[tr]{$u$}
\FAProp(10.,14.5)(10.,10.)(0.,){/Cycles}{0}
\FALabel(8.93,12.25)[r]{$g$}
\FAProp(10.,5.5)(10.,10.)(0.,){/Straight}{0}
\FALabel(9.18,7.75)[r]{$\tilde g$}
\FAVert(10.,14.5){0}
\FAVert(10.,5.5){0}
\FAVert(10.,10.){0}

\FADiagram{}
\FAProp(0.,15.)(10.,14.5)(0.,){/Straight}{1}
\FALabel(5.0774,15.8181)[b]{$u$}
\FAProp(0.,5.)(10.,5.5)(0.,){/Straight}{1}
\FALabel(5.0774,4.18193)[t]{$u$}
\FAProp(20.,17.)(10.,14.5)(0.,){/Straight}{0}
\FALabel(14.6847,16.5312)[b]{$\tilde g$}
\FAProp(20.,10.)(10.,5.5)(0.,){/ScalarDash}{-1}
\FALabel(17.2909,9.58086)[br]{$\tilde u_L$}
\FAProp(20.,3.)(10.,10.)(0.,){/Straight}{-1}
\FALabel(17.3366,3.77519)[tr]{$u$}
\FAProp(10.,14.5)(10.,10.)(0.,){/ScalarDash}{1}
\FALabel(8.93,12.25)[r]{$\tilde u_a$}
\FAProp(10.,5.5)(10.,10.)(0.,){/Straight}{0}
\FALabel(9.18,7.75)[r]{$\tilde g$}
\FAVert(10.,14.5){0}
\FAVert(10.,5.5){0}
\FAVert(10.,10.){0}

\FADiagram{}
\FAProp(0.,15.)(10.,14.5)(0.,){/Straight}{1}
\FALabel(5.0774,15.8181)[b]{$u$}
\FAProp(0.,5.)(10.,5.5)(0.,){/Straight}{1}
\FALabel(5.0774,4.18193)[t]{$u$}
\FAProp(20.,17.)(15.5,14.)(0.,){/Straight}{0}
\FALabel(17.5089,16.1017)[br]{$\tilde g$}
\FAProp(20.,10.)(10.,5.5)(0.,){/ScalarDash}{-1}
\FALabel(13.8364,8.43358)[br]{$\tilde u_L$}
\FAProp(20.,3.)(15.5,14.)(0.,){/Straight}{-1}
\FALabel(18.2486,4.40534)[r]{$u$}
\FAProp(10.,14.5)(10.,5.5)(0.,){/Straight}{0}
\FALabel(9.18,10.)[r]{$\tilde g$}
\FAProp(10.,14.5)(15.5,14.)(0.,){/ScalarDash}{1}
\FALabel(12.8903,15.3136)[b]{$\tilde u_a$}
\FAVert(10.,14.5){0}
\FAVert(10.,5.5){0}
\FAVert(15.5,14.){0}

\end{feynartspicture}
}

\\
\put(0,0){(b)}
\caption{Feynman diagrams for quark radiation via $uq_i \rightarrow \tilde{g} \tilde{u}_L q_i$, with 
$q_i = u,\,d,\,c,\,s,\,\bar{d},\,\bar{c},\,\bar{s}$. Only interference terms from EW (a) and QCD (b) diagrams contribute at $\mathcal{O}(\alpha_s^2\alpha)$. In panel (a), the diagrams of the second row contribute only for $q_i=u,d$ and the diagrams of the third row only for $q_i=\bar{d}$. In panel (b), the diagrams of the second row contribute only for $q_i = u$.}
\label{fig_qqchannels}
}

\FIGURE{
\vspace*{-1cm}
\scalebox{.9}{
\begin{feynartspicture}(12,3)(4,1)

\FADiagram{}
\FAProp(0.,15.)(10.,14.)(0.,){/Cycles}{0}
\FALabel(4.84577,16.2588)[b]{$g$}
\FAProp(0.,5.)(10.,6.)(0.,){/Straight}{1}
\FALabel(5.15423,4.43769)[t]{$u$}
\FAProp(20.,15.)(10.,14.)(0.,){/Straight}{0}
\FALabel(14.8706,15.3135)[b]{$\tilde g$}
\FAProp(20.,5.)(10.,6.)(0.,){/ScalarDash}{-1}
\FALabel(15.1542,6.56231)[b]{$\tilde u_L$}
\FAProp(10.,14.)(10.,6.)(0.,){/Straight}{0}
\FALabel(9.18,10.)[r]{$\tilde g$}
\FAVert(10.,14.){0}
\FAVert(10.,6.){1}

\FADiagram{}
\FAProp(0.,15.)(10.,14.)(0.,){/Cycles}{0}
\FALabel(4.84577,16.2588)[b]{$g$}
\FAProp(0.,5.)(10.,6.)(0.,){/Straight}{1}
\FALabel(5.15423,4.43769)[t]{$u$}
\FAProp(20.,15.)(10.,6.)(0.,){/Straight}{0}
\FALabel(16.98,13.02)[br]{$\tilde g$}
\FAProp(20.,5.)(10.,14.)(0.,){/ScalarDash}{-1}
\FALabel(17.6872,8.20582)[bl]{$\tilde u_L$}
\FAProp(10.,14.)(10.,6.)(0.,){/ScalarDash}{-1}
\FALabel(9.03,10.)[r]{$\tilde u_L$}
\FAVert(10.,14.){0}
\FAVert(10.,6.){1}

\FADiagram{}
\FAProp(0.,15.)(10.,14.)(0.,){/Cycles}{0}
\FALabel(4.84577,16.2588)[b]{$g$}
\FAProp(0.,5.)(10.,6.)(0.,){/Straight}{1}
\FALabel(5.15423,4.43769)[t]{$u$}
\FAProp(20.,15.)(10.,6.)(0.,){/Straight}{0}
\FALabel(16.98,13.02)[br]{$\tilde g$}
\FAProp(20.,5.)(10.,14.)(0.,){/ScalarDash}{-1}
\FALabel(17.6872,8.20582)[bl]{$\tilde u_L$}
\FAProp(10.,14.)(10.,6.)(0.,){/ScalarDash}{-1}
\FALabel(9.03,10.)[r]{$\tilde u_a$}
\FAVert(10.,6.){0}
\FAVert(10.,14.){1}

\FADiagram{ }
\FAProp(0.,15.)(10.,14.)(0.,){/Cycles}{0}
\FALabel(5.22388,16.2588)[b]{$g$}
\FAProp(0.,5.)(10.,6.)(0.,){/Straight}{1}
\FALabel(5.15423,4.43769)[t]{$u$}
\FAProp(20.,15.)(10.,6.)(0.,){/Straight}{0}
\FALabel(16.98,13.02)[br]{$\tilde g$}
\FAProp(20.,5.)(10.,14.)(0.,){/ScalarDash}{-1}
\FALabel(18.1872,7.70582)[bl]{$\tilde u_L$}
\FAProp(10.,10.)(10.,14.)(0.,){/ScalarDash}{1}
\FALabel(8.93,12.)[r]{$\tilde u_L$}
\FAProp(10.,10.)(10.,6.)(0.,){/ScalarDash}{-1}
\FALabel(8.93,8.)[r]{$\tilde u_a$}
\FAVert(10.,14.){0}
\FAVert(10.,6.){0}
\FAVert(10.,10.){1}

\end{feynartspicture}
}

\vspace*{-.7cm}
\scalebox{.9}{
\begin{feynartspicture}(9,3)(3,1)

\FADiagram{}
\FAProp(0.,15.)(6.,10.)(0.,){/Cycles}{0}
\FALabel(2.48771,11.7893)[tr]{$g$}
\FAProp(0.,5.)(6.,10.)(0.,){/Straight}{1}
\FALabel(3.51229,6.78926)[tl]{$u$}
\FAProp(20.,15.)(14.,10.)(0.,){/Straight}{0}
\FALabel(16.6478,13.0187)[br]{$\tilde g$}
\FAProp(20.,5.)(14.,10.)(0.,){/ScalarDash}{-1}
\FALabel(17.5123,8.21074)[bl]{$\tilde u_L$}
\FAProp(6.,10.)(14.,10.)(0.,){/Straight}{1}
\FALabel(10.,8.93)[t]{$u$}
\FAVert(6.,10.){0}
\FAVert(14.,10.){1}

\FADiagram{}
\FAProp(0.,15.)(6.,10.)(0.,){/Cycles}{0}
\FALabel(2.48771,11.7893)[tr]{$g$}
\FAProp(0.,5.)(6.,10.)(0.,){/Straight}{1}
\FALabel(3.51229,6.78926)[tl]{$u$}
\FAProp(20.,15.)(14.,10.)(0.,){/Straight}{0}
\FALabel(16.6478,13.0187)[br]{$\tilde g$}
\FAProp(20.,5.)(14.,10.)(0.,){/ScalarDash}{-1}
\FALabel(17.5123,8.21074)[bl]{$\tilde u_L$}
\FAProp(6.,10.)(14.,10.)(0.,){/Straight}{1}
\FALabel(10.,8.93)[t]{$u$}
\FAVert(14.,10.){0}
\FAVert(6.,10.){1}

\FADiagram{ }
\FAProp(0.,15.)(6.,10.)(0.,){/Cycles}{0}
\FALabel(2.48771,11.7893)[tr]{$g$}
\FAProp(0.,5.)(6.,10.)(0.,){/Straight}{1}
\FALabel(3.51229,6.78926)[tl]{$u$}
\FAProp(20.,15.)(14.,10.)(0.,){/Straight}{0}
\FALabel(16.6478,13.0187)[br]{$\tilde g$}
\FAProp(20.,5.)(14.,10.)(0.,){/ScalarDash}{-1}
\FALabel(17.5123,8.21074)[bl]{$\tilde u_L$}
\FAProp(10.,10.)(6.,10.)(0.,){/Straight}{-1}
\FALabel(8.,8.93)[t]{$u$}
\FAProp(10.,10.)(14.,10.)(0.,){/Straight}{1}
\FALabel(12.,11.07)[b]{$u$}
\FAVert(6.,10.){0}
\FAVert(14.,10.){0}
\FAVert(10.,10.){1}

\end{feynartspicture}
}

\caption{Counter term diagrams for the process $g\, u \rightarrow \tilde{g} \, \tilde{u}_L$.}
\label{fig_counterterms}
}

\clearpage

\bibliographystyle{JHEP}
\bibliography{references}

\providecommand{\href}[2]{#2}\begingroup\raggedright\begin{thebibliography}{10}

\bibitem{Wess}
J.~Wess and B.~Zumino, {\it {Supergauge Transformations in Four-Dimensions}},
  {\em Nucl. Phys.} {\bf B70} (1974) 39--50.

\bibitem{Wess2}
D.~V. Volkov and V.~P. Akulov, {\it {Is the Neutrino a Goldstone Particle?}},
  {\em Phys. Lett.} {\bf B46} (1973) 109--110.

\bibitem{MuonEX}
{\bf Muon g-2} Collaboration, G.~W. Bennett {\em et.~al.}, {\it {Measurement of
  the positive muon anomalous magnetic moment to 0.7-ppm}},  {\em Phys. Rev.
  Lett.} {\bf 89} (2002) 101804,
  [\href{http://xxx.lanl.gov/abs/hep-ex/0208001}{{\tt hep-ex/0208001}}].

\bibitem{MuonEX2}
{\bf Muon g-2} Collaboration, G.~W. Bennett {\em et.~al.}, {\it {Measurement of
  the negative muon anomalous magnetic moment to 0.7-ppm}},  {\em Phys. Rev.
  Lett.} {\bf 92} (2004) 161802,
  [\href{http://xxx.lanl.gov/abs/hep-ex/0401008}{{\tt hep-ex/0401008}}].

\bibitem{WMAP}
{\bf WMAP} Collaboration, J.~Dunkley {\em et.~al.}, {\it {Five-Year Wilkinson
  Microwave Anisotropy Probe (WMAP) Observations: Likelihoods and Parameters
  from the WMAP data}},  \href{http://xxx.lanl.gov/abs/0803.0586}{{\tt
  arXiv:0803.0586}}.

\bibitem{MSSM}
H.~P. Nilles, {\it {Supersymmetry, Supergravity and Particle Physics}},  {\em
  Phys. Rept.} {\bf 110} (1984) 1--162.

\bibitem{MSSM2}
H.~E. Haber and G.~L. Kane, {\it {The Search for Supersymmetry: Probing Physics
  Beyond the Standard Model}},  {\em Phys. Rept.} {\bf 117} (1985) 75--263.

\bibitem{MSSM3}
R.~Barbieri, {\it {Looking Beyond the Standard Model: The Supersymmetric
  Option}},  {\em Riv. Nuovo Cim.} {\bf 11N4} (1988) 1--45.

\bibitem{Ellis:2007fu}
J.~R. Ellis, S.~Heinemeyer, K.~A. Olive, A.~M. Weber, and G.~Weiglein, {\it
  {The Supersymmetric Parameter Space in Light of B-physics Observables and
  Electroweak Precision Data}},  {\em JHEP} {\bf 08} (2007) 083,
  [\href{http://xxx.lanl.gov/abs/0706.0652}{{\tt arXiv:0706.0652}}].

\bibitem{Buchmueller:2007zk}
O.~Buchm$\ddot{\mathrm u}$ller {\em et.~al.}, {\it {Prediction for the Lightest
  Higgs Boson Mass in the CMSSM using Indirect Experimental Constraints}},
  {\em Phys. Lett.} {\bf B657} (2007) 87--94,
  [\href{http://xxx.lanl.gov/abs/0707.3447}{{\tt arXiv:0707.3447}}].

\bibitem{Buchmueller}
O.~Buchm$\ddot{\mathrm u}$ller {\em et.~al.}, {\it {Predictions for
  Supersymmetric Particle Masses in the CMSSM using Indirect Experimental and
  Cosmological Constraints}},  \href{http://xxx.lanl.gov/abs/0808.4128}{{\tt
  arXiv:0808.4128}}.

\bibitem{Shamim:2007yy}
{\bf D0} Collaboration, M.~Shamim, {\it {Searches for Squarks and Gluinos with
  D0 Detector}},  \href{http://xxx.lanl.gov/abs/0710.2897}{{\tt
  arXiv:0710.2897}}.

\bibitem{D'Onofrio:2007pq}
{\bf CDF - Run II} Collaboration, M.~D'Onofrio, {\it {Inclusive Search for
  Squarks and Gluinos Production at CDF}},
  \href{http://xxx.lanl.gov/abs/0710.5114}{{\tt arXiv:0710.5114}}.

\bibitem{CDFnote9229}
{CDF Collaboration}, {\it {Search for Gluino and Squark Produkction in
  Multijets Plus Missing $E_T$ Final States}},  {\em CDF note} {\bf 9229}
  [\href{http://xxx.lanl.gov/abs/{See:
  http://www-cdf.fnal.gov/physics/exotic/r2a/20080214.squark gluino/squark
  gluino.html}}{{\tt {See:
  http://www-cdf.fnal.gov/physics/exotic/r2a/20080214.squark gluino/squark
  gluino.html}}}].

\bibitem{SUSY@lhc}
{\bf ATLAS} Collaboration, U.~De~Sanctis, {\it {Supersymmetry searches with
  ATLAS detector at LHC}},  {\em Nuovo Cim.} {\bf 121B} (2006) 761--770.

\bibitem{SUSY@lhc2}
{\bf CMS} Collaboration, A.~Tricomi, {\it {SUSY searches in early CMS data}},
  {\em J. Phys. Conf. Ser.} {\bf 110} (2008) 062026.

\bibitem{Randall}
L.~Randall and D.~Tucker-Smith, {\it {Dijet Searches for Supersymmetry at the
  LHC}},  \href{http://xxx.lanl.gov/abs/0806.1049}{{\tt arXiv:0806.1049}}.

\bibitem{Tree}
P.~R. Harrison and C.~H. Llewellyn~Smith, {\it {Hadroproduction of
  Supersymmetric Particles}},  {\em Nucl. Phys.} {\bf B213} (1983) 223.

\bibitem{Tree2}
G.~L. Kane and J.~P. Leveille, {\it {Experimental Constraints on Gluino Masses
  and Supersymmetric Theories}},  {\em Phys. Lett.} {\bf B112} (1982) 227.

\bibitem{Tree3}
E.~Reya and D.~P. Roy, {\it {Supersymmetric particle production at p anti-p
  collider energies}},  {\em Phys. Rev.} {\bf D32} (1985) 645.

\bibitem{Tree4}
S.~Dawson, E.~Eichten, and C.~Quigg, {\it {Search for Supersymmetric Particles
  in Hadron - Hadron Collisions}},  {\em Phys. Rev.} {\bf D31} (1985) 1581.

\bibitem{Tree5}
H.~Baer and X.~Tata, {\it {Component formulae for hadronproduction of
  left-handed and right-handed squarks}},  {\em Phys. Lett.} {\bf B160} (1985)
  159.

\bibitem{Beenakker:1996ch}
W.~Beenakker, R.~H$\ddot{\mathrm o}$pker, M.~Spira, and P.~M. Zerwas, {\it
  {Squark and gluino production at hadron colliders}},  {\em Nucl. Phys.} {\bf
  B492} (1997) 51--103, [\href{http://xxx.lanl.gov/abs/hep-ph/9610490}{{\tt
  hep-ph/9610490}}].

\bibitem{Beenakker1997}
W.~Beenakker, M.~Kr$\ddot{\mathrm a}$mer, T.~Plehn, M.~Spira, and P.~M. Zerwas,
  {\it {Stop production at hadron colliders}},  {\em Nucl. Phys.} {\bf B515}
  (1998) 3--14, [\href{http://xxx.lanl.gov/abs/hep-ph/9710451}{{\tt
  hep-ph/9710451}}].

\bibitem{Prospino}
W.~Beenakker, R.~H$\ddot{\mathrm o}$pker, and M.~Spira, {\it {PROSPINO: A
  program for the PROduction of Supersymmetric Particles In Next-to-leading
  Order QCD}},  \href{http://xxx.lanl.gov/abs/hep-ph/9611232}{{\tt
  hep-ph/9611232}}.

\bibitem{Hollik:2007wf}
W.~Hollik, M.~Kollar, and M.~K. Trenkel, {\it {Hadronic production of
  top-squark pairs with electroweak NLO contributions}},  {\em JHEP} {\bf 02}
  (2008) 018, [\href{http://xxx.lanl.gov/abs/0712.0287}{{\tt
  arXiv:0712.0287}}].

\bibitem{Beccaria:2008mi}
M.~Beccaria, G.~Macorini, L.~Panizzi, F.~M. Renard, and C.~Verzegnassi, {\it
  {Stop-antistop and sbottom-antisbottom production at LHC: a one-loop search
  for model parameters dependence}},
  \href{http://xxx.lanl.gov/abs/0804.1252}{{\tt arXiv:0804.1252}}.

\bibitem{Hollik:2008}
W.~Hollik and E.~Mirabella, {\it {Squark anti-squark pair production at the
  LHC: the electroweak contribution}},
  \href{http://xxx.lanl.gov/abs/0806.1433}{{\tt arXiv:0806.1433}}.

\bibitem{Bornhauser:2007bf}
S.~Bornhauser, M.~Drees, H.~K. Dreiner, and J.~S. Kim, {\it {Electroweak
  Contributions to Squark Pair Production at the LHC}},  {\em Phys. Rev.} {\bf
  D76} (2007) 095020, [\href{http://xxx.lanl.gov/abs/0709.2544}{{\tt
  arXiv:0709.2544}}].

\bibitem{Bozzi:2007me}
G.~Bozzi, B.~Fuks, B.~Herrmann, and M.~Klasen, {\it {Squark and gaugino
  hadroproduction and decays in non- minimal flavour violating supersymmetry}},
   {\em Nucl. Phys.} {\bf B787} (2007) 1--54,
  [\href{http://xxx.lanl.gov/abs/0704.1826}{{\tt arXiv:0704.1826}}].

\bibitem{MassR1}
F.~E. Paige, {\it {SUSY signatures in ATLAS at LHC}},
  \href{http://xxx.lanl.gov/abs/hep-ph/0307342}{{\tt hep-ph/0307342}}.

\bibitem{MassR2}
M.~Chiorboli and A.~Tricomi, {\it {Squark and gluino reconstruction in CMS}}, .
  CMS-NOTE-2004-029.

\bibitem{Kawagoe:2004rz}
K.~Kawagoe, M.~M. Nojiri, and G.~Polesello, {\it {A new SUSY mass
  reconstruction method at the CERN LHC}},  {\em Phys. Rev.} {\bf D71} (2005)
  035008, [\href{http://xxx.lanl.gov/abs/hep-ph/0410160}{{\tt
  hep-ph/0410160}}].

\bibitem{Kublbeck:1990xc}
J.~K$\ddot{\mathrm u}$blbeck, M.~B$\ddot{\mathrm o}$hm, and A.~Denner, {\it
  {FeynArts: Computer Algebraic Generation of Feynman Graphs and Amplitudes}},
  {\em Comput. Phys. Commun.} {\bf 60} (1990) 165--180.

\bibitem{Hahn:2000kx}
T.~Hahn, {\it {Generating Feynman diagrams and amplitudes with FeynArts 3}},
  {\em Comput. Phys. Commun.} {\bf 140} (2001) 418--431,
  [\href{http://xxx.lanl.gov/abs/hep-ph/0012260}{{\tt hep-ph/0012260}}].

\bibitem{Hahn:2001rv}
T.~Hahn and C.~Schappacher, {\it {The implementation of the minimal
  supersymmetric standard model in FeynArts and FormCalc}},  {\em Comput. Phys.
  Commun.} {\bf 143} (2002) 54--68,
  [\href{http://xxx.lanl.gov/abs/hep-ph/0105349}{{\tt hep-ph/0105349}}].

\bibitem{Hahn:1998yk}
T.~Hahn and M.~Perez-Victoria, {\it {Automatized one-loop calculations in four
  and D dimensions}},  {\em Comput. Phys. Commun.} {\bf 118} (1999) 153--165,
  [\href{http://xxx.lanl.gov/abs/hep-ph/9807565}{{\tt hep-ph/9807565}}].

\bibitem{Hahn:2006qw}
T.~Hahn and M.~Rauch, {\it {News from FormCalc and LoopTools}},  {\em Nucl.
  Phys. Proc. Suppl.} {\bf 157} (2006) 236--240,
  [\href{http://xxx.lanl.gov/abs/hep-ph/0601248}{{\tt hep-ph/0601248}}].

\bibitem{Denner:1991kt}
A.~Denner, {\it {Techniques for calculation of electroweak radiative
  corrections at the one loop level and results for W physics at LEP-200}},
  {\em Fortschr. Phys.} {\bf 41} (1993) 307--420,
  [\href{http://xxx.lanl.gov/abs/0709.1075}{{\tt arXiv:0709.1075}}].

\bibitem{Hollik:2003jj}
W.~Hollik and H.~Rzehak, {\it {The sfermion mass spectrum of the MSSM at the
  one-loop level}},  {\em Eur. Phys. J.} {\bf C32} (2003) 127--133,
  [\href{http://xxx.lanl.gov/abs/hep-ph/0305328}{{\tt hep-ph/0305328}}].

\bibitem{Heinemeyer:2004xw}
S.~Heinemeyer, W.~Hollik, H.~Rzehak, and G.~Weiglein, {\it {High-precision
  predictions for the MSSM Higgs sector at O(alpha(b) alpha(s))}},  {\em Eur.
  Phys. J.} {\bf C39} (2005) 465--481,
  [\href{http://xxx.lanl.gov/abs/hep-ph/0411114}{{\tt hep-ph/0411114}}].

\bibitem{Harris:2001sx}
B.~W. Harris and J.~F. Owens, {\it {The two cutoff phase space slicing
  method}},  {\em Phys. Rev.} {\bf D65} (2002) 094032,
  [\href{http://xxx.lanl.gov/abs/hep-ph/0102128}{{\tt hep-ph/0102128}}].

\bibitem{Catani:1996jh}
S.~Catani and M.~H. Seymour, {\it {The Dipole Formalism for the Calculation of
  QCD Jet Cross Sections at Next-to-Leading Order}},  {\em Phys. Lett.} {\bf
  B378} (1996) 287--301, [\href{http://xxx.lanl.gov/abs/hep-ph/9602277}{{\tt
  hep-ph/9602277}}].

\bibitem{Catani:1996jh2}
S.~Catani and M.~H. Seymour, {\it {A general algorithm for calculating jet
  cross sections in NLO QCD}},  {\em Nucl. Phys.} {\bf B485} (1997) 291--419,
  [\href{http://xxx.lanl.gov/abs/hep-ph/9605323}{{\tt hep-ph/9605323}}].

\bibitem{Catani:1996jh3}
S.~Catani, S.~Dittmaier, M.~H. Seymour, and Z.~Trocsanyi, {\it {The dipole
  formalism for next-to-leading order QCD calculations with massive partons}},
  {\em Nucl. Phys.} {\bf B627} (2002) 189--265,
  [\href{http://xxx.lanl.gov/abs/hep-ph/0201036}{{\tt hep-ph/0201036}}].

\bibitem{Dittmaier:1999mb}
S.~Dittmaier, {\it {A general approach to photon radiation off fermions}},
  {\em Nucl. Phys.} {\bf B565} (2000) 69--122,
  [\href{http://xxx.lanl.gov/abs/hep-ph/9904440}{{\tt hep-ph/9904440}}].

\bibitem{Dittmaier:2008md}
S.~Dittmaier, A.~Kabelschacht, and T.~Kasprzik, {\it {Polarized QED splittings
  of massive fermions and dipole subtraction for non-collinear-safe
  observables}},  \href{http://xxx.lanl.gov/abs/0802.1405}{{\tt
  arXiv:0802.1405}}.

\bibitem{Baur:1998kt}
U.~Baur, S.~Keller, and D.~Wackeroth, {\it {Electroweak radiative corrections
  to W boson production in hadronic collisions}},  {\em Phys. Rev.} {\bf D59}
  (1999) 013002, [\href{http://xxx.lanl.gov/abs/hep-ph/9807417}{{\tt
  hep-ph/9807417}}].

\bibitem{Dittmaier:2001ay}
S.~Dittmaier and M.~Kr$\ddot{\mathrm a}$mer, {\it {Electroweak radiative
  corrections to W-boson production at hadron colliders}},  {\em Phys. Rev.}
  {\bf D65} (2002) 073007, [\href{http://xxx.lanl.gov/abs/hep-ph/0109062}{{\tt
  hep-ph/0109062}}].

\bibitem{AguilarSaavedra:2005pw}
J.~A. Aguilar-Saavedra {\em et.~al.}, {\it {Supersymmetry parameter analysis:
  SPA convention and project}},  {\em Eur. Phys. J.} {\bf C46} (2006) 43--60,
  [\href{http://xxx.lanl.gov/abs/hep-ph/0511344}{{\tt hep-ph/0511344}}].

\bibitem{Wackeroth:1996hz}
D.~Wackeroth and W.~Hollik, {\it {Electroweak radiative corrections to resonant
  charged gauge boson production}},  {\em Phys. Rev.} {\bf D55} (1997)
  6788--6818, [\href{http://xxx.lanl.gov/abs/hep-ph/9606398}{{\tt
  hep-ph/9606398}}].

\bibitem{Diener:2003ss}
K.~P.~O. Diener, S.~Dittmaier, and W.~Hollik, {\it {Electroweak radiative
  corrections to deep-inelastic neutrino scattering: Implications for NuTeV?}},
   {\em Phys. Rev.} {\bf D69} (2004) 073005,
  [\href{http://xxx.lanl.gov/abs/hep-ph/0310364}{{\tt hep-ph/0310364}}].

\bibitem{Allanach:2001kg}
B.~C. Allanach, {\it {SOFTSUSY: A C++ program for calculating supersymmetric
  spectra}},  {\em Comput. Phys. Commun.} {\bf 143} (2002) 305--331,
  [\href{http://xxx.lanl.gov/abs/hep-ph/0104145}{{\tt hep-ph/0104145}}].

\bibitem{CDFtopmass}
See: {http://www-cdf.fnal.gov/physics/new/top/top.html}.

\end{thebibliography}\endgroup

\end{document}